\documentclass[11 pt]{article}
\pdfoutput=1 
\usepackage{jheppub} 
\usepackage{orcidlink}
\usepackage{lscape}
\usepackage{multicol}
\usepackage{feynman}
\usepackage{capt-of}
\usepackage{amsmath} % Added for split environment support
\usepackage{xcolor} % For \textcolor support
\usepackage{caption}
\usepackage{subcaption,longtable,stmaryrd}
\usepackage{slashed}
\usepackage{bigints}
\usepackage{cancel}
\usepackage{multicol}
\usepackage{blindtext}
\usepackage{tikz}
\usepackage{enumitem}
\usepackage[customcolors,shade]{hf-tikz}
\usepackage{graphicx}
\usepackage{cancel}
\usepackage{empheq}
\usepackage{tabularx}
\PassOptionsToPackage{table, svgnames, x11names, dvipsnames, hyperref}{xcolor}
\usepackage{xcolor}
\usetikzlibrary{decorations.pathmorphing}
\definecolor{phiColor}{HTML}{9A081B}
\definecolor{sigmaColor}{HTML}{417505}

\DeclareSymbolFont{usualmathcal}{OMS}{cmsy}{m}{n}
\DeclareMathAlphabet\boldsymbolcal{OMS}{cmsy}{b}{n}
\DeclareSymbolFontAlphabet{\mathcal}{usualmathcal}
\DeclareSymbolFont{rmlargesymbols}{OMX}{mdbch}{m}{n}
\DeclareMathSymbol{\rmointop}{\mathop}{rmlargesymbols}{72}
\definecolor{mygray}{gray}{0.5}

\title{\boldmath {Love beyond Einstein: Metric reconstruction and Love number in quadratic gravity using WEFT}}
\author[]{Arpan Bhattacharyya,}
\author[]{Saptaswa Ghosh,}
\author[]{Naman Kumar,}
\author[]{Shailesh Kumar,}
\author[]{Sounak Pal}
\affiliation[]{\it Indian Institute of Technology, Gandhinagar, Gujarat-382055, India}
\emailAdd{abhattacharyya@iitgn.ac.in}
\emailAdd{saptaswaghosh@iitgn.ac.in}
\emailAdd{namankumar5954@gmail.com}
\emailAdd{shailesh.k@iitgn.ac.in}
\emailAdd{palsounak@iitgn.ac.in}

\abstract{
We study tidal Love numbers of static black holes in four-dimensional quadratic theory of gravity, extending the result of GR. We use worldline effective field theory (WEFT) methods to compute metric perturbations from one-point functions, treating the higher-derivative terms perturbatively. We show that insertions of scalar fields on the worldline induce non-zero tidal tails, and the corresponding Love number displays no RG running. The same conclusion holds for the insertions of tensor fields. Furthermore, for scalar dipole perturbations, we derive a Yukawa-deformed Frobenius solution and match the asymptotic behavior to fix the UV charge, finding agreement with EFT predictions of Wilson coefficients.  Our work demonstrates that quadratic higher-curvature corrections induce non-zero but scale-independent tidal responses, offering a robust EFT framework to test deviations from GR in gravitational wave observations.
}
\makeatletter
\begin{document}
\maketitle
\flushbottom
%\newpage
\section{Introduction} \label{intro}

Recent advances in gravitational wave (GW) astronomy have yielded direct observations of binary black hole and neutron star mergers, offering valuable insights into theoretical modelling and deepening our understanding of strong-gravity regimes \cite{LIGOScientific:2016aoc, LIGOScientific:2016lio, LIGOScientific:2016sjg, LIGOScientific:2017bnn}. These observations not only act as a rigorous test of general relativity (GR) but also offer invaluable frameworks for probing intricate and foundational aspects of gravitational physics. Among the various observable quantities that arise within this line of investigation, a set of key variables is the tidal deformability parameters, commonly known as \textit{Love numbers}, which broadly describe how a massive compact object such as a black hole or star reacts under an external tidal field, resulting in modifications to gravitational waveforms. In other words, tidal deformability measures the compact objects' multipole moments in response to an external tidal field, with the Love number representing the real part of this response \cite{PhysRevD.34.991,Hinderer:2007mb, Binnington:2009bb, Damour:2009vw}. In the case of neutron stars, Love numbers provide a means of investigating the neutron star equation of state that reflects the influence of the star's internal structure on the waveform \cite{Vines:2011ud, Flanagan:2007ix}. For black holes in GR, it has been shown that Love numbers identically vanish in four-dimensional spacetime for both Schwarzschild and Kerr, giving interesting theoretical implications in relation to the no-hair theorem \cite{Landry:2015zfa, LeTiec:2020bos, Chia:2020yla, Charalambous:2021mea, Hui:2021vcv, BenAchour:2022uqo}. Along the same line, in the context of point-particle worldline effective field theory (EFT), Love numbers arise as Wilson coefficients, and their vanishing implies a fine-tuning problem analogous to the cosmological constant problem \cite{Porto:2016zng}. However, the behavior of Love numbers for static black holes in higher dimensions is considerably more nuanced in comparison to the four-dimensional case, where Love numbers can vanish, remain order one constant, or exhibit a classical renormalization group flow, depending on factors such as the multipole index ($\ell$) and spacetime dimensions, further highlighting the intricacies and richness of the dynamics involved \cite{Ivanov:2022hlo}. \vspace{1mm}

As the ongoing progress in the line of black hole Love numbers has put forward significant literature, a number of crucial issues still remain open, primarily concerning the ambiguity in defining Love numbers in GR and beyond, as well as challenges in extracting these from black hole perturbation theory. In this line of endeavor, the EFT description turns out to be the most prominent and useful approach to systematically address and resolve such problems \cite{Ivanov:2022hlo}. The point-particle EFT framework is a modern and emerging field that has become an essential tool for modelling the dynamics of black hole binaries, testing GR, and beyond, with considerable attention motivated by recent GW detections from binary black hole mergers. In the EFT formalism, each compact object in an inspiralling binary, such as a black hole or a neutron star, is approximated as a point particle whose internal structure, reflected in tidal interactions and multipolar deformations, is systematically incorporated through higher-order corrections along the worldlines. \textcolor{black}{Thus, the finite-size effects of compact objects in leading order are encoded in worldline operators of quadratic curvature \cite{Ivanov:2022hlo, Ivanov:2022qqt}}. These terms lead to Wilson coefficients, which, in the Newtonian limit, reduce to the classical Love numbers, characterizing the tidal deformability. However, vanishing Love numbers in the case of black holes imply the necessity of strong fine-tuning for worldline EFT, which further suggests a potential requirement for understanding the dynamics of compact objects in GR. \vspace{1mm}

The concept of Love numbers was first outlined within the framework of Newtonian gravity to estimate how an extended object, such as a star described by a fluid stress-tensor, gets deformed in response to the external gravitational field of another object. Let us say that we consider a spherically symmetric star that maintains the same symmetry in the absence of any external tidal field or perturbation; however, when a small object exerts a tidal field on the star, its geometry no longer remains spherically symmetric; instead, it gets deformed. Such deformations in the star can be characterized by internal multipole moments ($I_L$) that, under linear response theory, are proportional to the external perturbations. These induced moments depend on the structure and equation of state of the star and introduce measurable changes in the total gravitational potential around it. The quantity induced mass multipoles that, under the external tidal field, reduce the total potential in the following form \cite{Poisson_Will_2014}:
\begin{align}\label{potential2}
    \varphi_{p}(\textbf{x}) =  - \sum_{\ell = 2}\mathcal{E}_{L}n^{L}r^{\ell}\frac{(\ell-1)!}{\ell !} \Big(1 + k_{\ell}\frac{R^{2\ell+1}}{r^{2\ell+1}} \Big),
\end{align}
where, $n^{L}=n^{i_1}...n^{i_\ell}$ with unit direction vectors $n^{i}=x^{i}/r$, $L$ denotes the multi-index and $\mathcal{E}_{L}$ is the multipole moments. Here, $R$ represents the radius or size of the star, introduced for dimensional consistency. In the above expression, within Newtonian theory, the first term is the contribution of the source, whereas the second term indicates the response, and the coefficient of $r^{-(1+\ell)}$ gives the Love number. Note that we have subtracted the monopole term. Given that our discussion centres on black holes, we substitute $R$ with the Schwarzschild radius ($r_s$) in the analysis that follows. \vspace{1mm}

\noindent
\textbf{Love number in GR:} In GR, the idea of the Love number is notably more subtle than in the case of Newtonian theory. In the relativistic setting, the challenge arises in formulating a definition that maintains physical relevance while remaining independent of coordinate choices. The notion requires a methodology that preserves gauge invariance and transitions smoothly to the Newtonian limit, as discussed above in Eq. (\ref{potential2}). One of the approaches to address this in GR is by generalizing the classical expressions or constructing gauge-invariant quantities from metric perturbations in a locally asymptotic rest frame of the body. The related quantity, coming from the temporal component of the metric $h^{\textup{GR}}_{00} = \frac{1}{2}(g_{00}-1)$, can be written as \cite{Kol:2011vg}

\begin{align}\label{GR pot}
    h^{\textup{GR}}_{00}(\textbf{x}) = \sum_{\ell = 2} \mathcal{E}_{L}n^{L}r^{\ell} \frac{(\ell - 1)!}{\ell !} \Big[\Big\lbrace 1+c_{1}\Big(\frac{r_s}{r}\Big)+... \Big\rbrace + k_{\ell}\Big(\frac{R}{r} \Big)^{1+2\ell} \Big\lbrace 1+b_{1} \Big(\frac{r_s}{r}\Big)+...\Big\rbrace\Big].
\end{align}
In physical scenarios of interest, such as compact binaries, the external source responsible for the tidal field is typically a companion object. In such systems, the ratio $r_s/r$, where $r_s$ is the Schwarzschild radius of the object and $r$ is the orbital separation, scales as $v^{2}$ with $v$ being the characteristic orbital velocity. This scaling aligns naturally with the post-Newtonian (PN) approximation, and consequently, an expansion in the small parameter $r_s/r$ is conventionally referred to as a PN expansion. We must add that the terms in the first curly bracket present PN corrections to the source (termed source series) that arise due to the nonlinear nature of gravity. The terms in the second curly bracket of (\ref{GR pot}), including their coefficient, indicate the response contribution (termed response series) along with the associated PN correction. ($c_{1}, b_{1}$) are the coefficients that can be calculated explicitly. $k_{\ell}$ denotes the Love number for the multipolar configuration. However, a potential challenge with the discussion stated above lies in an ambiguity that arises when both the series source and response are not clearly separated, leading to an overlap between their respective series expansions \cite{Kol:2011vg, Charalambous:2021mea, LeTiec:2020spy}. In such cases, the coefficient accompanying the term $r^{-(1+\ell)}$ does not purely characterize the induced response but may also include contributions from the source itself that complicate the extraction of the Love number. One commonly used method to resolve the issue involves analytically continuing the multipole index $\ell$ from integers to real values \cite{Kol:2011vg, Chia:2020yla, Charalambous:2021mea, LeTiec:2020spy}, leveraging the fact that the source and response terms typically do not overlap for non-integer $\ell$. While this trick works in various cases, especially for black hole perturbations \cite{Kol:2011vg}, its validity remains uncertain due to the lack of a rigorous theoretical foundation, particularly in modified gravity theories \cite{Cardoso:2017cfl, Cai:2019npx,Nair:2024mya}. This highlights the need for a more systematic and well-defined framework independent of analytic continuation. The second problem relates to the gauge invariance, as Eq. (\ref{GR pot}) is written in a certain coordinate system, making it unclear whether the resulting definition is truly gauge-invariant \cite{Gralla:2017djj}. The next issue is connected to logarithmic corrections, where the structure of the field solutions can be more complex or can simply deviate from the form in Eq. (\ref{GR pot}), often involving logarithmic terms \cite{Kol:2011vg, Charalambous:2021mea}, which makes it harder to define Love numbers consistently without using the EFT approach. Another aspect deals with the interpretation of the coefficient $r^{-(1+\ell)}$ that leads to two distinct notions of the tidal response, i.e., conservative and dissipative: $k_{\ell}(\omega) \equiv k_{\ell}+i\nu_{\ell}\omega r_{s}+\mathcal{O}(\omega^{2}r_{s}^{2})$; where $\nu_{\ell}$ signifies the dissipative response coefficient and $\omega$ is frequency in the rest frame of the black hole. The related details have been extensively and nicely discussed in \cite{Fang:2005qq, Landry:2015zfa, LeTiec:2020bos, Chia:2020yla, LeTiec:2020spy, Poisson:2014gka, Poisson:2020vap, Kehagias:2024rtz, Combaluzier-Szteinsznaider:2024sgb,Gounis:2024hcm,DeLuca:2023mio,Riva:2023rcm,DeLuca:2022tkm} (see \cite{Perry:2023wmm} for discussion on dynamical Love number) and more recently \cite{Ivanov:2024sds,Ivanov:2025ozg,Caron-Huot:2025tlq}. 
%A parametrized classical framework has also been developed to compute tidal Love numbers and tidal dissipation numbers for static, spherically symmetric black holes, using a master equation of the Regge-Wheeler/Zerilli form with small correction to effective potential, connecting perturbative dynamics to GW observables in a theory-agnostic way \cite{kobayashi2025parametrizedtidaldissipationnumbers, Katagiri:2023umb}, including Love numbers with alternative approaches for non-GR theories \cite{Cano:2025zyk}.
\vspace{1mm}
\noindent

In the EFT framework, gravitational corrections are treated as perturbations around a flat background \cite{Goldberger:2004jt, Kol:2007rx, Kol:2009mj} (see also \cite{Goldberger:2020fot} for non-conservative effects) with post-Newtonian corrections representing graviton interactions with the external source. These corrections ($c_{2\ell+1}$) are computed explicitly, allowing for a matching of the external field profile to black hole perturbation theory results. This method bypasses the source/response ambiguity and avoids the need for analytic continuation of the multipole index. To extract Love numbers, a direct comparison of gauge-invariant observables, like gravitational wave scattering cross sections, is made between EFT and general relativity calculations \cite{Goldberger:2004jt, Porto:2016pyg, Ivanov:2022qqt}. In addition, to resolve the potential challenges associated with coordinate dependence, it is crucial to maintain consistency in the gauge choices used for one-point function calculations on both the EFT and UV sides. Additionally, logarithmic corrections can be seamlessly included within the EFT framework, where they are understood as a classical renormalization group (RG) running effect \cite{Kol:2011vg}. Thus, EFT plays an efficient and important role in determining key physical insights of compact binary systems.

%\textcolor{blue}{
%\textbf{Summary of results--GR and quadratic gravity in 4D:}
%We develop a WEFT analysis of four-dimensional quadratic-curvature gravity, perturbatively reconstructing the static background metric from tree-level one-point functions (after a 3+1 ADM decomposition and field redefinitions that diagonalize the massless graviton together with massive spin-2 and spin-0 modes), which yields Yukawa-tailed potentials already at 1PN. On the IR/EFT side we compute the response to external electric-type spin-0 and spin-2 tides and extract the associated Wilson coefficients (tidal Love numbers), finding them to be non-vanishing yet scale-independent i.e., showing no classical RG running, by contrast with GR where Love numbers vanish identically. On the UV side, for the scalar sector we solve the static dipole problem on the reconstructed background via a Yukawa-deformed Frobenius ansatz and fix the scalar Love number unambiguously by UV–IR matching. For gravitational (spin-2) tides, the EFT computation is completed but the UV analysis is deferred: an unambiguous determination of the gravitational Love number requires solving the modified Teukolsky equation (in line of \cite{Kumar:2025jsi}) appropriate to quadratic gravity, which is not reducible to the standard GR scalar master equation and will be addressed in future work.}\\\\
\textbf{Context of the paper \& overview:} This work presents a detailed study of the tidal Love number beyond the standard GR involving terms which are quadratic in curvatures in four dimensions, within the EFT framework. We know that black holes in four-dimensional vacuum GR exhibit vanishing Love numbers, implying that they remain unaffected by external tidal fields. This issue indicates key aspects of theoretical understanding and its observational consequences. If GW observations in the future detect any measurable non-vanishing tidal response, it will involve a notable deviation from the standard GR we know. This can possibly suggest that either GR is incomplete, indicating solutions arising in alternative theories of gravity \cite{Kumar:2024utz}, or that the objects are not black holes but exotic compact objects. %Such a discovery will be a remarkable step in our fundamental understanding of gravity and the true nature of compact astrophysical bodies. %Therefore, theories beyond GR can naturally lead to non-zero Love numbers, offering a clear avenue for testing deviations from GR. 
The obtained tidal response can thus serve as a sensitive probe of the underlying gravitational theory. This motivation to explore modified gravity theories with tidal Love numbers becomes particularly compelling, allowing us to make direct comparisons between theory and observation, potentially revealing the imprint of modifications to GR. In the context of black hole perturbation, there are some recent studies examining the tidal response beyond the GR case \cite{kobayashi2025parametrizedtidaldissipationnumbers, Katagiri:2023umb, Cano:2025zyk, Cardoso:2018ptl, Katagiri:2024fpn, Barbosa:2025uau}. However, in this pursuit, the EFT framework, adapted to include the new degrees of freedom, provides a systematic approach to capture finite-size effects and associated core findings. Unlike traditional perturbative techniques, EFT allows one to isolate and parametrize deviations from GR through gauge-invariant observables while maintaining control over the consistency with black hole perturbation \cite{Ivanov:2022qqt}. Apart from this, the EFT approach also enables us to efficiently address certain ambiguities, also pointed out above, related to source/response overlap and logarithmic corrections, etc \cite{Ivanov:2022hlo}. This makes EFT a powerful tool for interpreting tidal interactions and testing gravity in regimes previously inaccessible to direct measurement. \vspace{1mm}
The remainder of this work is organized as follows. In Section~(\ref{GR}), we briefly review the computation of the love number and metric reconstruction for GR using the EFT framework. In Section~(\ref{sec2}), we introduce the most general theory of gravity including correction terms to GR up to quadratic in curvatures in four dimensions, perform a 3+1 ADM decomposition, and carry out a field redefinition to diagonalize the propagating scalar ($\Phi$) and tensor ($\Sigma$) modes, subsequently deriving the propagators and Feynman rules used throughout the paper.  
In Section~(\ref{sec3}), we explain how classical solutions can be perturbatively reconstructed via the tree-level one-point function in quantum field theory.  
Section~(\ref{sec4}) revisits PN power counting in the presence of higher-derivative interactions, demonstrating that in $R^2$ gravity, each extra bulk vertex raises the PN order by two relative to pure GR.  In Section~(\ref{sec5}), we analyze spin-0 and spin-2 electric-type perturbations, showing that, unlike in GR, the induced Love numbers are non-vanishing but interestingly non-renormalizable. Section~(\ref{sec6}) presents the UV calculation for the static dipole ($\ell=1$) scalar perturbation on a Yukawa-deformed Schwarzschild background, derives a modified Frobenius ansatz capturing the $e^{-\mu r}/r^2$ tail, and matches this with the EFT prediction to fix the UV integration constant \( \tilde c \).  
%In Section (\ref{sec7}), we reformulate the problem in isotropic coordinates and numerically verify that our ansatz indeed cancels the leading Yukawa tail, confirming the gauge-invariance of the Love-number matching.  
%Section~(\ref{sec8}) completes the EFT–UV comparison by two complementary methods: matching Wilson coefficients in the worldline action and perturbatively transforming the UV metric into isotropic coordinates for a direct metric‐level match, both yielding \(\tilde c_1=-M/25\) when couplings satisfy $8\beta+3\gamma=0$ and the constrain on parameter space \(S_2=4S_0\) when couplings satisfy $2\beta+\gamma=0$. 
In Section~(\ref{sec9}), we comment on the Ladder symmetry and finally conclude with Section~(\ref{sec10}) by summarizing our main findings along with their future implications.\vspace{1mm}  

Technical details are collected in six Appendices: Appendix~\ref{app1} derives the master integrals needed for the reconstruction of one-point function, Appendix~\ref{app2} lists useful three-dimensional Fourier transforms. %with \(k_i k_j\) in the numerator.
Appendix~\ref{app_gen_integral} further provides a general strategy for solving some other types of integrals, and in Appendix~\ref{integrals}, we list integrals used in the reconstructed metric via EFT. Finally, Appendix~\ref{app_UV_fixing} provides a consistency check by matching the UV metric in Schwarzschild coordinates to the EFT metric in isotropic coordinates, where UV charges of the metric are determined by Wilsonian matching. Before proceeding further, we want to briefly pause and summarize the primary results of our work and compare them with those of GR.  \\\\
{\color{black}\textbf{Summary of results - GR versus quadratic gravity in 3+1D:}
It has been  established  that, the 3+1D black hole solutions of GR, i.e., Schwarzschild and Kerr black holes, exhibit vanishing static tidal Love numbers. Recent analyses have confirmed this result for scalar (spin-0) and electromagnetic (spin-1) gravitational (spin-2) field perturbations, extending the statement to all bosonic perturbations \cite{Binnington:2009bb, Kol:2011vg, Damour:2009vw, Charalambous:2021mea, Chia:2020yla, Ivanov:2022qqt}. Using the EFT formalism, where Love numbers are defined as Wilson coefficients of local worldline operators, one can cleanly separate the conservative and dissipative components of the response. Within this formulation, the conservative response coefficients—which correspond to the static Love numbers—are shown to vanish identically for Kerr black holes at all orders in spin \cite{Ivanov:2022qqt}. This explains that the absence of static tidal deformability is a generic property of black hole solutions in 3+1D vacuum GR. Further, the EFT analysis shows that these vanishing Love numbers do not acquire any classical RG running \cite{Ivanov:2022hlo}\footnote{Interestingly, while the conservative response vanishes, the dissipative part remains non-vanishing even for static external perturbations that arises from frame-dragging effects \cite{Chia:2020yla}.}. This behavior is closely connected to a special property of black hole solutions in vacuum GR under their static perturbations, called ladder symmetry \cite{Hui:2021vcv}. It has been shown that imposing ladder symmetry on any general spherically symmetric static spacetime requires the condition  $g_{tt}g_{rr}=-1$ (product of temporal and radial metric components) \cite{Hui:2021vcv,Hui:2022vbh,Rai:2024lho}. This condition underlies the cancellations that lead to vanishing Love numbers, providing a unified explanation for tidal deformability of black hole solutions in vacuum GR, i.e., Schwarzschild and Kerr. %Moreover, the dynamical Love number (frequency dependent) for Kerr black holes remains non-zero.

The present work extends the study of tidal Love numbers to 3+1D quadratic-curvature EFT of gravity. We develop a WEFT analysis of four-dimensional quadratic-curvature gravity, perturbatively reconstructing the static background metric from tree-level one-point functions (after a 3+1 ADM decomposition and field redefinitions that diagonalize the massless graviton together with massive spin-2 and spin-0 modes), which yields Yukawa-tailed potentials already at 1PN. On the IR/EFT side we compute the response to external electric-type spin-0 and spin-2 tides and extract the associated Wilson coefficients (tidal Love numbers), finding them to be \textit{non-vanishing} (except for some special values of quadratic-curvature couplings for which the ladder symmetry in the reconstructed metric gets restored) yet \textit{scale-independent} i.e., showing no classical RG running, by contrast with GR where Love numbers vanish identically. On the UV side, for the scalar sector we solve the static dipole problem on the reconstructed background via a Yukawa-deformed Frobenius ansatz and \textit{ fix the scalar Love number unambiguously by UV–IR matching}. For gravitational (spin-2) tides, the EFT computation is completed but the UV analysis is deferred: an unambiguous determination of the gravitational Love number requires solving the modified Teukolsky equation (in line of \cite{Li:2022pcy, Hussain:2022ins, Kumar:2025jsi}) appropriate to quadratic gravity, which is not reducible to the standard GR scalar master equation and will be addressed in future work.}

\section{A brief review of Metric reconstruction and computation of Love number in GR} \label{GR}
Computation of Love numbers in GR is quite significant because it establishes an important fact about the response of tidal perturbations. For black hole solutions in GR, the Love numbers always vanish, and this fact has also been supported by various kinds of considerations of Ladder symmetries \cite{Hui:2021vcv,Charalambous:2022rre}.
In this section, we briefly describe the metric reconstruction and calculation of the Love number in GR following the language of effective field theory. Considering a point particle to be coupled to gravity, the action can be written as follows,
\begin{align}
    S=S_{\textrm{EH}}+S_{\textrm{PP}}
\end{align}
where the $S_{\textrm{EH}}$ is the Einstein-Hilbert action and $S_{\textrm{PP}}$ denotes the point-particle action, and it can be cast as,
\begin{align}
    S_{PP}=-m\int d\tau \sqrt{\frac{dx^\mu}{d\tau}\frac{dx^\nu}{d\tau}g_{\mu\nu}}\,.
\end{align}
In the construction we use `static gauge', which fixes the worldline parameter $\tau=t$ and under this gauge choice the perturbative field expansions are given by\footnote{In the rest frame of the black hole (or equivalently, the static gauge), $v^i=0.$}
,
\begin{align}
    \phi=\bar{\phi}+\delta\phi,\,\,\,\,\,\,\, g_{\mu\nu}=\eta_{\mu\nu}+h_{\mu \nu};\, \,\,\,\,\,\frac{dx^{\mu}}{d\tau}=(1,v^i).
\end{align}

\noindent Within the EFT arena one treats the metric ($g_{\mu\nu}$) perturbatively above the Minkowski metric ($\eta_{\mu\nu}$). Now, after reconstructing the metric using EFT techniques of finding one-point functions, one recovers the Schwarzschild metric (when there is spherical symmetry) perturbatively. Then for different spins, i.e $s=0,1,2\cdots$, one computes the fluctuation profiles on the reconstructed background. Apart from the point particle description, finite size effects are quite important, accompanied by Wilson coefficients, which we discuss next as they will play an important role in this paper.

\subsection*{Finite size corrections:}
In the limit, the size of the body ($R$) is smaller than the external perturbation wavelength, $|\vec{q}|R<<1$, we can resort to the most general point particle EFT, incorporating the finite size effects. There are usually two types of moments, i.e., electric (even-parity) and magnetic (odd-parity). However, for GR, in the Schwarzschild spacetime, the electromagnetic duality in $D=4$ helps us to consider only the electric sector. Multipole moments of geometric bodies are given by,
\begin{align}
    E_{L}^s= E^s_{\mu_1\mu_2\mu_3\cdots \mu_\ell}\langle e^{\mu_1}_{a_1}\cdots e^{\mu_\ell}_{a_{\ell}}\rangle
\end{align}
where $\langle\cdots\rangle$ denotes the symmetric trace-free (STF) part of it and $e_{a_1}^{\mu_1}$ denotes the vierbein. For a few special cases, we have \cite{Ivanov:2022hlo},
\begin{align}
    \begin{split}
        &E^{s=0}_{a_1,\cdots a_\ell}=\nabla_{(a_1\cdots}\nabla_{a_\ell)}\phi\,,\\&
E^{s=0}_{a_1,\cdots a_\ell}=\nabla_{(a_1\cdots}\nabla_{a_\ell-1)}E_{a_l};\,\,\,\,\ E_{_a}=e^{\mu}_av^{\nu}F_{\mu\nu}\,,\\&
E^{s=0}_{a_1,\cdots a_\ell}=\nabla_{(a_1\cdots}\nabla_{a_\ell-2)}E_{a_{\ell-1}a_{\ell}};\,\,\,\,\,E_{ab}=v^{\alpha}v^{\mu}e^{\beta}_a e^{\nu}_b C_{\alpha\beta\mu\nu}    
    \end{split}
\end{align}
where $C_{\alpha\beta\mu\nu}$ denotes the Weyl tensor. Now we briefly describe how these give us the Love numbers in general.\\

\noindent
\textbullet \textbf{\,\,Love number in EFT:} Love numbers can be more effectively understood and resolved within the systematic framework of worldline EFT, offering a modern and robust approach to describing the gravitational response of compact objects \cite{Goldberger:2004jt, Goldberger:2007hy, Porto:2005ac, Porto:2016pyg, Levi:2018nxp} and in \cite{Bhattacharyya:2023kbh,Diedrichs:2023foj,Bernard:2023eul} for theories beyond GR. This formalism provides an efficient approach to modelling the finite-size effects, such as the tidal response, arising from the internal properties of compact objects in orbit. Within this approach, as stated previously, Love numbers emerge as Wilson coefficients that govern static finite-size interactions encoded in the effective action. 
\begin{align}\label{actioneft}
    S_{\textup{finite size}} = \sum_{\ell = 2}\frac{\lambda_{\ell}}{2\ell !}\int E^{a_{1}...a_{\ell}}E_{a_{1}...a_{\ell}} d\tau,
\end{align}
where $\langle\cdot\cdot\cdot\rangle$ refers to the symmetrization of indices followed by the subtraction of all possible trace terms to produce a trace-free symmetric tensor. Note that indices $a = (1, 2, 3)$ are spatial indices. A linear response analysis based on the effective action in Eq. (\ref{actioneft}) reveals that, by appropriately matching the response terms to the Newtonian case, one can identify the tidal coefficients $\lambda_{\ell}$ in terms of the black hole radius and the dimensionless Love numbers $k_\ell$. Specifically, the matching yields an expression of the form $\lambda_{\ell} = (-1)^{\ell}r_{s}^{1+2\ell}k_{\ell}\frac{2^{\ell}\sqrt{\pi}}{\Gamma(1/2-\ell)}$  \cite{Charalambous:2021mea}. %\vspace{1mm}

\tikzset {_hgzt6tvhp/.code = {\pgfsetadditionalshadetransform{ \pgftransformshift{\pgfpoint{89.1 bp } { -108.9 bp }  }  \pgftransformscale{1.32 }  }}}
\pgfdeclareradialshading{_kwum5zhfu}{\pgfpoint{-72bp}{88bp}}{rgb(0bp)=(1,1,1);
rgb(0bp)=(1,1,1);
rgb(25bp)=(0,0,0);
rgb(400bp)=(0,0,0)}

% Gradient Info
\tikzset {_2vs5wq0qm/.code = {\pgfsetadditionalshadetransform{ \pgftransformshift{\pgfpoint{89.1 bp } { -108.9 bp }  }  \pgftransformscale{1.32 }  }}}
\pgfdeclareradialshading{_vlmyk1bhk}{\pgfpoint{-72bp}{88bp}}{rgb(0bp)=(1,1,1);
rgb(0bp)=(1,1,1);
rgb(25bp)=(0,0,0);
rgb(400bp)=(0,0,0)}
\tikzset{every picture/.style={line width=0.75pt}} %set default line width to 0.75pt        
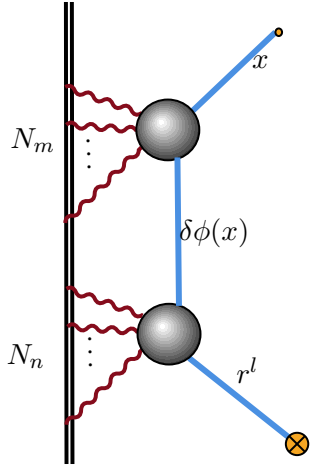
\begin{figure}
\centering
\begin{tikzpicture}
[x=0.75pt,y=0.75pt,yscale=-1.1,xscale=1]
%uncomment if require: \path (0,384); %set diagram left start at 0, and has height of 384

%Straight Lines [id:da290852224476425] 
\draw [line width=1.5]    (210.02,35.75) -- (210.32,247.75)(207.02,35.75) -- (207.32,247.75) ;
%Straight Lines [id:da8786580866378778] 
\draw [color={rgb, 255:red, 74; green, 144; blue, 226 }  ,draw opacity=1 ][line width=2.25]    (270.99,85.63) -- (312.69,49.63) ;
%Straight Lines [id:da3813285503172369] 
\draw [color={rgb, 255:red, 133; green, 12; blue, 28 }  ,draw opacity=1 ][line width=1.5]    (206.68,74.76) .. controls (208.77,73.68) and (210.36,74.19) .. (211.44,76.28) .. controls (212.53,78.37) and (214.12,78.88) .. (216.21,77.8) .. controls (218.3,76.72) and (219.89,77.23) .. (220.97,79.32) .. controls (222.05,81.41) and (223.64,81.92) .. (225.73,80.84) .. controls (227.82,79.76) and (229.41,80.27) .. (230.5,82.36) .. controls (231.58,84.45) and (233.17,84.96) .. (235.26,83.88) .. controls (237.35,82.8) and (238.94,83.31) .. (240.02,85.4) .. controls (241.11,87.49) and (242.7,88) .. (244.79,86.92) -- (245.15,87.03) -- (245.15,87.03) ;
%Straight Lines [id:da2524190021509931] 
\draw [color={rgb, 255:red, 133; green, 12; blue, 28 }  ,draw opacity=1 ][line width=1.5]    (207.26,92.15) .. controls (209.03,90.59) and (210.7,90.7) .. (212.25,92.47) .. controls (213.8,94.24) and (215.47,94.35) .. (217.24,92.8) .. controls (219.01,91.25) and (220.67,91.35) .. (222.23,93.12) .. controls (223.78,94.89) and (225.45,95) .. (227.22,93.45) .. controls (228.99,91.9) and (230.66,92.01) .. (232.21,93.78) .. controls (233.77,95.55) and (235.43,95.65) .. (237.2,94.1) .. controls (238.97,92.55) and (240.64,92.66) .. (242.19,94.43) -- (242.53,94.45) -- (242.53,94.45) ;
%Straight Lines [id:da6245604773318241] 
\draw [color={rgb, 255:red, 133; green, 12; blue, 28 }  ,draw opacity=1 ][line width=1.5]    (206.68,137.15) .. controls (206.8,134.8) and (208.04,133.68) .. (210.39,133.8) .. controls (212.74,133.92) and (213.98,132.8) .. (214.1,130.45) .. controls (214.22,128.1) and (215.46,126.98) .. (217.81,127.09) .. controls (220.16,127.21) and (221.4,126.09) .. (221.52,123.74) .. controls (221.63,121.39) and (222.87,120.27) .. (225.22,120.39) .. controls (227.57,120.51) and (228.81,119.39) .. (228.93,117.04) .. controls (229.05,114.69) and (230.29,113.57) .. (232.64,113.68) .. controls (234.99,113.8) and (236.23,112.68) .. (236.35,110.33) .. controls (236.47,107.98) and (237.71,106.86) .. (240.06,106.98) .. controls (242.41,107.1) and (243.65,105.98) .. (243.77,103.63) -- (245.15,102.38) -- (245.15,102.38) ;
%Shape: Ellipse [id:dp061939804464581694] 
\path  [shading=_kwum5zhfu,_hgzt6tvhp] (242.53,94.45) .. controls (242.53,86.75) and (249.64,80.51) .. (258.42,80.51) .. controls (267.19,80.51) and (274.3,86.75) .. (274.3,94.45) .. controls (274.3,102.15) and (267.19,108.39) .. (258.42,108.39) .. controls (249.64,108.39) and (242.53,102.15) .. (242.53,94.45) -- cycle ; % for fading 
 \draw   (242.53,94.45) .. controls (242.53,86.75) and (249.64,80.51) .. (258.42,80.51) .. controls (267.19,80.51) and (274.3,86.75) .. (274.3,94.45) .. controls (274.3,102.15) and (267.19,108.39) .. (258.42,108.39) .. controls (249.64,108.39) and (242.53,102.15) .. (242.53,94.45) -- cycle ; % for border 

%Straight Lines [id:da3032128477531796] 
\draw [color={rgb, 255:red, 74; green, 144; blue, 226 }  ,draw opacity=1 ][line width=2.25]    (323.6,238.54) -- (269.2,199.16) ;
%Flowchart: Summing Junction [id:dp9729026022536856] 
\draw  [fill={rgb, 255:red, 245; green, 166; blue, 35 }  ,fill opacity=1 ] (317.94,238.54) .. controls (317.94,235.71) and (320.47,233.42) .. (323.6,233.42) .. controls (326.72,233.42) and (329.25,235.71) .. (329.25,238.54) .. controls (329.25,241.38) and (326.72,243.67) .. (323.6,243.67) .. controls (320.47,243.67) and (317.94,241.38) .. (317.94,238.54) -- cycle ; \draw   (319.6,234.92) -- (327.59,242.17) ; \draw   (327.59,234.92) -- (319.6,242.17) ;
%Shape: Ellipse [id:dp14899664810927848] 
\draw  [fill={rgb, 255:red, 245; green, 166; blue, 35 }  ,fill opacity=1 ] (312.69,49.63) .. controls (312.69,48.67) and (313.44,47.9) .. (314.36,47.9) .. controls (315.28,47.9) and (316.03,48.67) .. (316.03,49.63) .. controls (316.03,50.59) and (315.28,51.37) .. (314.36,51.37) .. controls (313.44,51.37) and (312.69,50.59) .. (312.69,49.63) -- cycle ;
%Straight Lines [id:da31467590906317877] 
\draw [color={rgb, 255:red, 133; green, 12; blue, 28 }  ,draw opacity=1 ][line width=1.5]    (207.28,167.33) .. controls (209.37,166.25) and (210.96,166.76) .. (212.04,168.85) .. controls (213.12,170.94) and (214.71,171.45) .. (216.8,170.37) .. controls (218.89,169.29) and (220.48,169.8) .. (221.57,171.89) .. controls (222.65,173.98) and (224.24,174.49) .. (226.33,173.41) .. controls (228.42,172.33) and (230.01,172.84) .. (231.09,174.93) .. controls (232.18,177.02) and (233.77,177.53) .. (235.86,176.45) .. controls (237.95,175.36) and (239.54,175.87) .. (240.62,177.96) .. controls (241.7,180.05) and (243.29,180.56) .. (245.38,179.48) -- (245.75,179.6) -- (245.75,179.6) ;
%Straight Lines [id:da5579238049588245] 
\draw [color={rgb, 255:red, 133; green, 12; blue, 28 }  ,draw opacity=1 ][line width=1.5]    (207.86,184.72) .. controls (209.63,183.16) and (211.3,183.27) .. (212.85,185.04) .. controls (214.4,186.81) and (216.07,186.92) .. (217.84,185.37) .. controls (219.61,183.82) and (221.27,183.92) .. (222.83,185.69) .. controls (224.38,187.46) and (226.05,187.57) .. (227.82,186.02) .. controls (229.59,184.47) and (231.25,184.57) .. (232.81,186.34) .. controls (234.36,188.11) and (236.03,188.22) .. (237.8,186.67) .. controls (239.57,185.12) and (241.23,185.23) .. (242.78,187) -- (243.13,187.02) -- (243.13,187.02) ;
%Straight Lines [id:da3982130896828814] 
\draw [color={rgb, 255:red, 133; green, 12; blue, 28 }  ,draw opacity=1 ][line width=1.5]    (207.28,229.72) .. controls (207.4,227.37) and (208.64,226.25) .. (210.99,226.37) .. controls (213.34,226.49) and (214.58,225.37) .. (214.69,223.02) .. controls (214.81,220.67) and (216.05,219.55) .. (218.4,219.66) .. controls (220.75,219.78) and (221.99,218.66) .. (222.11,216.31) .. controls (222.23,213.96) and (223.47,212.84) .. (225.82,212.96) .. controls (228.17,213.08) and (229.41,211.96) .. (229.53,209.61) .. controls (229.65,207.26) and (230.89,206.14) .. (233.24,206.25) .. controls (235.59,206.37) and (236.83,205.25) .. (236.95,202.9) .. controls (237.07,200.55) and (238.31,199.43) .. (240.66,199.55) .. controls (243.01,199.66) and (244.25,198.54) .. (244.37,196.19) -- (245.75,194.94) -- (245.75,194.94) ;
%Shape: Ellipse [id:dp9921560874368213] 
\path  [shading=_vlmyk1bhk,_2vs5wq0qm] (243.13,188.17) .. controls (243.13,180.47) and (250.24,174.23) .. (259.01,174.23) .. controls (267.79,174.23) and (274.9,180.47) .. (274.9,188.17) .. controls (274.9,195.87) and (267.79,202.1) .. (259.01,202.1) .. controls (250.24,202.1) and (243.13,195.87) .. (243.13,188.17) -- cycle ; % for fading 
 \draw   (243.13,188.17) .. controls (243.13,180.47) and (250.24,174.23) .. (259.01,174.23) .. controls (267.79,174.23) and (274.9,180.47) .. (274.9,188.17) .. controls (274.9,195.87) and (267.79,202.1) .. (259.01,202.1) .. controls (250.24,202.1) and (243.13,195.87) .. (243.13,188.17) -- cycle ; % for border 

%Straight Lines [id:da348997265436385] 
\draw [color={rgb, 255:red, 74; green, 144; blue, 226 }  ,draw opacity=1 ][line width=2.25]    (263.27,107.05) -- (264.03,175.19) ;

% Text Node
\draw (222.57,96.11) node [anchor=north west][inner sep=0.75pt]  [rotate=-88.44]  {$\cdots $};
% Text Node
\draw (299.47,59.07) node [anchor=north west][inner sep=0.75pt]    {$x$};
% Text Node
\draw (177.59,92.35) node [anchor=north west][inner sep=0.75pt]    {$N_{m}$};
% Text Node
\draw (223.33,186.97) node [anchor=north west][inner sep=0.75pt]  [rotate=-88.44]  {$\cdots $};
% Text Node
\draw (291.5,197.66) node [anchor=north west][inner sep=0.75pt]    {$r^{l}$};
% Text Node
\label{fig.1}
\draw (175.66,191.41) node [anchor=north west][inner sep=0.75pt]    {$N_{n}$};
% Text Node
\draw (262.61,133.68) node [anchor=north west][inner sep=0.75pt]    {$\delta \phi ( x)$};
\end{tikzpicture}
\caption{A general figure of Pyramid-like diagrams appearing in the evaluation of the response coefficients.}
\label{fiig1}
\end{figure}
\textcolor{black}{
\begin{align}
    \textbf{UV fall-off} = \textrm{Source coefficient} + \textrm{Tidal Love number vertex (Response coefficient)\,.}
\end{align}}
To determine whether a non-zero Love number arises, one begins by analyzing the asymptotic behavior of the radial equation in the ultraviolet (UV) regime. Specifically, one examines whether the solution exhibits a \(1/r^{\ell+1}\) fall-off at large radius. If such a fall-off is absent, then the sum of the source and response coefficients must vanish. \textcolor{black}{Consequently, if the response coefficient either vanishes identically or the source coefficient precisely matches the UV fall-off behavior, the corresponding Love number is zero in both cases. In the latter scenario, the vanishing of the Love number is tied to the absence of a fall-off term in the radial solution that mirrors the structure of the response coefficient, often referred to as the \emph{Love number vertex}.}

\noindent
This UV-Effective Field Theory (EFT) matching plays a crucial role, particularly in theories involving higher-derivative or curvature corrections. Even within the framework of GR, the calculation becomes increasingly intricate for higher multipolar tidal perturbations or when higher post-Newtonian (PN) corrections to the Love number vertex are taken into account. These complexities arise from the possibility of potential bulk interaction vertices. In Fig.~(\ref{fiig1}) we draw pyramid diagrams for possible bulk interactions with background field insertion shown (in yellow cross) at the end of the blue propagator. In the following section, we extend this analysis to the case of quadratic EFTs in $D=4$. 

\section{EFT of Gravity: Quadratic corrections in $D=4$}\label{sec2}
The most general quadratic EFT of gravity in four dimensions has the following form:
\begin{align}
    \begin{split}
        S=\int d^4x\,\sqrt{-g}\left(\frac{m_p^2}{2}R+\frac{1}{6f_0^2}R^2-\frac{1}{2\xi^2}C_{\mu\nu\alpha\beta}C^{\mu\nu\alpha\beta}\right)\label{1.1a}
    \end{split}
\end{align}
where $m_p^2=\frac{1}{8\pi G}$, $f_0$ and $\xi$ are the theory coupling constants and $C_{\mu\nu\alpha\beta}$ is the Weyl tensor (traceless part of Riemann tensor $R_{\mu\nu\alpha\beta}$).  The motivation behind including quadratic curvature terms comes from the form of one-loop corrections \cite{tHooft:1974toh}. Including these higher-curvature invariants renders the theory \textcolor{black}{renormalizable} \cite{Stelle:1976gc} (the cost is the appearance of ghost-like modes). \textcolor{black}{Moreover, the system is also asymptotically free \cite{Fradkin:1981hx,Fradkin:1981iu} in the sense of gauge theories}.  The spectrum of this theory contains a massless graviton, a massive spin-2 ghost excitation with $m_1^2=\frac{m_p^2\xi^2}{2}$ and a massive non-ghost spin-0 excitation with $m_2^2=\frac{m_p^2f_0^2}{2}$ \cite{Stelle:1977ry}. Also, for the special case of $D=4$, quadratic gravity admits Ricci flat ($R=0$) solution and both Schwarzschild and non-Schwarzschild black hole solutions exist \cite{Lu:2015cqa,Lu:2015psa}, which is crucial for our study since we are interested in the existence of a horizon or a black hole solution. Therefore, we restrict our study to the $D=4$ case. % \textcolor{red}{In $D>4$, the situation is somewhat different, with no exact Ricci flat solution. However, one can attempt to study linearized theory where $\delta R=0$ at the linearized level \cite{Lu:2017kzi}: AB: Not needed}. 
To summarize, whether one approaches from pure EFT reasoning, explicit loop calculations, or an explicit UV theory like string theory, higher-curvature invariants are inevitable. They encode the fingerprints of whatever UV physics completes GR\footnote{\textcolor{black}{  GR can be regarded as the leading-order term in an EFT expansion of gravity, ultimately expected to be completed by a UV-complete theory. In the context of our work, solving the equations of motion in the UV regime corresponds to analyzing perturbative dynamics within  GR or higher-derivative extensions such as quadratic gravity. On the other hand, the EFT perspective pertains to the worldline effective field theory, wherein the point-particle action is organized as a Wilsonian expansion.}}. All of this makes the study of quadratic gravity in $D=4$ both important and interesting. \\\\
%\textcolor{red}{AB: Needless to say, the EFT approach to quadratic gravity allows us to extract the physically relevant modifications to general relativity without having to deal with the problematic aspects of higher-derivative theories like ghost modes: Not needed. Misleading}. \\
Discarding the surface terms (arising due to the integral over the topological Gauss-Bonnet combination in $D=4$ spacetime) in (\ref{1.1a}), we can rewrite the third term as,
\begin{align}
    \begin{split}
        \int d^4x\,\sqrt{-g}\left(-\frac{1}{2\xi^2}C_{\mu\nu\alpha\beta}C^{\mu\nu\alpha\beta}\right)=-\frac{1}{\xi^2}\int d^4x \sqrt{-g}\,\left(R_{\mu\nu}R^{\mu\nu}-\frac{1}{3}R^2\right)
    \end{split}
\end{align}
Then we have,
\begin{align}
    \begin{split}
        S=\int d^4x \sqrt{-g}\left(\frac{m_p^2}{2}R+\left(\frac{1}{6f_0^2}+\frac{1}{3\xi^2}\right)R^2-\frac{1}{\xi^2}R_{\mu\nu}R^{\mu\nu}\right)\,. 
    \end{split} 
\end{align}
Therefore, the action is given by:
\begin{align}
\begin{split}\label{action}
S &=\frac{m_p^2}{2} \int dt d^3x \sqrt{-g}\Big(\underbrace{R}_{\textcolor{black}{S_1}}+\beta \underbrace{R^2}_{\textcolor{black}{S_2}}+\gamma \underbrace{R_{\mu\nu}R^{\mu\nu}}_{{S_3}}\Big)
\end{split}
\end{align}
    where $R_{\mu\nu}$ and $R$ are the Ricci tensor and the scalar calculated from the 3+1 decomposed metric $g_{ab}$ given by \footnote{We follow the metric signature ($-+++$) throughout the paper.}
\begin{equation}
    ds^2=-e^{2\phi}(dt-\mathcal{A}_idx^i)^2+e^{-2\phi}\gamma_{ij}dx^idx^j,\label{2.5t}
\end{equation}
    where $\gamma_{ij}=\delta_{ij}+\sigma_{ij}$ is the spatial metric with $\sigma_{ij}$ as the fluctuating part, $\phi$ is the dilaton field (Newtonian potential), and $\mathcal{A}_i$ is the gravitomagnetic vector. Since it is necessary to fix the gauge for correct matching between EFT (IR) and \textcolor{black}{quadratic gravity (UV)} observables, we work in the ``consistent'' static gauge given by \footnote{An EFT is called gauge ``consistent'' if we can perturbatively reconstruct the background metric via the off-shell graviton one point function \cite{Ivanov:2022hlo}.}:
\begin{equation}
    \sigma_{ij}=(1+\sigma)\delta_{ij},\ \mathcal{A}_i=0.
\end{equation}
One can also directly match gauge-independent results, but matching of one-point functions requires the choice of a gauge consistent with the background UV solution. Therefore, working in the static gauge, let us proceed with $S_1$, which is given order by order (in $\sigma$ and $\phi$) as follows:
\begin{equation}
\begin{aligned}\label{EHAction}
\textcolor{black}{S_1^{(2)}}=& \frac{m_p^2}{2}\int dt d^3x \Big(\frac{1}{2}\partial_i\sigma\partial^i\sigma-2\delta^{ij}\partial_i\phi \partial_j\phi\Big)\,,\\
\textcolor{black}{S_1^{(3)}}=& \frac{m_p^2}{2}\int dt d^3x \Big(\frac{-3}{4}\sigma\partial_i\sigma\partial^i\sigma{-}\sigma\delta^{ij}\partial_i\phi \partial_j\phi\Big)\,.
\end{aligned}
\end{equation}
Here, the superscripts denote the perturbation order. We next proceed to evaluate the \textcolor{black}{$S_2$}. Up to quadratic order we get,
\begin{align}
\begin{split}\label{S22}
&\textcolor{black}{S_2^{(2)}}=4\beta\frac{m_p^2}{2}\int dt d^3x\Big(\sigma \partial_{i}^2\partial_{j}^2\sigma + \phi\partial_{i}^2\partial_{j}^2\phi - 2
\phi\partial_{i}^2\partial_{j}^2\sigma\Big)\,.\\&
\end{split}
\end{align}
The cubic part takes the following form,  
\begin{align}
\begin{split}\label{S23}
&\textcolor{black}{S_2^{(3)}}
=\beta\frac{m_p^2}{2}\int dt d^3x\Big(-6\partial_i^2\sigma\partial_j\sigma\partial_j\sigma-4\partial_i^2\sigma\partial_j\sigma\partial_j\phi +8\partial_i^2\sigma\partial_j\phi\partial_j\phi-10 \sigma \partial_i^2\sigma\partial_j^2\sigma +8\phi \partial_i^2\sigma\partial_j^2\sigma\\&\hspace{3.5cm}-16 \phi \partial_i^2\sigma\partial_j^2\phi+ 12\sigma\partial_i^2\sigma\partial_j^2\phi +6\partial_i^2\phi\partial_j\sigma\partial_j\sigma+4\partial_i^2\phi\partial_j\sigma\partial_j\phi -8 \partial_i^2\phi\partial_j\phi\partial_j\phi +8\phi\partial_i^2\phi\partial_j^2\phi\\&\hspace{3.5cm}-2\sigma \partial_i^2\phi\partial_j^2\phi      \Big)\,.
\end{split}
\end{align}
Finally, we proceed to evaluate \textcolor{black}{$S_3$}, which is order by order takes the following form:
\begin{align}
\begin{split}
{S_3^{(2)}}=&\gamma\frac{m_p^2}{2}\int dt d^3x\Big(4\partial_i^2\phi \partial_j^2\phi-4\partial_i^2\sigma\partial_j^2\phi+\frac{3}{2}\partial_i^2\sigma\partial_j^2\sigma\Big)\,,\label{S32}
\end{split}
\end{align}
and,
\begin{align}
\begin{split}
S_3^{(3)}=&\gamma\frac{m_p^2}{2}\int dt d^3x \Bigg[\sigma\Big(-\frac{25}{8}\partial^{2}_{i}\sigma\partial^{2}_{j}\sigma-\frac{5}{8}(\partial_{i}\partial_{j}\sigma)^{2} + 6\partial_{i}^{2}\sigma\partial_{j}^{2}\phi-2\partial_{i}^{2}\phi\partial_{j}^{2}\phi \Big) +\\& \hspace{2cm}\phi \Big(\frac{5}{2}\partial^{2}_{i}\sigma\partial^{2}_{j}\sigma+\frac{1}{2}(\partial_{i}\partial_{j}\sigma)^{2} +8\partial^{2}_{i}\phi\partial_{j}^{2}\phi-8\partial^{2}_{i}\sigma\partial_{j}^{2}\phi\Big) -\frac{7}{4}\partial^{2}_{i}\sigma (\partial_{j}\sigma)^{2}-\frac{3}{4}\partial_{i}\sigma\partial_{j}\sigma\partial_{i}\partial_{j}\sigma  \\&
\hspace{2cm}-2\partial^{2}_{i}\sigma\partial_{j}\sigma\partial_{j}\phi +2\partial_{i}^{2}\sigma(\partial_{j}\phi)^{2} +2\partial_{i}\phi\partial_{j}\phi\partial_{i}\partial_{j}\sigma + 3\partial^{2}_{i}\phi (\partial_{j}\sigma)^{2}+4\partial^{2}_{i}\phi\partial_{j}\sigma {\partial_{j}}\phi-4\partial_{i}^{2}\phi(\partial_{j}\phi)^{2}\Bigg]\,.\label{S33}
\end{split}
\end{align}
\subsection{Diagonalization and field redefinition}
We see that in the quadratic part of the action, we have cross terms coming from the higher derivative corrections. So, one can choose a basis such that the Lagrangian has no cross-term. We have the  quadratic part of the total action,
\begin{align}
    \begin{split}
S^{(2)}&= S^{(2)}_1+S^{(2)}_2+S^{(2)}_3\,,
\\&= \frac{m_p^2}{2}\int dt d^3x \Big[\left(\frac{1}{2}\partial_i\sigma\partial_i\sigma-2\partial_i\phi \partial_i\phi\right)+4\beta \left(\partial_{i}^2\sigma \partial_{j}^2\sigma + \partial_{i}^2\phi\partial_{j}^2\phi - 2
\phi\partial_{i}^2\partial_{j}^2\sigma\right)\\ &
\hspace{2.5cm}+\gamma \left(4\partial_i^2\phi \partial_j^2\phi-4\partial_i^2\sigma\partial_j^2\phi+\frac{3}{2}\partial_i^2\sigma\partial_j^2\sigma\right)\Big]\,.\label{1.6k}
    \end{split}
\end{align}
For convenience, we define the Fourier modes as \footnote{Three vectors are denoted by the boldfaced symbols.},
\begin{align}
    \begin{split}
        \phi(x)=\int d^3\boldsymbol{k}\,d\omega\,e^{i(\boldsymbol{k}\cdot\boldsymbol{x}-\omega\,t)}\tilde\phi(-\boldsymbol{k},\omega),\,\quad \sigma(x)=\int d^3\boldsymbol{k}\,d\omega\,e^{i(\boldsymbol{k}\cdot\boldsymbol{x}-\omega\,t)}\tilde\sigma(-\boldsymbol{k},\omega)\,.\label{1.7}
    \end{split}
\end{align}
Now putting \eqref{1.7} into \eqref{1.6k} we will get,
\begin{align}
    \begin{split}
       S^{(2)}=\frac{m_p^2}{2}\int d^3\boldsymbol{k}\,d\omega\Bigg[&\left(\frac{1}{2}\boldsymbol{k}^2\,\tilde\sigma(-k)\tilde\sigma(k)-2\boldsymbol{k}^2\,\tilde\phi(-k)\tilde\phi(k)\right)+4\beta\Big(\boldsymbol{k}^4\,\tilde\sigma(-k)\tilde\sigma(k)+\boldsymbol{k}^4\,\tilde\phi(-k)\tilde\phi(k)\\ &+2\boldsymbol{k}^4\tilde\phi(-k)\tilde\sigma(k)\Big)+\gamma\left(4\,\boldsymbol{k}^4\tilde\phi(-k)\tilde\phi(k)+\frac{3}{2}\boldsymbol{k}^4\,\tilde\sigma(-k)\tilde{\sigma}(k)-4\boldsymbol{k}^4\,\tilde{\phi}(-k)\tilde\sigma(k)\right)\Bigg]\,.\label{1.8}
    \end{split}
\end{align}
The action in \eqref{1.8} can be written as,
\begin{align}
    \begin{split}
        S^{(2)}=\frac{m_p^2}{2}\int d^3\boldsymbol{k}\,d\omega\,\begin{bmatrix}
\tilde\phi(k)& \tilde\sigma(k)  
\end{bmatrix}\boldsymbolcal{A}_{2\times2}[k]\begin{bmatrix}
\tilde\phi(-k) \\
\tilde\sigma(-k) 
\end{bmatrix}\,
    \end{split}
\end{align}
where the matrix $\boldsymbolcal{A}_{2\times 2}$ is given by,
\begin{align}
    \begin{split}
      \boldsymbolcal{A}_{2\times 2}[k]=  
    \left[
\begin{array}{cc}
-2 k^2+ 4 k^4   (\beta +\gamma ) & -k^4   (2 \beta +\gamma ) \\
 -k^4   (2 \beta +\gamma ) &  \frac{k^2}{2}+ 4 k^4 \left(\beta +\frac{3 \gamma }{8}\right) \\
\end{array}
\right]\,.
    \end{split}
\end{align}
Now, we would like to find a basis transformation,
\begin{align}
    \begin{bmatrix}
\tilde\phi(-k) \\
\tilde\sigma(-k) 
\end{bmatrix}\to \begin{bmatrix}
\tilde\Phi(-k) \\
\tilde\Sigma(-k) 
\end{bmatrix}=\boldsymbolcal{O} \begin{bmatrix}
\tilde\phi(-k) \\
\tilde\sigma(-k) 
\end{bmatrix},
\end{align}
such that there will be no cross-term in the quadratic action.
One can see that the matrix  $\boldsymbolcal{A}_{2\times 2}$ is a symmetric matrix and therefore can be diagonalized by an orthogonal transformation:
\begin{align}
    \begin{split}
        \boldsymbolcal{A}_{2\times 2}=\boldsymbolcal{O}^{T}\cdot\boldsymbolcal{D}_{2\times 2}\cdot \boldsymbolcal{O}
    \end{split}
\end{align}
where,
\begin{align}
    \begin{split}
        \boldsymbol{O}=\left(
\begin{array}{cc}
 1 & \frac{2}{5} k^2   (2 \beta +\gamma ) \\
- \frac{2}{5}  k^2  (2 \beta +\gamma ) & 1 \\
\end{array}
\right)+\cdots,\,\,\boldsymbolcal{D}_{2\times 2}=\left(
\begin{array}{cc}
-2 k^2+  4 k^4   (\beta +\gamma )& 0 \\
 0 & \frac{k^2}{2}+4 k^4  \left(\beta +\frac{3 \gamma }{8}\right) \\
\end{array}
\right)+\cdots
    \end{split}
\end{align}
Therefore, in terms of the new variable, the action can be written as,
\begin{align}
    \begin{split} S^{(2)}=\frac{m_p^2}{2}\int d^3\boldsymbol{k}\,d\omega\,\begin{bmatrix}
\tilde\Phi(k)& \tilde\Sigma(k)  
\end{bmatrix}\boldsymbolcal{D}_{2\times2}[k]\begin{bmatrix}
\tilde\Phi(-k) \\
\tilde\Sigma(-k) 
\end{bmatrix}\,.
    \end{split}
\end{align}
Now, the old fields can be written in terms of new variables as,
\begin{align}
    \begin{split}
      &  \tilde\phi(-k)\to \tilde\Phi(-k) -\frac{2}{5}  \boldsymbol{k}^2   (2\beta + \gamma )\tilde\Sigma(-k)+\cdots,\\ &
      \tilde\sigma(-k)\to \tilde\Sigma(-k) +\frac{2}{5} \boldsymbol{k}^2   (2\beta +\gamma )\tilde\Phi(-k)+\cdots.
    \end{split}
\end{align}
In position space, the relation reads to,
\begin{align}\label{trans}
    \begin{split}
    &    \phi(x)\to \Phi(x) +\frac{2}{5}(2\beta+\gamma)\nabla^2\,\Sigma(x)+\cdots,\\ &
    \sigma(x)\to \Sigma(x)-\frac{2}{5}(2\beta+\gamma)\nabla^2\Phi(x)+\cdots.
    \end{split}
\end{align}
%Similarly, in position space, the diagonal matrix (which becomes a differential operator) reads,
%\begin{align}
   % \begin{split}
    %   \hat{ \boldsymbolcal{D}}\left(\frac{\partial}{\partial x}\right)=\begin{pmatrix}
            %4(\gamma+\beta)\partial_{i}^2\pa%rtial_j^2 -2\partial_i^2& 0\\
      %      0 & \frac{1}{2}(3\gamma+8\beta)\partial_i^2\partial_j^2+\frac{1}{2}\partial_i^2
   %     \end{pmatrix}
 %   \end{split}
%\end{align}
\textbf{\textit{Constraints on parameters:}}
\\ 
\begin{align}
   \gamma+\beta<0 \,\,\,\,\,\,\& \,\,\,\,\,\,3\gamma+8\beta>0.\label{parameter_constraints}
\end{align}
So, to avoid tachyonic instability, one can choose $\gamma<0$ and $\frac{3}{8}|\gamma|<\beta<|\gamma|$.
With the transformations mentioned in (\ref{trans}), the action (\ref{action}) can be written in terms of the new field variables ($\Sigma(x), \Phi(x)$) linearized in ($\beta, \gamma$). As a result, we notice that in the action corresponding to $R^{2}$ and $R_{\mu\nu} R^{\mu\nu}$, the field variables will be replaced directly by ($\phi(x)\rightarrow \Phi(x), \sigma(x)\rightarrow \Sigma(x)$), since Eqs. (\ref{S22}), (\ref{S23}), (\ref{S32}) and (\ref{S33}) are already linearized in ($\gamma, \beta$). However, the action (\ref{EHAction}) can be written in terms of the transformed fields in the following manner:
%\begin{equation}
%\begin{aligned}\label{EHActionNewfield}
  %  \textcolor{black}{S_{1}^{(2)}} =& \alpha \frac{m_p^2}{2}\int dt d^3x \Big[\frac{1}{2}\partial_{i}\Sigma\partial^{i}\Sigma-2\delta^{ij}\partial_{i}\Phi\partial_{j}\Phi-\frac{4}{5}(\beta+2\gamma) \Big\lbrace\partial_{i}(\nabla^{2}\Phi)\partial^{i}\Sigma+4\delta^{ij}\partial_{i}(\nabla^{2}\Sigma)\partial_{j}\Phi \Big\rbrace\Big] + \mathcal{O}(\beta^{2}, \gamma^{2}), \\
   % \textcolor{black}{S_{1}^{(3)}} =& \alpha \frac{m_p^2}{2}\int dt d^3x \Big[-\frac{3}{4}\Sigma\partial_{i}\Sigma\partial^{i}\Sigma-\delta^{ij}\Sigma\partial_{i}\Phi\partial_{j}\Phi+\frac{3}{5}(\beta+2\gamma)\Big\lbrace 2\Sigma\partial_{i}(\nabla^{2}\Phi)\partial^{i}\Sigma+\nabla^{2}\Phi\partial_{i}\Sigma\partial^{i}\Sigma\Big\rbrace \\
    %& -\frac{4}{5}\delta^{ij}(\beta+2\gamma) \Big\lbrace 2\Sigma\partial_{i}\Phi\partial_{j}(\nabla^{2}\Sigma)-\nabla^{2}\Phi\partial_{i}\Phi\partial_{j}\Phi \Big\rbrace\Big] + \mathcal{O}(\beta^{2}, \gamma^{2}).
%\end{aligned}
%\end{equation}
% Quadratic part (S_1^(2)) after field redefinition:
\[
\begin{aligned}
\textcolor{black}{S_1^{(2)}} = \frac{ m_p^2}{2} \int dt\, d^3x\, \Biggl[
&\frac{1}{2}\,(\partial_i\Sigma)(\partial^i\Sigma) - 2\,(\partial_i\Phi)(\partial^i\Phi)\\[1mm]
& - \; \frac{2}{5}(2\beta+\gamma)\Bigl(
(\partial_i\Sigma)(\partial^i(\nabla^2\Phi))
+ 4\,(\partial_i\Phi)(\partial^i(\nabla^2\Sigma))
\Bigr)
\Biggr]\ + \mathcal{O}((\beta,\gamma)^2)\,,
\end{aligned}
\]
% Cubic part (S_1^(3)) after field redefinition:
\[
\begin{aligned}
\textcolor{black}{S_1^{(3)}} = \frac{ m_p^2}{2} \int dt\, d^3x\, \Biggl[& 
-\frac{3}{4}\,\Sigma\,(\partial_i\Sigma)(\partial^i\Sigma)
-\Sigma\,(\partial_i\Phi)(\partial^i\Phi)\\[1mm]
& + \frac{2}{5}(2\beta+\gamma)\Biggl(
\frac{3}{4}\,(\nabla^2\Phi)(\partial_i\Sigma)(\partial^i\Sigma)
+\frac{3}{2}\,\Sigma\,(\partial_i\Sigma)(\partial^i(\nabla^2\Phi))\\[1mm]
&\quad\quad\quad - 2\,\Sigma\,(\partial_i\Phi)(\partial^i(\nabla^2\Sigma))
+ (\nabla^2\Phi)(\partial_i\Phi)(\partial^i\Phi)
\Biggr)
\Biggr] + \mathcal{O}((\beta,\gamma)^2)\,.
\end{aligned}
\]%In terms of the parametrization given in \eqref{1.1a}, identifying $\beta\to \frac{1}{6f_0^2}-\frac{1}{\xi^2},\,\gamma\to -\frac{1}{\xi^2}$ the constraints looks like,  
As we are dealing with static black holes, the worldline action takes the form,
\begin{align}
    \begin{split}
        S_{\textrm{BH}}=-m\int dt\,\left[1+\frac{2}{5}(2\beta+\gamma)\nabla^2\Sigma(x)+\sum_{n=1}^{\infty}\frac{\Phi^n}{n!}\right]+\mathcal{O}\Big((2\beta+\gamma)^2\Big).
    \end{split}\label{pp_action}
\end{align}
\textbf{\textit{Bulk vertices in quadratic EFT:}}\\\\
We list down all the possible bulk 3-point vertices in the theory under consideration,
\begin{align}
    \begin{split}
       & \boldsymbolcal{V}_1[\Phi\Phi\Phi]
        %=\frac{4}{5}(\beta+2\gamma) \nabla^2\Phi\,\partial_i\Phi\partial_i\Phi+8\beta\, \nabla^2\,\Phi \partial_i\Phi\partial_i\Phi-8\beta \,\Phi\nabla^2\Phi\nabla^2\Phi-8\gamma\Phi\nabla^2\Phi\nabla^2\Phi+4\gamma \nabla^2\Phi \partial_i\Phi\partial_{i}\Phi
        =-\frac{18}{5} (2 \beta + \gamma )\nabla^2\Phi\,\partial_i\Phi\partial_i\Phi+8(\gamma+\beta)\Phi\nabla^2 \Phi\nabla^2\Phi,\\ &
        \boldsymbolcal{V}_2[\Sigma\Sigma\Sigma]=-\frac{3}{4}\Sigma \partial_i\Sigma\partial_i\Sigma-\left(6\beta+\frac{7\gamma}{4} \right)\,\nabla^2\Sigma\,\partial_i\Sigma\partial_i\Sigma-\left(10\beta+\frac{25}{8}\gamma\right)\,\Sigma\,\nabla^2\Sigma\,\nabla^2\Sigma\\&\hspace{8cm}-\frac{5\gamma}{8}\Sigma (\partial_i\partial_j\Sigma)^2-\frac{3\gamma}{4}\partial_i\Sigma\partial_i\Sigma\partial_i\partial_j\Sigma,\\ &
\boldsymbolcal{V}_3[\Sigma\Sigma\Phi]=\frac{3}{5}(2\beta+\gamma)\Sigma(\partial_i\Sigma)(\partial^i(\nabla^2\Phi))+\frac{33}{10}(2\beta+\gamma)\nabla^2\Phi\,\partial_i\Sigma\partial_i \Sigma+6(2\beta+\gamma)\Sigma\nabla^2\Sigma\nabla^2\Phi,\\&
        \boldsymbolcal{V}_4[\Sigma\Phi\Phi]=-\Sigma\partial_i\Phi\partial_i\Phi+2(4\beta+\gamma)\nabla^2\Sigma\partial_i\Phi\partial_i\Phi-2(\beta+\gamma)\Sigma\nabla^2\Phi\nabla^2\Phi+4(\beta+\gamma)\nabla^2\Phi\partial_i\Sigma\partial_i\Phi\\&\hspace{10 cm}-8(2\beta+\gamma)\nabla^2\Sigma\Phi\nabla^2\Phi.
    \end{split}
\end{align}
\textbf{\textit{Some key differences between GR and quadratic gravity:} }
\begin{itemize}
    \item After field redefinition, also in static gauge, the worldline action now has both the modes: $\Phi, \,\Sigma$.
    \item The bulk interaction vertex now consists of $\Sigma\Sigma\Phi,\,\Phi^3$ unlike GR.
\end{itemize}
 \subsection{Propagators and Feynman rules }
    \begin{itemize}
        \item The propagators for different bulk fields can now be derived from the quadratic part of the action (\ref{1.8}). Therefore, the propagator for $\Phi,\,\Sigma$ are given by \footnote{In vacuum, it is an expectation that quadratic curvature operators can be removed by local field redefinitions and therefore do not affect on-shell observables~\cite{Burgess:2003jk}.  
In contrast, in the presence of electromagnetic sources, recent analyses show that quadratic (higher–derivative) terms do modify the metric already at linear order in the corresponding couplings~\cite{Kats:2006xp,Cheung:2020gbf}. These studies are perturbative and typically retain only the massless branch of the solution; consequently, in the limit \(Q\to 0\) (vanishing electric charge) the higher–curvature corrections to the metric disappear—consistent with our explicit solution. When the bulk action is supplemented by the infinite tower of worldline Wilson coefficients, field redefinitions act on both the bulk fields and the worldline sector; accordingly, they induce shifts in the geodesic equation and in derived observables. Physical quantities remain invariant once the induced redefinitions are taken into account, but intermediate expressions can differ across field bases. The complementary, massive branch generates short–range (Yukawa–suppressed) effects that reduce in the classical limit to contact interactions represented by delta–function potentials. Despite their distributional nature, exponentiation of the eikonal phase in the classical limit permits the extraction of finite, physically meaningful observables from such contact terms with suitable regularization~\cite{Bhattacharyya:2025nfp}.
   }, 
        \vspace{1 cm}
        \hfsetfillcolor{white!10}
\hfsetbordercolor{black}
\begin{align}
\begin{split}
\tikzmarkin[disable rounded corners=true]{esp1qo}(0.5,-0.8)(-0.4,0.8) 
    &\langle\Phi(x_1)\,\Phi(x_2)\rangle
    =\frac{1}{m_p^2}\,\delta(t_1-t_2)\!\int_{\boldsymbol k}
      \frac{e^{i\boldsymbol k\cdot (\boldsymbol x_1-\boldsymbol x_2)}}
           {-4\boldsymbol k^4(\gamma+\beta)+2\boldsymbol k^2}
    \quad\equiv\quad
    \begin{tikzpicture}[
      baseline={(0,0)},           
      decoration={
        snake,
        amplitude=1mm,
        segment length=4mm
      }
    ]
      \draw[
        decorate,
        draw=phiColor,
        line width=2pt
      ] (0,0) -- (2,0) node[midway, above] {$\Phi$};
    \end{tikzpicture}\,,
    \\[2ex]
    &\langle\Sigma(x_1)\,\Sigma(x_2)\rangle
    =-\frac{2}{m_p^2}\,\delta(t_1-t_2)\!\int_{\vec k}
      \frac{e^{i\boldsymbol k \cdot (\boldsymbol x_1-\boldsymbol x_2)}}
           {(8\beta+3\gamma)\,\boldsymbol k^4+\boldsymbol k^2}
    \quad\equiv\quad
    \begin{tikzpicture}[
      baseline={(0,0)},
      decoration={
        snake,
        amplitude=1mm,
        segment length=4mm
      }
    ]
      \draw[
        decorate,
        draw=sigmaColor,
        line width=2pt
      ] (0,0) -- (2,0);
      \draw[decorate,
      draw=sigmaColor,
      line width=2pt
      ](0,0.1)--(2,0.1) node[midway, above]{$\Sigma$};
    \end{tikzpicture}\,.
    \tikzmarkend{esp1qo}\label{2.26m}
\end{split}
\end{align}

%\begin{align}
%\begin{split}
  %  \tikzmarkin[disable rounded corners=true]{esp1qo}(0.7,-0.5)(-0.4,0.8) &\langle\Phi(x_1)\,\Phi(x_2)\rangle=\hspace{ 0.1 cm}\begin{minipage}[h]{0.15\linewidth}
	%\vspace{-4pt}
	%\scalebox{0.19}{\begin{feynman}
    %\gluon[lineWidth=6, showArrow=false, color=9A081B, label=$\Phi$]{4.00, 4.00}{7.40, 4.00}
%\end{feynman}
%}
%\end{minipage}\hspace{ -0.3 cm}=\frac{1}{m_p^2}\delta(t_1-t_2)\int_{\vec k}\frac{e^{i\boldsymbol k\cdot (\boldsymbol{x}_1-\boldsymbol{x}_2)}}{-4\boldsymbol{k}^4(\gamma+\beta)+2\boldsymbol{{k}}^2}\\ &
%\langle\bold\Sigma(x_1)\,\bold\Sigma(x_2)\rangle=\hspace{ 0.1 cm}\begin{minipage}[h]{0.15\linewidth}
%	\vspace{-4pt}
%	\scalebox{0.19}
 %   {\begin{feynman}
    %\gluon[lineWidth=4, color=417505, label=$\Sigma$]%{4.00, 4.00}{7.40, 4.00}
%\end{feynman}
%}
%\end{minipage}\hspace{ -0.3 cm}=-\frac{2}{m_p^2}\delta(t_1-t_2)\int_{\boldsymbol{ k}}\frac{e^{i\boldsymbol k\cdot (\boldsymbol{x}_1-\boldsymbol{x}_2)}}{(8\beta+3\gamma)\boldsymbol{k}^4+\boldsymbol{k}^2}  \tikzmarkend{esp1qo}\label{2.26m}\end{split}
%\end{align}
\vspace{1cm}
    \item We also show the Feynman rules in Fig.~(\ref{fig0}) for the insertion of $\Phi$ and $\Sigma$ into the worldline as they will be useful later. They are as follows: \begin{figure}[ht!]
        \centering
\includegraphics[width=0.6\linewidth]{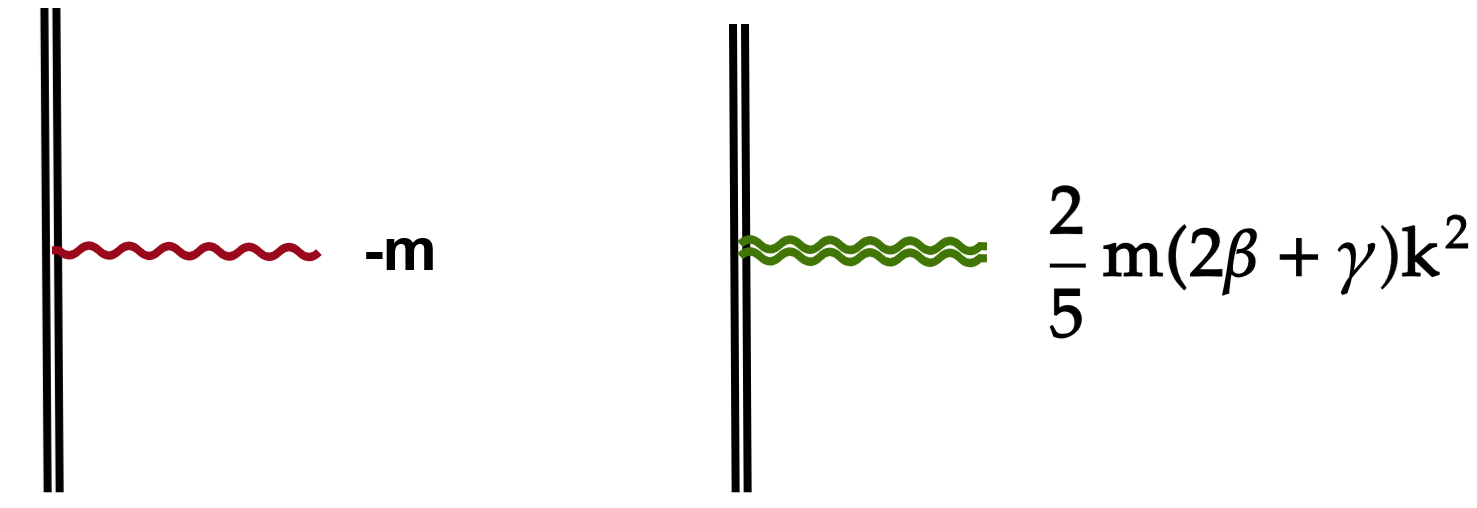}
        \caption{ Figure showing the Worldline insertions of the redefined $\Phi,\Sigma$ fields with specific vertex factors.}
        \label{fig0}
    \end{figure}
    \end{itemize}
    \section{PN counting} \label{sec3}
Before proceeding with one-point function computation, we briefly discuss the post‐Newtonian (PN) counting as we will perform the metric reconstruction only up to some PN order. For a worldline EFT with the bulk theory described by GR, each insertion of the minimal point‐mass coupling on the worldline \footnote{For simplicity, the discussion in this section proceeds in a unit where $8\pi G=c=1$ so that mass and length carry the same dimension.}
\[
S_{\rm pp}\supset -\int m\,\phi
\]
carries exactly one factor of $m/r\sim v^2$, so that an $N$‐fold insertion generates  $N$‑PN effect, and the overlap with an $r^{-(\ell+1)}$ “response” term always comes from the $2\ell+1$ PN diagrams
with $2\ell+1$ worldline vertices. In a modified‐gravity theory with additional higher‑curvature or extra‐field interactions, however, the PN power‐counting must be revisited:
\begin{itemize}
  \item For GR, only the insertion of $m\,\phi$ on the worldline  generates the $m/r$ suppression, hence
    \[
      {\rm PN\ order} \;=\; \#\{\text{worldline insertions of }m\,\phi\}\,.
    \]
    To get a fall off like $r^{-(\ell+1)}$ one still needs exactly $2\ell+1$ such insertions.

  \item For higher-curvature gravity, each new higher‐derivative bulk interaction introduces its own $v$-scaling independent of the worldline insertions. A diagram with fewer than $2\ell+1$ point mass vertices and one (or more) of these new bulk vertices can still scale like $(m/r)^{2\ell+1}\,$ and hence mimic the same $r^{-(\ell+1)}$ profile.
\end{itemize}
\subsection{``Modified'' power counting for quadratic gravity}
For quadratic gravity gravity it is evident from the action (\ref{action}) that the dimension of the couplings $\beta\,,\gamma$ is of $L^2$ ($\textrm{Length}^2$). Keeping this in mind, we now revisit the PN scaling:
\begin{figure}[htbp]
    \centering
\includegraphics[width=0.58\linewidth]{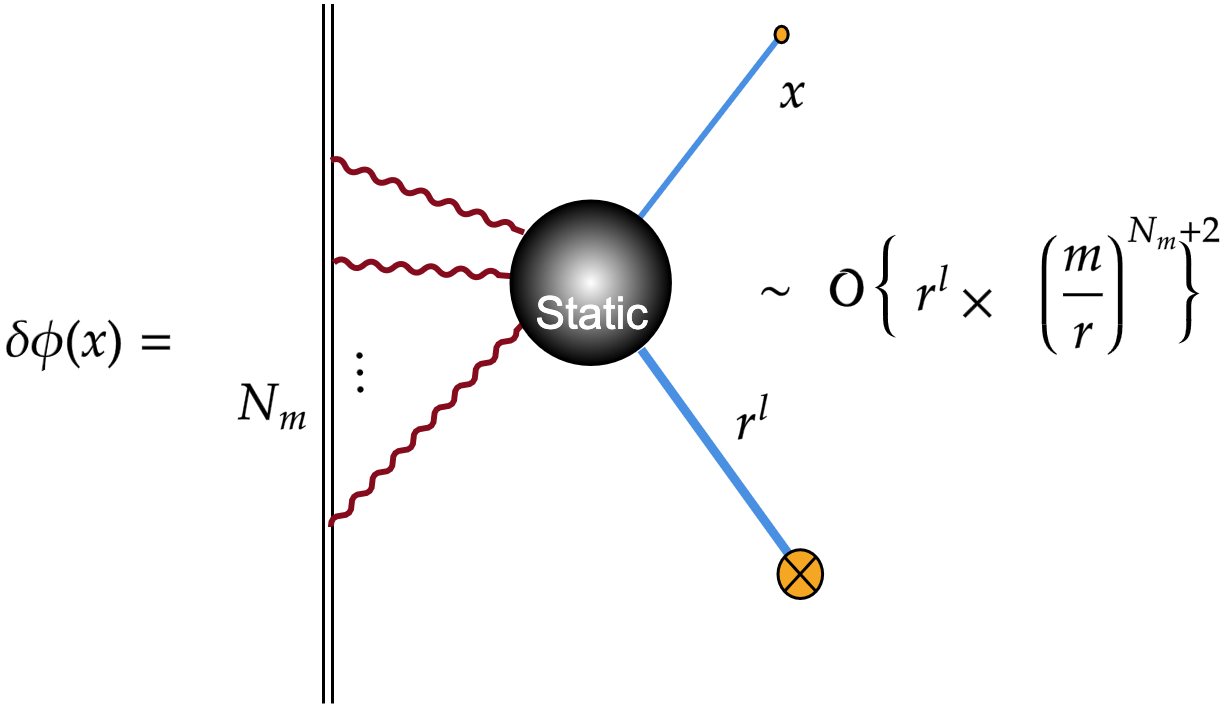}
    \caption{Schematic representation of diagrams that we compute to reconstruct the metric. The blob is made of various higher derivative terms, which are connected by propagators in the static gauge. The  GR power counting is modified by the higher derivative vertices in the bulk.}
    \label{fig5}
\end{figure}
\begin{enumerate}
  \item \textbf{Propagators:}  Each internal graviton or scalar line scales as
  \(\sim1/r\).
  \item \textbf{GR vertices (\(R\) term):}  Two spatial derivatives \((\partial\partial)\sim1/r^2\)
  and \(\int d^3x\sim r^3\) combine to give a net scaling \(\sim r\).
  \item \textbf{\(R^2\) type vertices:}  
    Four spatial derivatives \((\partial)^4\sim1/r^4\) and \(\int d^3x\sim r^3\) give
    \(\sim1/r\).  Each such vertex carries one factor of \(\alpha\).
  \item \textbf{Worldline insertions:}  Each insertion of \(-m\!\int dt\,\varphi\)
  contributes one factor of \(m\).
  \item \textbf{Classical topology constraint:}  For tree‐level diagrams,
  \(P_h+P_{\delta\varphi}=V-1\), where \(P_h,P_{\delta\varphi}\) are internal lines
  and \(V=V_{\rm GR}+V_{R^2}\) the total number of bulk vertices.
\end{enumerate}
Combining these factors, the “blob” (before the external field propagator)
scales as
\[
m^{N_m}\;\frac{1}{r^{P_h+P_{\delta\varphi}}}
\;r^{V_{\rm GR}}\;\frac{1}{r^{V_{R^2}}}
\;=\;
m^{N_m}\;r^{\,V_{\rm GR}-(P_h+P_{\delta\varphi})}\,\frac{1}{r^{V_{R^2}}}
\;=\;
m^{N_m}\;r^{\,1-V_{R^2}}.
\]
Attaching the external propagator \(1/r\) to reach the field point gives
\[
\delta\varphi(x)\sim
r^\ell\;
\frac{m^{N_m}\,r^{1-V_{R^2}}}{r}
=
r^\ell\;(m/r)^{N_m}\;
\frac{1}{r^{V_{R^2}}}.
\]
Since each \(R^2\) vertex introduces an extra factor \(1/r\) and one power of \(\alpha\),
and \(\alpha/r^2\sim(m/r)^2\) in PN counting, each \(R^2\) insertion raises the PN order by two:
\[
\delta\varphi(x)\sim
r^\ell\;(m/r)^{N_m+2\,V_{R^2}}.
\]
Therefore, for one insertion to the worldline $N_m=1$ and each \(R^2\) bulk vertex, the PN order goes as $(m/r)^3$, i.e., 3PN order (in contrast to 1PN order in GR) as schematically shown in Fig.~\eqref{fig5}. \\\\
\noindent
Now, having the power counting rules in hand, we proceed to compute the one-point functions of $\phi(x)$ and $\sigma(x)$ for reconstructing the metric via the methods of effective field theory \cite{Ivanov:2022hlo}.

 \section{Perturbative metric reconstruction from one-point function in QFT}\label{sec4}
 In this section, we discuss the methodology used to perturbatively reconstruct the metric.
\subsection*{Tree level one-point function as a solution of classical e.o.m:}
Let us consider an action of the form:
\begin{align}
    \begin{split}
        S[g,T]=S_{g}[g]+S_{\textrm{int}}[g,T],
    \end{split}
\end{align}
where $T^{\mu\nu}$ is the physical source. Our goal is to solve the classical equation of motion (e.o.m.), i.e.,
\begin{align}
    \frac{\delta S[g,T]}{\delta g}=0.
\end{align}
Quantum field theory gives a very efficient way to solve this equation of motion by defining the generating functional $\mathbb{W}$ as,
\begin{align}
    \begin{split}
        e^{\frac{i}{\hbar}\mathbb{W}}=\int [\mathcal{D}g_{\mu\nu}]\,e^{\frac{i}{\hbar}S[g,T]+\frac{i}{\hbar}\int g_{\mu\nu}\hat{T}^{\mu\nu}},
    \end{split}
\end{align}
where $\hat T$ is the fiducial source. Now, the one-point function in the in-out formalism of QFT is defined as,
\begin{align}
    \langle g\rangle_{\textrm{in-out}}=\frac{\delta \mathbb{W}[\hat T]}{\delta \hat T}\Big|_{\hat T=0}\,.
\end{align}
Furthermore, one can define quantum effective action as a Legendre transformation of the generating functional and also find the quantum effective e.o.m.
\begin{align}
    A_{\textrm{eff}}=\frac{i}{\hbar}\int g_{\mu\nu}\hat T^{\mu\nu}-\mathbb{W}[\hat{T}]\to \frac{\delta A_{\textrm{eff}}}{\delta g}\Big|_{g=\langle g(x)\rangle}=0\,.
\end{align}
Now, the one-point function $\langle g(x)\rangle$ at tree level solves the classical e.o.m.

\subsection*{Computing the one-point function:}

\begin{figure}[htb!]
\centering
\includegraphics[width=0.9\linewidth]{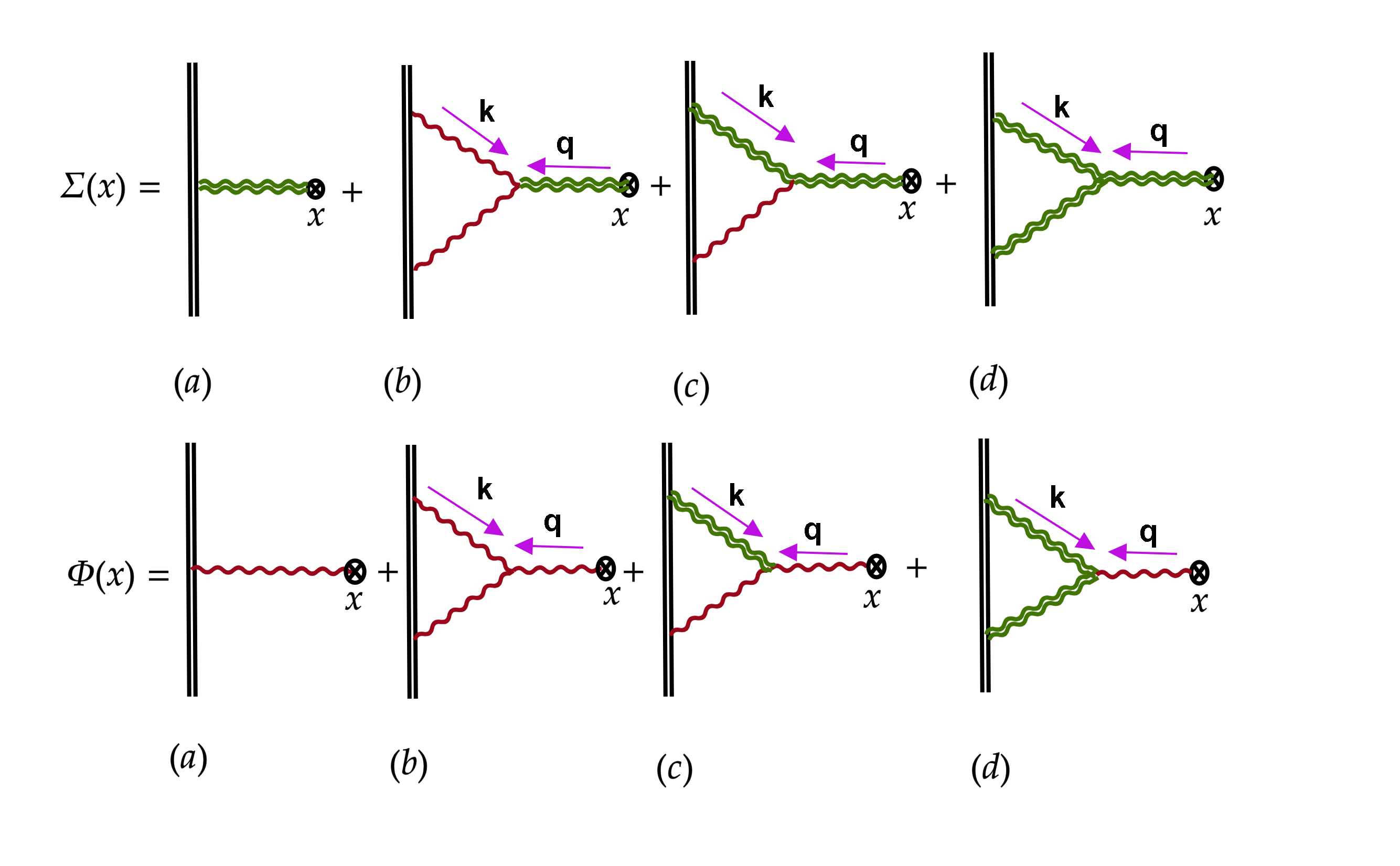}
\caption{Feynman diagrams contributing to the computation of the one-point function of $\Sigma$ and $\Phi$ up to 1PN. We need to sum up these diagrams to perturbatively reconstruct the metric. }\label{fig1}
\end{figure}

In $3+1$-decomposition, the path integral reduces to (for the static case),
\begin{align}
   \int \mathcal{D}g_{\mu\nu}\to \int\mathcal{D}\Phi\,\,\mathcal{D}\Sigma_{ij},
\end{align}
and the one-point function we intend to compute is given by,
\begin{align}
    \begin{split}
       \langle \Phi,\Sigma_{ij}\rangle=\frac{1}{Z}\int [\mathcal{D}\Phi\,\,\mathcal{D}\Sigma_{kl}]\,\{\Phi(x),\Sigma_{ij}(x)\}\exp{\left(iS[\Phi,\Sigma,T]\right)}.
    \end{split}
\end{align}
Let us calculate the one-point function of $\Phi(x)$. Before that, let us reflect on the Feynman diagrams (Fig.(\ref{fig1})) that contribute at \textit{1PN} order. We note that a $\Sigma$ insertion to the worldline contributes a factor of $(2\beta+\gamma)$. Therefore, at the linearized level, we cannot have diagrams with two insertions of $\Sigma$ to the worldline. Similarly, for the $\Sigma\Phi\Phi$ vertex, only the first term will contribute, namely $-\Sigma\partial_i\Phi\partial_i\Phi$ to the one-point function of $\Phi$ since other terms already carry $(\beta,\gamma)$ factors which multiplied with $(2\beta+\gamma)$ factor of the worldline (coming from the insertion of $\Sigma$) will become higher order in $(\beta,\gamma)$. This means that for the one-point function $\Phi$, only the first three diagrams will contribute, while for $\Sigma$, only the first two diagrams will contribute.
We now proceed with the calculations\footnote{Note that $\int_q$ is a shorthand for $\int\frac{d^3q}{(2\pi)^3}\,.$ Also, $|\boldsymbol{ r}|=r\,.$}.
\begin{align}
    \begin{split}
        \text{\textbullet}\,\,\langle\Phi(x)\rangle_{(a)}=-m\int dt_1 \left\langle\Phi(\boldsymbol{x}_1(t_1))\,\Phi(\boldsymbol{x},t)\right\rangle&=-\frac{m}{m_p^2}\int dt_1\delta(t_1-t)\int_{\boldsymbol{q}}\frac{e^{i\boldsymbol{q}\cdot(\boldsymbol{x}_1-\boldsymbol{x})}}{-4(\gamma+\beta)\boldsymbol{q}^4+2\boldsymbol{q}^2}\,,\\ &
        =-\frac{m}{m_p^2}\int_{\boldsymbol{q}}\frac{e^{i\boldsymbol{q}\cdot \boldsymbol{r}}}{-4(\gamma+\beta)\boldsymbol{q}^4+2\boldsymbol{q}^2}\,,\\ &
        =-\frac{m}{2m_p^2} \int_{\boldsymbol{q}}e^{i\boldsymbol{q}\cdot \boldsymbol{r}}\left(\frac{1}{\boldsymbol{q}^2}-\frac{1}{\boldsymbol{q}^2+\delta_1^2}\right)\,,\\&=-\frac{m}{8\pi m_p^2}\frac{\left(1-e^{-\delta_1\,r}\right)}{r},\,\delta_1^2=-\frac{1}{2(\gamma+\beta)}\,.
    \end{split}
\end{align}
The second diagram renders the following contribution to the one-point function,
\begin{align}
    \begin{split}
\text{\textbullet}\langle\Phi(x)\rangle_{(b)}&=\int \boldsymbolcal{D}[\Phi,\Sigma]\,\Phi(\boldsymbol{x},t)\left(\prod_{i=1}^2\int dt_i \Phi(\boldsymbol{x}_1,t_i)\right)\\&\hspace{3.5 cm}\times\int d^3\boldsymbol{z}\,ds \left[-\frac{9}{5} (2 \beta + \gamma )\nabla^2\Phi\,\partial_i\Phi\partial_i\Phi+8(\gamma+\beta)\Phi\nabla^2 \Phi\nabla^2\Phi\right]\,,\\ &
        =\int dt_1 dt_2\,d^3\boldsymbol{z}\,ds\,\Bigg[ -\frac{9}{5}(2\beta+\gamma)\nabla_{\boldsymbol{(z)}}^2\langle\Phi(\boldsymbol{x},t)\Phi(\boldsymbol{z},s)\rangle \partial^{\boldsymbol{(z)}}_{i}\langle\Phi(\boldsymbol{x}_1,t_1)\Phi(\boldsymbol{z},s)\rangle \partial^{\boldsymbol{(z)}}_{i}\langle\Phi(\boldsymbol{x}_1,t_2)\Phi(\boldsymbol{z},s)\rangle \\ &\hspace{3.5 cm}+4(\gamma+\beta)\langle\Phi(\boldsymbol{x},t)\Phi(\boldsymbol{z},s)\rangle\,\nabla_{\boldsymbol{(z)}}^2\langle\Phi(\boldsymbol{x}_1,t_1)\Phi(\boldsymbol{z},s)\rangle \,\nabla_{\boldsymbol{(z)}}^2\langle\Phi(\boldsymbol{x}_1,t_2)\Phi(\boldsymbol{z},s)\rangle \Bigg]\,.\label{fig2_integral}
    \end{split}
\end{align}
This can be written as
\begin{align}
    \begin{split}
        &\langle\Phi(x)\rangle_{(b)}=\frac{1}{m_p^4}\int\,dt_1dt_2dsd^3z\delta(t_1-s)\delta(t_2-s)\delta(t-s)\Bigg[-\frac{9}{5}(2\beta+\gamma)\int_{k,q}\Big\{\partial_i\frac{e^{ik\cdot(x_1-z)}}{-4k^4(\gamma+\beta)+2k^2}\\&\hspace{3cm}\partial_i\frac{e^{-i(k+q)\cdot(x_1-z)}}{-4(k+q)^4(\gamma+\beta)+2(k+q)^2}\partial_i^2\frac{e^{iq\cdot(x-z)}}{-4q^4(\gamma+\beta)+2q^2}\Big\}\\&\hspace{3cm}+4(\gamma+\beta)\int_{k,q}\Big\{\partial_i^2\frac{e^{ik\cdot(x_1-z)}}{-4k^4(\gamma+\beta)+2k^2}\partial_i^2\frac{e^{-i(k+q)\cdot(x_1-z)}}{-4(k+q)^4(\gamma+\beta)+2(k+q)^2}\\&\hspace{6cm}\times\frac{e^{iq\cdot(x-z)}}{-4q^4(\gamma+\beta)+2q^2}\Big\}+4(\gamma+\beta)\Big\}\Bigg]\,.
    \end{split}
\end{align}
The integral evaluates to
\begin{equation}
   \langle\Phi(x)\rangle_{(b)}=-\frac{9}{5}(2\beta+\gamma)I_1+4(\beta+\gamma)I_2
\end{equation}
where, 
\begin{align}
\begin{split}
    I_1&=\frac{m_1^2}{16}\int_{\boldsymbol q}\frac{e^{i{\boldsymbol q}\cdot {\boldsymbol r}}}{({\boldsymbol q}^2+m_1^2)}\mathcal{I},\,\,\,\,\,\,\,
     I_2=\frac{m_1^4}{16}\int_{\boldsymbol q}\frac{e^{i {\boldsymbol q}\cdot {\boldsymbol r}}}{{\boldsymbol q}^2({\boldsymbol q}^2+m_1^2)}\mathcal{I}
     \end{split}
\end{align}
Here $m_1^2=-\frac{1}{2(\gamma+\beta)}$. Furthermore, for calculation purposes, we split the integral $\mathcal{I}$ into massless, mixed and massive types and then sum these integrals to get the total contribution as
\begin{align}
\begin{split}
     &\mathcal{I}(\text{total})= \mathcal{I}(\text{massless type})+\mathcal{I}(\text{mixed type})+\mathcal{I}(\text{massive type}).%\\&
   % \hspace{1.3cm}=\mathcal{I}_1+\mathcal{I}_2+(\mathcal{I}_3+\mathcal{I}_4).
\end{split}
\end{align}
The integral $\mathcal{I}$ with both legs massive is given by (refer to Appendix~\ref{app1} for explicit calculations of the Master integral $\mathcal{D}_{ab}$):
\begin{equation}
\mathcal{I}(\text{massive type})=2\mathcal{D}_{01}-(2m_1^2+\boldsymbol{q}^2)\mathcal{D}_{11}.
\end{equation}
We substitute (in $\mathcal{I}(\text{massive type)}$) $m_1=m_2=0$ to get the integral with both legs massless ($\mathcal{M}$-type) and $m_1=0$ or $m_2=0$ to obtain the integral with one massless and one massive leg ($\mathcal{K}$-type). However, note that $m_1\neq0$ for $I_2$, which follows from the structure of the one-point function. The relevant integrals are listed in Appendix~\ref{integrals}.
%\hfsetfillcolor{white!14}
%\hfsetbordercolor{black}

%The massless type integral can be readily calculated as,
%\begin{align}
 %   J_1(\text{massless}) =-\frac{m_1^2}{128\pi^2m_p^4}\Big[\frac{1}{r^2}-\frac{m_1\pi e^{-m_1r}}{2r}\Big].
%\end{align}
The one-point function is given as,
\begin{equation}
    \langle\Phi(x)\rangle_{(b)}=-\frac{9}{5}(2\beta+\gamma)(I_{1a}+I_{1b}+I_{1c})+4(\beta+\gamma)I_2.
\end{equation}
The remaining diagrams can be cast as (up to linear order in couplings),
\begin{align}
    \begin{split}
\text{\textbullet}\,\langle\Phi(x)\rangle_{(c)}&=\int dt_1 dt_2\,d^3\boldsymbol{z}\,ds\,\Bigg[-\frac{2}{5}(2\beta+\gamma)\langle\Sigma(\boldsymbol{x}_1,t_1)\Sigma(\boldsymbol{z},s)\rangle \\&\hspace{2cm}\times\partial^{\boldsymbol{(z)}}_{i}\langle\Phi(\boldsymbol{x}_1,t_2)\Phi(\boldsymbol{z},s)\rangle \partial^{\boldsymbol{(z)}}_{i}\langle\Phi(\boldsymbol{x},t)\Phi(\boldsymbol{z},s)\rangle\Bigg]+\mathcal{O}\Big((\beta,\gamma)^2\Big)\,. %\\ &\hspace{3.5 cm}%-2(\gamma+\beta)\langle\Sigma(\boldsymbol{x}_1,t_1)\Sigma(\boldsymbol{z},s)\rangle\,\nabla_{\boldsymbol{(z)}}^2\langle\Phi(\boldsymbol{x}_1,t_2)\Phi(\boldsymbol{z},s)\rangle \,\nabla_{\boldsymbol{(z)}}^2\langle\Phi(\boldsymbol{x},t)\Phi(\boldsymbol{z},s)\rangle \\&\hspace{3.5 cm}+2(
        %\gamma+4\beta)\nabla_{\boldsymbol{(z)}}^2\langle\Sigma(\boldsymbol{x}_1,t_1)\Sigma(\boldsymbol{z},s)\rangle\partial^{\boldsymbol{(z)}}_{i}\langle\Phi(\boldsymbol{x}_1,t_2)\Phi(\boldsymbol{z},s)\rangle \partial^{\boldsymbol{(z)}}_{i}\langle\Phi(\boldsymbol{x},t)\Phi(\boldsymbol{z},s)\rangle\\&+4(\beta+\gamma)\partial_i^{\boldsymbol{(z)}}\langle\Sigma(\boldsymbol{x}_1,t_1)\Sigma(\boldsymbol{z},s)\rangle\nabla^2_{\boldsymbol{(z)}}\langle\Phi(\boldsymbol{x}_1,t_2)\Phi(\boldsymbol{z},s)\rangle \partial^{\boldsymbol{(z)}}_{i}\langle\Phi(\boldsymbol{x},t)\Phi(\boldsymbol{z},s)\rangle\Bigg]
    \end{split}
\end{align}
This can be written as
\begin{align}
    \begin{split}
&\langle\Phi(x)\rangle_{(c)}=\frac{1}{m_p^4}\int\,dt_1dt_2dsd^3z\delta(t_1-s)\delta(t_2-s)\delta(t-s)\\&\Bigg[-\frac{2}{5}(2\beta+\gamma)\Big\{\int_{k,q}\frac{e^{ik\cdot(x_1-z)}}{k^4(3\gamma+8\beta)+k^2}\partial_i\frac{e^{-i(k+q)\cdot(x_1-z)}}{-4(k+q)^4(\gamma+\beta)+2(k+q)^2}\partial_i\frac{e^{iq\cdot(x-z)}}{-4q^4(\gamma+\beta)+2q^2}\Big\}\Bigg]\,.
    \end{split}
\end{align}
The integral evaluates to
\begin{equation}
\langle\Phi(x)\rangle_{(c)}=-\frac{2}{5}(2\beta+\gamma)J_1;\,\,\, J_1=\frac{m_1^2}{8}\int_q\frac{e^{i {\boldsymbol q}\cdot {\boldsymbol r}}}{{\boldsymbol q}^2({\boldsymbol q}^2+m_1^2)}\mathcal{H}\,.
\end{equation}
Again, we split the integral $\mathcal{H}$ into massless, mixed and massive types and then sum these integrals to get the total contribution as,
\begin{align}
\begin{split}
     &\mathcal{H}(\text{total})= \mathcal{H}(\text{massless type})+\mathcal{H}(\text{mixed type})+\mathcal{H}(\text{massive type}).
\end{split}
\end{align}
The integral $\mathcal{H}$ with both legs massive is given by, 
\begin{equation}
\mathcal{H}(\text{massive type)}=\mathcal{J}_{10}+\mathcal{J}_{01}-(-m_1^2+m_2^2+{\boldsymbol q}^2)\mathcal{J}_{11}.
\end{equation}
Here, $m_2^2=1/(8\beta+3\gamma)$. We substitute (in $\mathcal{H}(\text{massive type})$) $m_1=m_2=0$ to get the integral with both legs massless ($\mathcal{M}$-type) and $m_1=0$ or $m_2=0$ to obtain the integral with one massless and one massive leg ($\mathcal{K}$-type). 
This gives (refer to Appendix~\ref{app1} for more details about the Master integrals),
Therefore, the one-point function reads
\begin{equation}
   \langle\Phi(x)\rangle_{(c)}=-\frac{2}{5}(2\beta+\gamma)(J_{1a}+J_{1b}+J_{1c}). 
\end{equation}
Let us now evaluate the one-point of $\Sigma(x)$. The re-defined field $\Sigma$ can enter the worldline since the point-particle action (\ref{pp_action}) now also contains the redefined field $\Sigma$  with the following contribution  to the one-point function given by,
\begin{align}
    \begin{split}
        \text{\textbullet}\,\,\langle\Sigma(x)\rangle_{(a)}&=\frac{2}{5}(2\beta+\gamma)\int dt_1 \left\langle\Sigma(\boldsymbol{x}_1(t_1))\,\Sigma(\boldsymbol{x},t)\right\rangle\,,\\&=-\frac{4}{5m_p^2}m(2\beta+\gamma)\int dt_1\delta(t_1-t)\int_{\boldsymbol{\ell}}\frac{e^{i\boldsymbol{q}\cdot(\boldsymbol{x}_1-\boldsymbol{x})}\,\boldsymbol{q}^2}{(8\beta+3\gamma)\boldsymbol{q}^4+\boldsymbol{q}^2},\\ &
        =-\frac{4}{5m_p^2}m(2\beta+\gamma)\delta^2\int_{\boldsymbol{q}}\frac{e^{i\boldsymbol{q}\cdot \boldsymbol{r}}}{\boldsymbol{q}^2+\delta^2},\,\delta^2=\frac{1}{(8\beta+3\gamma)}\,,\\ &
        =-\frac{4m\delta^2(2\beta+\gamma)}{5m_p^2}\frac{\left(e^{-\delta\,r}\right)}{4\pi r}.
    \end{split}
\end{align}
\begin{align}
    \begin{split}
\text{\textbullet}\,\langle\Sigma(x)\rangle_{(b)}&=\int dt_1 dt_2\,d^3\boldsymbol{z}\,ds\\&\times\Bigg[-\langle\Sigma(\boldsymbol{x},t)\Sigma(\boldsymbol{z},s)\rangle\partial^{\boldsymbol{(z)}}_{i}\langle\Phi(\boldsymbol{x}_1,t_1)\Phi(\boldsymbol{z},s)\rangle \partial^{\boldsymbol{(z)}}_{i}\langle\Phi(\boldsymbol{x}_1,t_2)\Phi(\boldsymbol{z},s)\rangle\\&+2(4\beta+\gamma)\nabla^2_{\boldsymbol{(z)}}\langle\Sigma(\boldsymbol{x},t)\Sigma(\boldsymbol{z},s)\rangle\partial^{\boldsymbol{(z)}}_{i}\langle\Phi(\boldsymbol{x}_1,t_1)\Phi(\boldsymbol{z},s)\rangle \partial^{\boldsymbol{(z)}}_{i}\langle\Phi(\boldsymbol{x}_1,t_2)\Phi(\boldsymbol{z},s)\rangle\\&-2(\beta+\gamma)\langle\Sigma(\boldsymbol{x},t)\Sigma(\boldsymbol{z},s)\rangle\nabla^2_{\boldsymbol{(z)}}\langle\Phi(\boldsymbol{x}_1,t_1)\Phi(\boldsymbol{z},s)\rangle \nabla^2_{\boldsymbol{(z)}}\langle\Phi(\boldsymbol{x}_1,t_2)\Phi(\boldsymbol{z},s)\rangle\\&+4(\beta+\gamma) \partial^{\boldsymbol{(z)}}_{i}\langle\Sigma(\boldsymbol{x},t)\Sigma(\boldsymbol{z},s)\rangle \partial^{\boldsymbol{(z)}}_{i}\langle\Phi(\boldsymbol{x}_1,t_1)\Phi(\boldsymbol{z},s)\rangle \nabla^2_{\boldsymbol{(z)}}\langle\Phi(\boldsymbol{x}_1,t_2)\Phi(\boldsymbol{z},s)\rangle\\&-8(2\beta+\gamma)\nabla^2_{\boldsymbol{(z)}}\langle\Sigma(\boldsymbol{x},t)\Sigma(\boldsymbol{z},s)\rangle \langle\Phi(\boldsymbol{x}_1,t_1)\Phi(\boldsymbol{z},s)\rangle \nabla^2_{\boldsymbol{(z)}}\langle\Phi(\boldsymbol{x}_1,t_2)\Phi(\boldsymbol{z},s)\rangle\Bigg]\,.
    \end{split}
\end{align}
The integral evaluates to
\begin{align}
\begin{split}
        \langle\Sigma(x)\rangle_{(b)}&=-K_1+2(4\beta+\gamma)K_2-2(\beta+\gamma)K_3+4(\beta+\gamma)K_4\\&\hspace{5mm}-8(2\beta+\gamma)K_5,
\end{split}
\end{align}
where 
\begin{align}
\begin{split}
    &K_1=\frac{m_2^2}{8}\int_{\boldsymbol q}\frac{e^{i {\boldsymbol q}\cdot {\boldsymbol r}}}{{\boldsymbol q}^2({\boldsymbol q}^2+m_2^2)}\mathcal{G},\,\,\,\,\,\,\,
     K_2=\frac{m_2^2}{8}\int_{\boldsymbol q}\frac{e^{i{\boldsymbol q}\cdot {\boldsymbol r}}}{({\boldsymbol q}^2+m_2^2)}\mathcal{G},\\&K_3=\frac{m_2^2m_1^4}{4}\int_{\boldsymbol q}\frac{e^{i{\boldsymbol q}\cdot {\boldsymbol r}}}{{\boldsymbol q}^2({\boldsymbol q}^2+m_2^2)}\mathcal{D}_{11},\,\,\,\,\, K_4=\frac{m_2^2m_1^2}{8}\int_{\boldsymbol q}\frac{e^{i{\boldsymbol q}\cdot {\boldsymbol r}}}{{\boldsymbol q}^2({\boldsymbol q}^2+m_2^2)}\widetilde{\mathcal{G}},\\&
     K_5=\frac{m_2^2m_1^2}{4}\int_{\boldsymbol q}\frac{e^{i{\boldsymbol q}\cdot{\boldsymbol  r}}}{({\boldsymbol q}^2+m_2^2)}(\mathcal{K}_{11}+\mathcal{D}_{11}).
     \end{split}
\end{align}
The integral $\mathcal{G}$ and $\widetilde{\mathcal{G}}$ are given as
\begin{align}
    \begin{split}
        \mathcal{G}=\mathcal{D}_{01}+\mathcal{D}_{10}-({\boldsymbol q}^2+2m_1^2)\mathcal{D}_{11},\,\,\,\,\,\widetilde{\mathcal{G}}=\mathcal{D}_{01}+\mathcal{D}_{10}-{\boldsymbol q}^2\mathcal{D}_{11}.
    \end{split}
\end{align}
As before, we split the integrals $\mathcal{G},\widetilde{\mathcal{G}}$ as massless, mixed and massive type
\begin{align}
    \begin{split}
        &\mathcal{G}(\text{total})=\mathcal{G}(\text{massless)}+\mathcal{G}(\text{mixed)}+\mathcal{G}(\text{massive),}\\&
        \widetilde{\mathcal{G}}(\text{total})=\widetilde{\mathcal{G}}(\text{massless})+\widetilde{\mathcal{G}}(\text{mixed})+\widetilde{\mathcal{G}}(\text{massive}).
    \end{split}
\end{align}
This gives (see Appendix \ref{app_gen_integral} for a general strategy for solving these integrals)
Therefore, the one-point function takes the following form
\begin{align}
\begin{split}
\langle\Sigma(x)\rangle_{(b)}&=-(K_{1a}+K_{1b}+K_{1c})+2(4\beta+\gamma)(K_{2a}+K_{2b}+K_{2c})-2(\beta+\gamma)K_3\\&\hspace{8mm}+4(\beta+\gamma)(K_{4a}+K_{4b}+K_{4c})-8(2\beta+\gamma)K_5\,.
\end{split}
\end{align}
\subsection*{\underline{Metric Reconstruction}}
To reconstruct the metric perturbatively, we evaluate the metric by expansion in `original fields', $\phi(x)$  and $\sigma(x)$. Considering the metric as $g_{\mu \nu} $, its temporal and spatial components, i.e, $g_{00}$ and $g_{ij}$ respectively\footnote{considering $\gamma_{ij}=\delta_{ij}+\sigma_{ij}$} can be casted as,
\begin{align}
    &g_{00}=-(1+2\langle\phi(x)\rangle)=-1-2\bigg(\langle\Phi(x)\rangle+\frac{2}{5}(2\beta+\gamma)\nabla^2\langle\Sigma(x)\rangle\bigg)\,,\\&
    g_{ij}=\delta_{ij}(1-2\langle\phi(x) \rangle+\langle\sigma(x)\rangle)=\delta_{ij}\bigg[1-2\Bigg(\langle\Phi(x) \rangle +\frac{2}{5}(2\beta+\gamma)\nabla^2\langle \Sigma(x)\rangle \Bigg)\\&\hspace{7cm}+\langle\Sigma(x)\rangle-\frac{2}{5}(2\beta+\gamma)\nabla^2\langle\Phi(x)\rangle]
\end{align}
where $\langle \Phi(x)\rangle = \langle \Phi(x)\rangle_a+ \langle \Phi(x)\rangle_b+ \langle \Phi(x)\rangle_c$ and $\langle \Sigma(x)\rangle=\langle \Sigma(x)\rangle_a+ \langle \Sigma(x)\rangle_b$ for metric reconstruction upto 1PN order. Now, having the one-point functions in hand for both $\Phi(x)$ and $\Sigma(x)$, we sum them up, and hence the perturbative expansion becomes,
\begin{align}
\begin{split}
    &g_{00}=-1-2\Bigg(-{Gm}\frac{\left(1-e^{-\frac{1}{\sqrt{-2(\beta+\gamma)}}\,r}\right)}{r}-\frac{9}{5}(2\beta+\gamma)(I_{1a}+I_{1b}+I_{1c})+4(\beta+\gamma)I_2+\cdots\Bigg),\\&
g_{ij}=\delta_{ij}\Bigg(1+\frac{2Gm}{r}+\Bigg(-2Gm+ \frac{Gm(2\beta+\gamma)}{5(\beta+\gamma)}\Bigg)\frac{e^{-\frac{1}{\sqrt{-2(\beta+\gamma})}\,r}}{ r}-\frac{8G m(2\beta+\gamma)}{5(8\beta+3\gamma)}\frac{e^{-\frac{1}{\sqrt{8\beta+3\gamma}}\,r}}{ r}\\&\hspace{2cm} -(K_{1a}+K_{1b}+K_{1c})+2(4\beta+\gamma)(K_{2a}+K_{2b}+K_{2c})-2(\beta+\gamma)K_3\\&\hspace{3cm}+4(\beta+\gamma)(K_{4a}+K_{4b}+K_{4c})-8(2\beta+\gamma)K_5+\cdots \Bigg).
\label{4.36t}
    \end{split}
\end{align}
Here in \eqref{4.36t} \{I\} integrals are defined in \eqref{4.15 m} and \eqref{4.16y}. Similarly, the \{K\} integrals in the spatial component of the metric are given in \eqref{4.32n}. Now we proceed to \textcolor{black}{compute the PN corrections to the source term of the perturbation} and check the running of the Love number for the theory under consideration.\vspace{1mm}

For computational simplicity, we will mainly focus on the two phenomenologically interesting choices $2\beta+\gamma=0$ and $8\beta+3\gamma=0\,.$ These cases are particularly useful since the expression for $g_{ij}$ can be made to collapse to a single Yukawa tail instead of two. Moreover, in the large $r$-limit that we are interested in, the higher PN order terms contained in the integrals $\{I, K\}$ do not contribute, and the metric is only given up to the leading PN order.
\begin{itemize}
    \item {{\bf Case I}}: $2\beta+\gamma=0$\\
    In this case, the two Yukawa masses are equal, and the metric takes the simpler form
    \begin{align}
    \begin{split}
    &f(r)=1+2\phi=1-\frac{2GM}{r}+\frac{2GMe^{-\mu r}}{r},\\&
    g(r)=1-2\phi+\sigma=1+\frac{2GM}{r}-\frac{2GMe^{-\mu r}}{r}. \label{M1}   
    \end{split}
\end{align}
Here $\mu=1/\sqrt{|\gamma|}$.
\item {{\bf Case II}}: $8\beta+3\gamma=0$\\
    In this case, the fourth term in $g_{ij}$ tends to zero, and the metric takes the form
    \begin{align}
    \begin{split}
   & f(r)=1+2\phi=1-\frac{2GM}{r}+\frac{2GMe^{-\tilde\mu r}}{r},\\&
    g(r)=1-2\phi+\sigma=1+\frac{2GM}{r}-\frac{48GM}{25}\frac{e^{-\tilde\mu r}}{r}.
    \end{split}\label{M2}
    \end{align}
where $\tilde\mu=\frac{2}{\sqrt{5}}\mu$.
%\newpage
Finally, one can consider a bit more general scenario as shown in the following, 
\item {{\bf Case III}}: $\beta+k\gamma=0$\\
    Finally, we focus on this general case for $\frac{3}{8}<k<1$. In this case, the metric takes the form
    \begin{align}
    \begin{split}
    f(r) &= 1 - \frac{2GM}{r} +\frac{2GM e^{-\mu_A r}}{r}, \\
    g(r) &= \Bigg\{1 + \frac{2GM}{r} -\frac{2GMe^{-\mu_Ar}}{r}-\frac{GM(1-2k)}{5(1-k)}\frac{e^{-\mu_A r}}{r} \\&\hspace{2cm}- \frac{8GM(1-2k)}{5(3-8k)}\frac{e^{-\mu_B r}}{r}\Bigg\},
    \end{split}\label{M3}
\end{align}
where the two distinct mass scales, $\mu_A$ and $\mu_B$, are given by:
\begin{align}
\begin{split}
    \mu_A = \frac{1}{\sqrt{-2\gamma(1-k)}},  \,\,\,\,
    \mu_B = \frac{1}{\sqrt{\gamma(3-8k)}}.
    \end{split}
\end{align}
{\textbf{\textit{Dominant term analysis at large \(r\):}}}
At a large radius (\(r\)), the Yukawa term with the smaller mass scale (\(\mu\)) in its exponent will decay more slowly and therefore dominate the expression. The relative size of \(\mu_A\) and \(\mu_B\) depends on the specific value of the parameter \(k\). The crossover point occurs at $k=1/2$.

The two cases for dominance are as follows:

\begin{description}
    \item[Case 1: Dominance of the $\mu_A$ term]: 
    when \(\boldsymbol{\frac{3}{8} < k < \frac{1}{2}}\), the mass scale \(\mu_A\) is smaller than \(\mu_B\). Therefore, the terms containing \(e^{-\mu_A r}\) decay more slowly and will dominate the Yukawa corrections at large \(r\).

    \item[Case 2: Dominance of the $\mu_B$ term]: 
    when \(\boldsymbol{\frac{1}{2} < k < 1}\), the mass scale \(\mu_B\) is smaller than \(\mu_A\). In this regime, the term containing \(e^{-\mu_B r}\) is the dominant one at large \(r\).
\end{description}
\end{itemize}

%textcolor{red}{AB: Motivate the two choices $8\beta+3\gamma=0$ and $2\beta+\gamma=0$ here, i.e what happens to the metric for these two choices. Is this $\gamma$ or $|\gamma|$?}
\section{Spin-0/2 perturbation and (re)-normalizability of Love number?}\label{sec5}
%\textcolor{red}{AB: Shall we call it Love number of Wilson coefficient, as the Love number will come later after the UV matching?}
In this section, we evaluate the one-point function for spin-0 and spin-2 perturbations and check whether the Love numbers run under RG or not.

\subsubsection*{Spin-0 perturbation}

\begin{figure}[ht!]
    \centering
\includegraphics[width=0.22\linewidth]{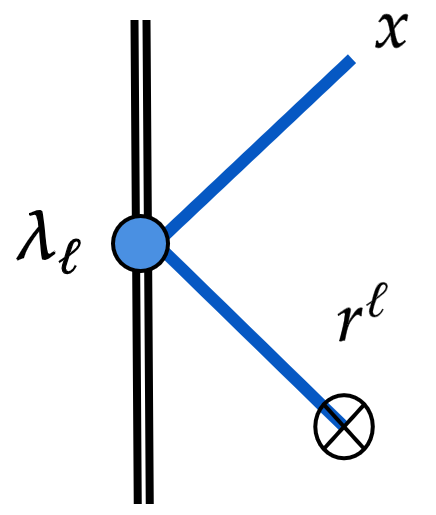}
    \caption{Diagrammatic representation of Love number vertex. The cross denotes the insertion of the tidal field. The other line representing the response has a propagator associated with it (in the static limit).}
    \label{fig:love_vertex}
\end{figure}

\begin{figure}[ht!]
    \centering
\includegraphics[width=0.6\linewidth]{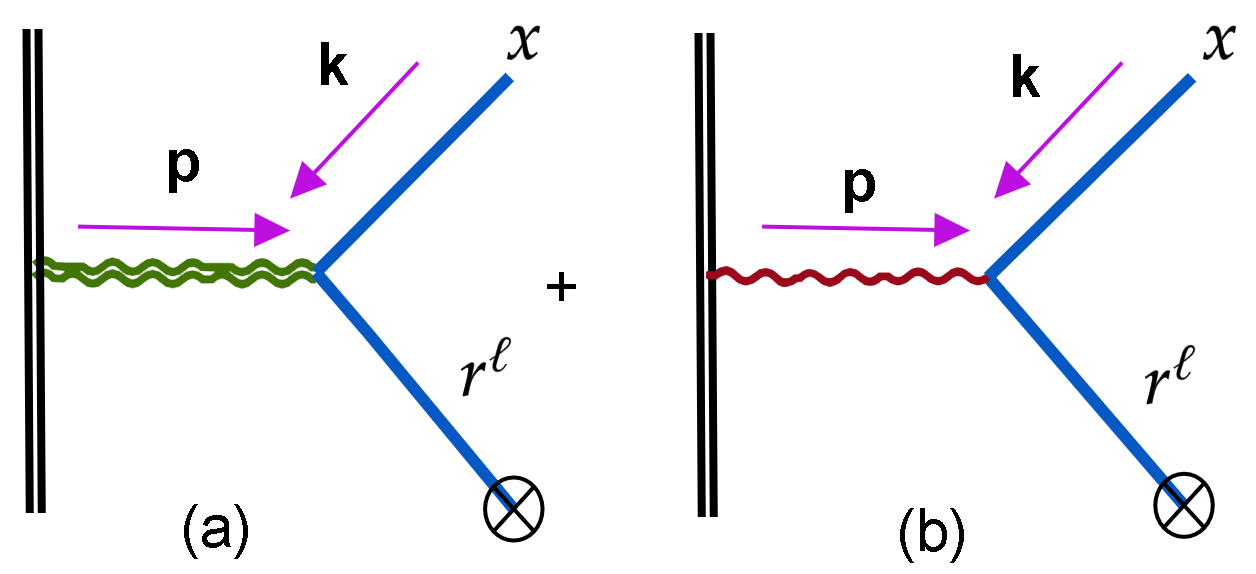}
    \caption{Diagrams contributing to the Love number for spin-0 perturbation. The wavy lines denote the insertion of re-defined fields-$\Sigma$ and $\Phi$ to the worldline. The cross denotes the insertion of the tidal field (external source). }
    \label{fig2}
\end{figure}

Let us first analyze spin-0 %and spin-2 
perturbation. %together. 
The action of a bulk scalar field $\chi$ is 
\begin{equation}
    S_{\chi}=-\frac{1}{2}\int dtd^3x\sqrt{\gamma}\gamma^{ij}\partial_i\chi\partial_j\chi\label{scalar_action}
\end{equation}
where $\gamma_{ij}=(1+\sigma)\delta_{ij}$ is the spatial part of the metric in the 3+1 decomposition of the spacetime as mentioned before. Note that $\chi$ only couples to $\sigma$ for the gauge that we have chosen. For the case of GR, this leads to the vanishing of the Love number for the scalar perturbations since only $\phi$ can be inserted in the worldline. The choice of the specific gauge (isotropic) makes this interpretation direct owing to the structure of the Einstein-Hilbert action. In our case, the redefined field $\Sigma$ can also be inserted in the worldline, giving a non-zero contribution to the Love number, which is again manifest in the gauge choice. To see this explicitly and evaluate the non-zero Love number, let us expand the action (\ref{scalar_action}) of the scalar field to quadratic and cubic order (in the dynamical fields). For this, we decompose $\chi$ into a background and response term as $\chi=\bar\chi+\delta\chi$. Then, the action reads
\begin{align}
\begin{split}
S_{\chi}^{(2)}&=\int dtd^3x\bigg(-\frac{1}{2}(\partial_i\delta\chi\partial^i\delta\chi)-(\partial_i\bar\chi\partial^i\delta\chi)\bigg),\\
S_{\chi}^{(3)}&=\int dtd^3x\bigg(-\frac{1}{4}\sigma(\partial_i\delta\chi\partial^i\delta\chi)-\frac{1}{2}\sigma(\partial_i\bar\chi\partial^i\delta\chi)\bigg).
\end{split}
\end{align}
Here, $\sigma\to\Sigma+c\nabla^2 \Phi(x)$ with $c=-\frac{2}{5}(2\beta+\gamma)$. Before proceeding further, we consider the interaction of the point particle with an external scalar field.  The effective field theory logic dictates that we should write down all possible couplings to $\chi$ on the worldline that are consistent with the symmetries of the problem.  Using the particle’s spacetime velocity $v^\mu\equiv x'^\mu$, which satisfies $v_\mu v^\mu=-1$, we construct the transverse projector
\begin{equation}
  P^\nu_{\phantom{\nu}\mu} \equiv \delta^\nu_{\phantom{\nu}\mu} + v_\mu v^\nu\,.
\end{equation}
This projector allows us to separate derivatives into temporal and spatial parts.  Since we are interested in the black hole’s static response, we will ignore operators with time derivatives of the scalar and work to leading order in the particle’s velocity.  The most general action (up to field redefinitions) up to second order in the bulk scalar field and leading order in velocities is \cite{Hui:2020xxx}
\begin{equation}
S = -\frac{1}{2}\int d^Dx\,(\partial\chi)^2
  + \int d\tau\,\mathcal{E}
    \left[
      \frac12\,\mathcal{E}^{-2}x'^\mu x'_\mu - \frac{m^2}{2}
      - g\,\chi
      + \sum_{\ell=1}^\infty \frac{\lambda_\ell}{(2\ell)!}\,
        \bigl(\partial_{(a_1}\cdots\partial_{a_\ell)_T}\chi\bigr)^2
    \right],
\end{equation}
where $(\cdots)_T$ denotes the symmetrized traceless component of the enclosed indices.  The $g\chi$ coupling would render the point particle a scalar charge, but for an uncharged black hole, $g=0$ (this is the famous no-hair theorem in GR). The coefficients $\lambda_\ell$ are Wilson coefficients encoding the object’s static scalar response (scalar Love numbers). To match these to the EFT calculation, we imagine an external, long-wavelength tidal field,
\begin{equation}
  \chi^{(0)}(x) = c_{a_1\cdots a_\ell}\,x^{a_1}\cdots x^{a_\ell},
\end{equation}
with $c_{a_1\cdots a_\ell}$ symmetric and traceless.  Expanding
\begin{equation}
  \chi = \chi^{(0)} + \varepsilon\,\chi^{(1)} + \cdots,
\end{equation}
the worldline operator acts as an effective source in
\begin{equation}
  \Box\,\chi^{(1)}(x)
  = \lambda_\ell\,(-1)^\ell
    \int d\tau\;c_{a_1\cdots a_\ell}\,
    \partial^{a_1}\!\cdots\partial^{a_\ell}\,
    \delta^{(D)}\bigl(x-x(\tau)\bigr)\,.
\end{equation}
Working in Fourier space and placing the particle at the origin,
\begin{equation}
  \chi^{(1)}(p)
  = \lambda_\ell\;(-i)^\ell\,
  \frac{c_{a_1\cdots a_\ell}\,p^{a_1}\cdots p^{a_\ell}}{p^2}\,.
\end{equation}
Transforming back to position space using
\begin{align}\int d^dp\,\frac{e^{i\boldsymbol p\cdot\boldsymbol x}}{p^2}
 = \frac{\Gamma\bigl(\tfrac{d}{2}-1\bigr)}{(4\pi)^{d/2}}\,|\boldsymbol x|^{2-d}\end{align}
one finds
\begin{equation}
  \chi(x)
  = c_{a_1\cdots a_\ell}x^{a_1}\cdots x^{a_\ell}
  \Biggl[
    1
    + \lambda_\ell\,(-1)^\ell\,
      \frac{\Gamma(\tfrac{d}{2}-1)\,\Gamma(2-\tfrac{d}{2})}
           {2^{2\ell}\,\pi^{d/2}\,\Gamma(2-\ell-\tfrac{d}{2})}\,
      |\boldsymbol x|^{2-d-2\ell}
  \Biggr].
  \label{eq:scalar-response}
\end{equation}
From the EFT point of view, this translates to evaluating the diagram shown in Fig.(\ref{fig:love_vertex})
Therefore, in our case, we need to evaluate the following diagram (Fig.(\ref{fig2})) with a single insertion to the worldline to compute the Love number or equivalently, evaluate the response at $x$ due to the tidal field.
The vertex contributing to the diagrams is given as
\begin{equation}
    \boldsymbolcal{V}=-\frac{1}{2}\Sigma\partial_i\delta\bar\chi\partial^i\delta\chi+\frac{1}{5}(2\beta+\gamma)\nabla^2\Phi\partial_i\delta\bar\chi\partial^i\delta\chi.
\end{equation}
The propagators for different bulk fields can be derived by identifying the quadratic part of the action. Therefore, the propagator for $\delta\chi$
is given by, 
\begin{align}
    \begin{split}
        &\langle\delta\chi(x_1)\,\delta\chi(x_2)\rangle=\hspace{ 0.1 cm}\begin{minipage}[h]{0.15\linewidth}
	\vspace{-13pt}
	\scalebox{0.32}{\begin{feynman}
    \fermion[lineWidth=6, showArrow=false, color=0658C3, label=$\Huge{\delta\chi}$]{4.00, 4.00}{7.40, 4.00}
\end{feynman}
}
\end{minipage}\hspace{ 0.6 cm}=\delta(t_1-t_2)\int_{{\boldsymbol  k}}\frac{e^{i\boldsymbol k\cdot (\boldsymbol{x}_1-\boldsymbol{x}_2)}}{\boldsymbol{{k}}^2}\,.
        \end{split}
    \end{align}
The diagrams can be readily evaluated as follows:
\begin{equation}
    \langle\delta\chi\rangle_{(a)}=-2\times\frac{2}{5m^2m_p^2}(-i)^\ell M\mathcal{E}_i\int_{{\boldsymbol k}}k^i e^{i\boldsymbol{k}\cdot \boldsymbol{r}}\bigg[\frac{1}{{\boldsymbol k}^2}-\frac{1}{{\boldsymbol k}^2+m_1^2}\bigg].
\end{equation}
The tidal field has a profile considered as $\bar\chi=\mathcal{E}_{i_1...i_l}x^{i_1...i_l}\propto r^lY_{lm}(\theta,\phi)$ where $\mathcal{E}_{i_1...i_l}$ is a constant tidal moments' tensor, $m^2=\frac{1}{(2\beta+\gamma)}$ and $m_1^2=\frac{1}{(8\beta+3\gamma)}$. We take $l=1$ so that the tidal field has one index as $\mathcal{E}_ix^i$ to evaluate spin-0 perturbations. The integral evaluates to
\begin{equation}
    \langle\delta\chi\rangle_{(a)}=-\frac{\mathcal{E}_ix^iM}{5m^2m_p^2\pi r^3}\Big[1-e^{-m_1r}(1+m_1r)\Big].
\end{equation}
The second diagram is computed as
\begin{equation}
    \langle\delta\chi\rangle_{(b)}=\frac{(-i)^\ell M\mathcal{E}_i}{2m_p^2}\int_{{\boldsymbol k}}k^i e^{i\boldsymbol{k}\cdot \boldsymbol{r}}\bigg[\frac{1}{{\boldsymbol k}^2}-\frac{1}{{\boldsymbol k}^2+m_2^2}\bigg]
\end{equation}
where $m_2^2=-\frac{1}{2(\beta+\gamma)}$. This integral evaluates to
\begin{equation}
   \langle\delta\chi\rangle_{(b)}=\frac{\mathcal{E}_ix^iM}{8\pi m_p^2 r^3}\Big[1-e^{-m_2r}(1+m_2r)\Big]. 
\end{equation}
Therefore, the total contribution is given by adding both diagrams,
\begin{equation}
    \langle\delta\chi\rangle=\langle\delta\chi\rangle_{(a)}\times\Big(-\frac{1}{2}\Big)+\langle\delta\chi\rangle_{(b)}\times\Big(\frac{1}{5}(2\beta+\gamma)\Big).
\end{equation}
This results in
\begin{equation}
    \langle\delta\chi\rangle=\frac{\mathcal{E}_ix^iM}{10m^2m_p^2\pi r^3}\Big[1-e^{-m_1r}(1+m_1r)\Big]+\frac{\mathcal{E}_ix^iM}{40m_p^2m^2\pi r^3}\Big[1-e^{-m_2r}(1+m_2r)\Big].\label{one_point_fn_spin_0}
\end{equation}
\textcolor{black}{The dimensionless coefficient $c_3$\footnote{Recall that $c_{2\ell+1}$ is the EFT coefficient at $(2\ell+1)$-PN order capturing the $\ell$-th correction to the tidal source. For example, in the case of monopole ($\ell=0$), $c_{2\ell+1}$ has only one term, namely $c_1$. For dipole, $c_{2\ell+1}$ will have three terms, namely, $\{c_1,c_2,c_3\}$ and for quadrupole, we will have terms from $c_1$ to $c_5$.} can be read as the coefficient of the leading $1/r^3$ tail at $r\to\infty$ (power-law tail) in the one-point function and is given by (in terms of the couplings)
\begin{align}
&c_3=\frac{M}{4\mu},\hspace{1.9cm}8\beta+3\gamma=0,\\&\nonumber
c_3=0,\hspace{2.2cm}2\beta+\gamma=0,\\&\nonumber
c_3=\frac{(1-2k)M}{\mu},\hspace{0.4cm}\beta+k\gamma=0.
\end{align}
where $\mu=\frac{1}{\sqrt{|\gamma|}}$} and $\frac{3}{8}<\beta<\frac{1}{2}$. As a consistency check, for GR  ($m,m_1,m_2\to\infty$ or equivalently $\beta,\gamma\to0$), we get $\langle\delta\chi\rangle=0$ from (\ref{one_point_fn_spin_0}).

\subsubsection*{Spin-2 electric-type perturbation}

\begin{figure}[ht!]
    \centering
\includegraphics[width=0.5\linewidth]{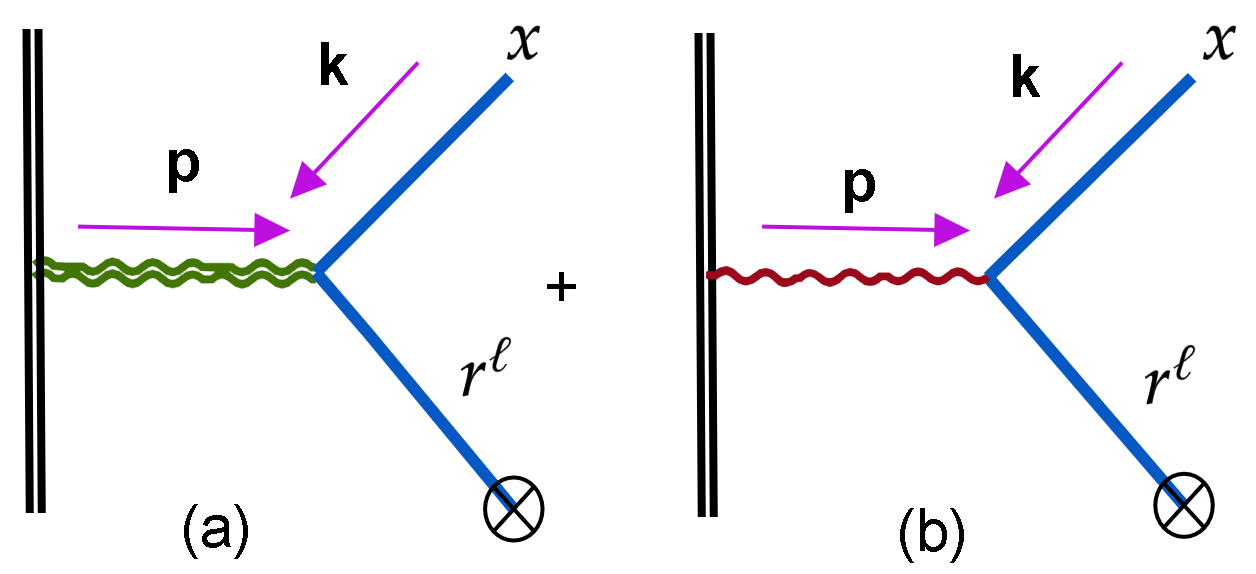}
    \caption{Diagram contributing to Love number for spin-2 perturbation. The wavy lines denote the fields $\Sigma$ and $\Phi$, while the blue line denotes the separation of the perturbation of the dilaton field ($\delta\Phi)$ as the response $\delta\bar\Phi$ and the tidal field $\delta\Phi_B$, which is denoted by the cross.}
    \label{fig3}
\end{figure}

To evaluate the spin-2 electric-type perturbation, we first decompose the dilaton $\Phi$ as
\begin{equation}
    \Phi=\Phi_{\text{BH}}+\delta\Phi.
\end{equation}
Here, $\Phi_{\text{BH}}$ is the background solution that matches the (static) black hole solution, and $\delta\Phi$ is essentially the perturbation of the Newtonian potential. Note that, for this type of perturbation $\delta\sigma(x)$ vanishes. But unlike GR, the perturbation of the redefined field $\Sigma$ will not vanish; rather, it will give the following. 
\begin{equation}
    \delta\sigma(x)\rightarrow\delta\Sigma(x)-\frac{2}{5}(2\beta+\gamma)\nabla^2\delta\Phi(x).
\end{equation}
Since $\delta\sigma$ vanishes, we have the relation:
\begin{equation}
    \delta\Sigma(x)=\frac{2}{5}(2\beta+\gamma)\nabla^2\delta\Phi(x).\label{perturbation_relation}
\end{equation}
The tidal effects for this case are controlled by the electric tidal field, which is related to the Weyl tensor as
\begin{equation}
    E_{ij}\equiv C_{0i0j}=-\Big(\partial_i\partial_j-\frac{1}{3}\delta_{ij}\partial^2\Big)\delta\phi.
\end{equation}
This tells that the electric perturbation depends on $\delta\phi$. However, using (\ref{perturbation_relation}), we can write
\begin{equation}
    \delta\phi=\hat{\mathcal{D}}\delta\Phi,\label{operator_relation}
\end{equation}
where the differential operator is given as
\begin{equation}
    \hat{\mathcal{D}}=\Big[1+\Big(\frac{2}{5}(2\beta+\gamma)\Big)^2(\nabla^2)^2\Big].
\end{equation}
Therefore, at the linearized order in $\gamma,\beta$ that we are working in, we have
\begin{equation}
    \delta\phi\equiv\delta\Phi+\mathcal{O}\Big((\beta,\gamma)\Big)^2.
\end{equation}
This allows us to work with the re-defined field perturbation $\delta\Phi$ instead of the original field perturbation $\delta\phi$. We now proceed to evaluate the action for $\delta\Phi$ for quadratic gravity to the quadratic in perturbation $\delta\Phi$ (and upto cubic order in the  dynamical fields) which takes the following form:
\begin{equation}
\begin{split}
  S_{\delta\Phi}^{(3)}&=-\frac{m_p^2}{2}\int dtd^3x\ \Sigma\partial_i\delta\Phi\partial^i\delta\Phi\,\,\,-(\beta+\gamma)m_p^2\int dtd^3x\ \Big(\Sigma\nabla^2\delta\Phi\nabla^2\delta\Phi-4\Phi_{\text{BH}}\nabla^2\delta\Phi\nabla^2\delta\Phi\\&-2\nabla^2\delta\Phi\partial_i\Sigma\partial_i\delta\Phi\Big)-4(2\beta+\gamma)m_p^2\int dtd^3x\Phi\nabla^2\Phi\nabla^2\Sigma-2(2\beta+\gamma)m_p^2\int dtd^3x\ (\nabla^2\delta\Phi\partial_i\Phi_{\text{BH}}\partial^i\delta\Phi).  
\end{split}
\label{spin_2_action}
\end{equation}
The first term is reminiscent of the Einstein-Hilbert action, while the second and third terms arise from higher-curvature corrections. Unlike GR, the action of the spin-2 perturbation does not coincide with the action of the bulk scalar field. By virtue of (\ref{perturbation_relation}), the terms containing $\delta\Sigma$ in action will contribute at $\mathcal{O}\Big((\beta,\gamma)^2\Big)$. Since we are working only at the linearized order in $\beta$ and $\gamma$, we ignore these terms in the computation of the action of spin-2 perturbation. However, the contribution from the GR term, namely $-\Sigma\partial_i\Phi\partial^i\Phi$ survives since it gives a contribution in the first order in $\gamma,\beta$ for the perturbation $\delta\Sigma\,.$
%, leading to a factor of $\frac{10}{5}=2$ instead of $\frac{9}{5}$ in the last term of the action (\ref{spin_2_action}). 
We now decompose the perturbation as the tidal field (external source) $\delta\Phi_B$ and the response $\delta\bar{\Phi}$:
\begin{equation}
    \delta\Phi = \delta\Phi_B + \delta\bar{\Phi}.
\end{equation}
Then, the derivatives act linearly:
\begin{equation}
   \partial_i \delta\Phi = \partial_i \delta\Phi_B + \partial_i \delta\bar{\Phi}, \quad
\nabla^2 \delta\Phi = \nabla^2 \delta\Phi_B + \nabla^2 \delta\bar{\Phi}. 
\end{equation}
Substituting into the action (\ref{spin_2_action}), we get:
\begin{align}
S^{(3)}_{\delta\Phi}&= -\frac{m_p^2}{2} \int dt\,d^3x\, \Sigma \left[
(\partial_i \delta\Phi_B)^2 + 2\, \partial_i \delta\Phi_B\, \partial^i \delta\bar{\Phi} + (\partial_i \delta\bar{\Phi})^2
\right] \nonumber \\
&- (\beta + \gamma)m_p^2 \int dt\,d^3x\, \Big[
\Sigma \left((\nabla^2 \delta\Phi_B)^2 + 2\, \nabla^2 \delta\Phi_B\, \nabla^2 \delta\bar{\Phi} + (\nabla^2 \delta\bar{\Phi})^2 \right) \nonumber \\
&- 4\Phi_{\text{BH}}\left((\nabla^2 \delta\Phi_B)^2 + 2\, \nabla^2 \delta\Phi_B\, \nabla^2 \delta\bar{\Phi} + (\nabla^2 \delta\bar{\Phi})^2 \right)\Big]\nonumber\\&+2(\beta+\gamma)m_p^2\int dtd^3x\Big[\nabla^2\delta\Phi_B\partial_i\Sigma\partial^i\delta\Phi_B+\nabla^2\delta\Phi_B\partial_i\Sigma\partial^i\delta\bar\Phi\nonumber\\&+\nabla^2\delta\bar\Phi\partial_i\Sigma\partial^i\delta\Phi_B+\nabla^2\delta\bar\Phi\partial_i\Sigma\partial^i\delta\bar\Phi\Big]\nonumber\\&-\frac{9}{5}(2\beta+\gamma)m_p^2\int dtd^3x\Big[\nabla^2\delta\Phi_B\partial_i\Phi_{\text{BH}}\partial^i\delta\Phi_B+\nabla^2\delta\Phi_B\partial_i\Phi_{\text{BH}}\partial^i\delta\bar\Phi\nonumber\\&+\nabla^2\delta\bar\Phi\partial_i\Phi_{\text{BH}}\partial^i\delta\Phi_B+\nabla^2\delta\bar\Phi\partial_i\Phi_{\text{BH}}\partial^i\delta\bar\Phi\Big]-4(2\beta+\gamma)m_p^2\int dtd^3x\delta\Phi_B\nabla^2\delta\bar\Phi\nabla^2\Sigma\,.
\end{align}
From the EFT perspective, we need to sum the following diagrams (Fig.~(\ref{fig3})) (equivalently, measure the response at $x$ due to the tidal field) with a single insertion to the worldline to extract the Love number for spin-2 perturbation.
The vertex contribution to the diagram is
\begin{equation}
\begin{split}
\boldsymbolcal{V} ={}& 
-\,\tfrac12
\underbrace{%
\Sigma\,\partial_i\delta\Phi_B\,\partial^i\delta\bar\Phi
-2(\beta+\gamma)\,\Sigma\,\nabla^2\delta\Phi_B\,\nabla^2\delta\bar\Phi
+2(\beta+\gamma)\,\nabla^2\delta\Phi_B\,\partial_i\Sigma\,\partial^i\delta\bar\Phi
}_{\textcolor{black}{\text{contribution from }\Sigma\text{-insertion}}}
\\[6pt]
&\quad
\,\underbrace{+2(\beta+\gamma)\,\nabla^2\delta\bar\Phi\,\partial_i\Sigma\,\partial^i\Phi_B-
4(2\beta+\gamma)\;\delta\Phi_B\,\nabla^2\delta\bar\Phi\,\nabla^2\Sigma
}_{\textcolor{black}{\text{contribution from }\Sigma\text{-insertion}}}
\\[6pt]
&\quad
-\,\underbrace{%
\frac{9}{5}(2\beta+\gamma)\,\nabla^2\delta\bar\Phi\,\partial_i\Phi_{\mathrm{BH}}\,\partial^i\delta\Phi_B
+8(\beta+\gamma)\,\Phi_{\mathrm{BH}}\,\nabla^2\delta\Phi_B\,\nabla^2\delta\bar\Phi
}_{\textcolor{black}{\text{contribution from }\Phi\text{-insertion}}}
\\[6pt]
&\quad
-\,\underbrace{%
\frac{1}{5}(2\beta+\gamma)\,\nabla^2\delta\bar\Phi\,\partial_i\Phi_{\mathrm{BH}}\,\partial^i\delta\Phi_B
}_{\textcolor{black}{\delta\Sigma\text{ contribution from first term}}}\,.
\end{split}
\end{equation}
%\textcolor{red}{AB: Got confused. Is this full $\delta \bar{\Phi}$ or the source or tidal  field?}
The propagator for $\delta\bar{\Phi}$ is given by, 
    \begin{align}
        \begin{split}
    &\langle\delta\bar{\Phi}(x_1)\,\delta\bar{\Phi}(x_2)\rangle=\hspace{ -0.1 cm}\begin{minipage}[h]{0.15\linewidth}
	\vspace{-14pt}
	\scalebox{0.28}{\begin{feynman}
    \fermion[lineWidth=6, showArrow=false, color=0658C3, label=$\Huge{\delta\bar\Phi}$]{4.00,4.00}{7.40, 4.00}
\end{feynman}
}\end{minipage}\,\,\,\,\,=\hspace{0.3 cm}\frac{1}{m_p^2}\delta(t_1-t_2)\int_{{\boldsymbol  k}}\frac{e^{i\boldsymbol k\cdot (\boldsymbol{x}_1-\boldsymbol{x}_2)}}{-4\boldsymbol{k}^4(\gamma+\beta)+2\boldsymbol{{k}}^2}
        \end{split}
    \end{align}
For spin-2 perturbation, the leading order contribution comes from $\ell=2$ ($\ell=0,1$ are absent). This can also be understood from the structure of the Weyl tensor $C_{0i0j}$, which kills off $l=0,1$ modes and only the $l=2$ quadrupole moment survives. Therefore, the tidal field will have a 2-index structure as $\mathcal{E}_{ij}x^ix^j$ where $\mathcal{E}_{ij}$ is a symmetric trace-free tensor (STF). %Also, the derivative of the tidal field in the momentum space reads
%\begin{equation}
 %   \int d^3x\,E_{ij}\,x^i\,e^{-i\boldsymbol{k}\cdot\boldsymbol{x}}
%= i\,(2\pi)^3\,E_{ij}\,\frac{\partial}{\partial k_i}\,\delta^{(3)}(\boldsymbol{k})\,.
%\end{equation}
Taking these factors into account, we now proceed with the evaluation of the diagrams as follows:
\begin{equation}
    \langle\delta\Phi\rangle_{(a)}[\boldsymbolcal{V}_1]=-\frac{(m_1m_2)^2}{5m^2m_p^2}\times (-4)M\mathcal{E}_{ij}\int_{{\boldsymbol k}}k^ik^j e^{ik\cdot x}\frac{1}{{\boldsymbol k}^2}\bigg[\frac{1}{{\boldsymbol k}^2+m_1^2}\bigg]\bigg[\frac{1}{{\boldsymbol k}^2+m_2^2}\bigg],
\end{equation}
where $m^2=\frac{1}{(2\beta+\gamma)}$, $m_1^2=\frac{1}{(8\beta+3\gamma)}$ and $m_2^2=-\frac{1}{2(\gamma+\beta)}$ as before. The integral evaluates to
\begin{equation}
\langle\delta\Phi\rangle_{(a)}[\boldsymbolcal{V}_1]
= C\,\mathcal{E}_{ij}\,
\frac{x^i x^j}{4\pi\,(m_2^2 - m_1^2)\,r^5}
\biggl[
\frac{e^{-m_1 r}\,(m_1 r + 1)\,(m_1 r + 3)}{m_1^2}
\;-\;
\frac{e^{-m_2 r}\,(m_2 r + 1)\,(m_2 r + 3)}{m_2^2}
\biggr],
\label{eq:dPhi_final}
\end{equation}
\begin{equation}
C = \frac{4\,(m_1\,m_2)^2\,M}{5\,m^2\,m_p^2}\,, 
\qquad
r = \sqrt{x^k x^k}\,.
\label{eq:C_def}
\end{equation}
%\begin{equation}
%\begin{split}
 %   \langle\delta\Phi\rangle_{(a)}[\boldsymbolcal{V}_1]=\frac{2\mathcal{E}_{ij}x^ix^jM}{5m^2m_p^2\pi}\frac{1}{r^3}%-\frac{2\mathcal{E}_{ij}x^ix^jM}{5(m_2m)^2\pi}\frac{1-e^{-m_2r}(1+m_2r)}{r^3}\\+\frac{2\mathcal{E}_{ij}x^ix^jM}{5m^2\pi}\frac{1}{r}
   % -\frac{2\mathcal{E}_{ij}x^ix^jM}{5\pi m^2m_p^2(m_1^2-m_2^2)r^3}\Big[e^{-m_1r}(1+m_1r)m_2^2-e^{-m_2r}(1+m_2r)m_1^2\Big].
%\end{split}
%\end{equation}
Also, we have
\begin{equation}
    \langle\delta\Phi\rangle_{(a)}[\boldsymbolcal{V}_2]=\langle\delta\Phi\rangle_{(a)}[\boldsymbolcal{V}_3]=\langle\delta\Phi\rangle_{(a)}[\boldsymbolcal{V}_5],
\end{equation}
which is a result of the fact that $\mathcal{E}_{ij}$ is an STF\footnote{This is because for an STF, $\nabla^2(\mathcal{E}_{ij}x^ix^j)=2\mathcal{E}^i_i=0$.}. Finally,
\begin{equation}
    \langle\delta\Phi\rangle_{(a)}[\boldsymbolcal{V}_4]=-\frac{4M}{5m^2m_p^2}\mathcal{E}_{ij}(m_1m_2)^2\int_{\vec{k}}\frac{k^ik^j e^{i\boldsymbol{k}\cdot \boldsymbol{r}}}{\boldsymbol k^2+m_2^2}\bigg[\frac{1}{\boldsymbol k^2+m_1^2}\bigg].
\end{equation}
This gives
\begin{equation}
 \langle\delta\Phi\rangle_{(a)}[\boldsymbolcal{V}_4]
=
-(\boldsymbol{x}\cdot \mathcal{E}\cdot \boldsymbol x)\frac{\,M\,(m_1m_2)^2}{20\pi r^5\,m^2 m_p^2\,(m_2^2 - m_1^2)}
\Bigl[
(m_1^2r^2 + 3m_1r + 3)\,e^{-m_1r}
\;-\;
(m_2^2r^2 + 3m_2r + 3)\,e^{-m_2r}
\Bigr]\,.
\end{equation}
%\begin{equation}
 %   \langle\delta\Phi\rangle_{(a)}[\boldsymbolcal{V}_4]=%-\frac{2\mathcal{E}_{ij}x^ix^jM}{5m_2^2\pi}\frac{1-e^{-m_2r}(1+m_2r)}{r^3}+
    %-\frac{2\mathcal{E}_{ij}x^ix^jM(m_1m_2)^2}{5\pi m^2m_p^2(m_2^2-m_1^2)r^3}\Big[e^{-m_1r}(1+m_1r)-e^{-m_2r}(1+m_2r)\Big].
%\end{equation}
Therefore, the total contribution from the first diagram is
\begin{equation}
\begin{split}
\langle\delta\Phi(\boldsymbol x)\rangle
&=\;
\frac{C\,(\boldsymbol{x}\cdot \mathcal{E}\cdot \boldsymbol x)}{4\pi\,(m_2^2 - m_1^2)\,r^5}\,
\Biggl\{\,
-\tfrac12\Bigl[
\frac{(m_1r+1)(m_1r+3)}{m_1^2}\,e^{-m_1r}
\;-\;
\frac{(m_2r+1)(m_2r+3)}{m_2^2}\,e^{-m_2r}
\Bigr]\\
&\quad\qquad
+\frac{1}{m_2^2}\,
\Bigl[
(m_1^2r^2 + 3m_1r + 3)\,e^{-m_1r}
\;-\;
(m_2^2r^2 + 3m_2r + 3)\,e^{-m_2r}
\Bigr]
\Biggr\}
\end{split}
\label{eq:deltaPhi_total}
\end{equation}
where,
\[
C = \frac{4\,(m_1m_2)^2\,M}{5\,m^2\,m_p^2}\,.
\]
%\begin{equation}
%\begin{split}
 % \langle\delta\Phi\rangle_{(a)}=-\frac{\mathcal{E}_{ij}x^ix^jM}{5m^2m_p^2\pi}\frac{1}{r^3}+\frac{\mathcal{E}_{ij}x^ix^jM}{5\pi m^2m_p^2(m_1^2-m_2^2)r^3}\Big[e^{-m_1r}(1+m_1r)m_2^2-e^{-m_2r}(1+m_2r)m_1^2\Big]\\-\frac{4\mathcal{E}_{ij}x^ix^jM(m_1m_2)^2}{5\pi m^4m_p^2(m_2^2-m_1^2)r^3}\Big[e^{-m_1r}(1+m_1r)-e^{-m_2r}(1+m_2r)\Big].
%\end{split}
%\end{equation}
Similarly, the second diagram is computed as
\begin{equation}
    \langle\delta\Phi\rangle_{(b)}[\boldsymbolcal{V}_6]= \langle\delta\Phi\rangle_{(b)}[\boldsymbolcal{V}_8]=-\frac{Mm_2^2\mathcal{E}_{ij}}{2m_p^2}\int_{\boldsymbol{k}}\frac{k^ik^j e^{i\boldsymbol k\cdot \boldsymbol r}}{\boldsymbol k^2+m_2^2}\bigg[\frac{1}{\boldsymbol k^2}-\frac{1}{\boldsymbol k^2+m_2^2}\bigg].
\end{equation}
This integral evaluates to
\begin{equation}
\begin{split}
\bigl\langle\delta\Phi\bigr\rangle_{(b)}
&= -\frac{M\,m_2^2}{2m_p^2}\,\mathcal{E}_{ij}
\int\!\frac{d^3k}{(2\pi)^3}\,
\frac{k^i k^j\,e^{i\boldsymbol k\cdot\boldsymbol r}}{{\boldsymbol k}^2 + m_2^2}
\Bigl[\tfrac1{{\boldsymbol k}^2} - \tfrac1{{\boldsymbol k}^2 + m_2^2}\Bigr]\,,\\
&= \frac{M\,m_2^2}{2m_p^2}\,\mathcal{E}_{ij}\,
\Bigl[-\partial_i\partial_j\Bigr]\,
\underbrace{\int\!\frac{d^3k}{(2\pi)^3}\,
\Bigl[\tfrac{1}{{\boldsymbol k}^2({\boldsymbol k}^2+m_2^2)} - \tfrac{1}{({\boldsymbol k}^2+m_2^2)^2}\Bigr]\,e^{i\boldsymbol k\cdot\boldsymbol r}
}_{\displaystyle J(r)}\,,\\
&= \frac{M}{16\pi\,m_p^2}\,\mathcal{E}_{ij}\,
\frac{x^i x^j}{r^5}\,
\Bigl[
6 \;-\; e^{-m_2 r}\,(m_2^3 r^3 + 3 m_2^2 r^2 + 6 m_2 r + 6)
\Bigr]
\end{split}
\end{equation}
where,
\[
J(r)=\frac{2 - (2 + m_2 r)e^{-m_2 r}}{8\pi\,m_2^2\,r}
\]
and we used
\(\displaystyle
-\partial_i\partial_j\,J(r)
=\frac{x^i x^j}{r^2}\,\bigl[J''(r)-\tfrac{J'(r)}r\bigr]
\)
with \(J''-\tfrac1rJ'=\bigl[6 - e^{-m_2r}(m_2^3r^3+3m_2^2r^2+6m_2r+6)\bigr]/(8\pi\,m_2^2r^3)\).  \\%\begin{equation}
  % \langle\delta\Phi\rangle_{(b)}=-\frac{\mathcal{E}_{ij}x^ix^jM}{2\pi m_p^2r^3}\Big[1-e^{-m_2r}(1+m_2r)\Big]+\frac{\mathcal{E}_{ij}x^ix^jMm_2^2}{4\pi m_p^2}\frac{e^{-m_2r}}{r}. 
%\end{equation}
The seventh vertex contribution vanishes since $\mathcal{E}_{ij}$ is symmetric and trace-free. Therefore, the total contribution by summing the diagrams is given as
\begin{equation}
\begin{split}
\langle\delta\Phi(\boldsymbol x)\rangle
&=\frac{C\,(\boldsymbol{x}\cdot \mathcal{E}\cdot \boldsymbol x)}{4\pi\,(m_2^2 - m_1^2)\,r^5}
\Biggl\{
-\tfrac12\Bigl[
\frac{(m_1r+1)(m_1r+3)}{m_1^2}\,e^{-m_1r}
-\frac{(m_2r+1)(m_2r+3)}{m_2^2}\,e^{-m_2r}
\Bigr]\\
&\qquad\quad
+\frac{1}{m_2^2}\,
\Bigl[
(m_1^2r^2 + 3m_1r + 3)\,e^{-m_1r}
-(m_2^2r^2 + 3m_2r + 3)\,e^{-m_2r}
\Bigr]
\Biggr\}\\[8pt]
&\quad
\hspace{2cm}-\;\frac{2}{m^2}\,\frac{M}{16\pi\,m_p^2}\,\mathcal{E}_{ij}\,\frac{x^i x^j}{r^5}
\bigl[
6 - e^{-m_2 r}\,(m_2^3 r^3 + 3m_2^2 r^2 + 6m_2 r + 6)
\bigr]\,,
\end{split}
\end{equation}
\[
C = \frac{4\,(m_1m_2)^2\,M}{5\,m^2\,m_p^2}\,.  
\]
%\begin{equation}
%\begin{split}
    %\langle\delta\Phi\rangle&=\langle\delta\Phi\rangle_{(a)}+\langle\delta\Phi\rangle_{(b)}\\&=-\frac{\mathcal{E}_{ij}x^ix^jM}{5m^2m_p^2\pi}\frac{1}{r^3}-\frac{\mathcal{E}_{ij}x^ix^jMm_2^2}{2m^2m_p^2\pi}\frac{e^{-m_2r}}{r}+\frac{\mathcal{E}_{ij}x^ix^jM}{m^2m_p^2\pi}\frac{1-e^{-m_2r}(1+m_2r)}{r^3}\\&+\frac{\mathcal{E}_{ij}x^ix^jM}{5\pi m^2m_p^2(m_1^2-m_2^2)r^3}\Big[e^{-m_1r}(1+m_1r)m_2^2-e^{-m_2r}(1+m_2r)m_1^2\Big]\\&-\frac{4\mathcal{E}_{ij}x^ix^jM(m_1m_2)^2}{5\pi m^4m_p^2(m_2^2-m_1^2)r^3}\Big[e^{-m_1r}(1+m_1r)-e^{-m_2r}(1+m_2r)\Big].
%\end{split}
%\end{equation}
Again, as a consistency check, in the limit $m,m_1,m_2\to\infty$, we recover the GR case, $\langle\delta\Phi\rangle=0$. \\\\ Like the Spin-0 case, the dimensionless coefficient $c_{2\ell+1}$ can be read as the coefficient of the leading $1/r^{2l+1}$ tail. For our case, it is the coefficient of $1/r^5$ and is given as (in terms of the couplings)
\begin{equation}
c_5=-6M\sqrt{2\beta+\gamma}.
\end{equation}
%The non-vanishing of the Love number suggests that the Love numbers in this case are renormalizable. 
\subsection*{\textit{Comment on (re)-normalizability of Love number:}}
In \cite{Ivanov:2022hlo}, a classical RG equation for $\lambda_\ell$ has been derived by demanding scale-invariance of observables, here, physical one-point functions. One can have a non-vanishing $\nu\frac{d}{d\nu}\lambda_\ell(\nu)$ if there are logarithmic terms (coming from the loop corrections) present in the one-point functions, where $\nu$ is a renormalization scale.  
%The classical RG flow (in the presence of loops as in GR) is given by \cite{Ivanov:2022hlo}
%\begin{equation}
 %   \nu\frac{d}{d\nu}\lambda_\ell(\nu)=-c_{2\ell+1}m^{2\ell+1}\frac{4\pi}{(2\ell-1)!!}.
%\end{equation}
%Here, $\nu$ is a renormalization scale. 
Hence, the running of $\lambda_\ell$ exactly cancels the $\nu$-dependence of the logarithmic terms.\par But for our case, the term containing $c_5$ arises at the tree level itself (owing to the modified PN counting in higher curvature gravity) and there are no logarithmic dependence. Therefore, 
\begin{equation}
\frac{d\lambda_\ell(\nu)}{d\nu}=0.
\end{equation}
implying no RG-flow of Love number, although the presence of the $c_5$ coefficient indicates a non-vanishing Love number, which we will show (for the scalar perturbation) in the next section by performing a matching with the UV computation.   
%This essentially means,
\begin{equation}
c_{2\ell+1}\neq0\,\,\, \text{but no log-dependence (tree-level)}\implies\text{no RG running} \nonumber. 
\end{equation}
% \textcolor{red}{AB: I think the equation (6.45) is a bit deceptive without additional explanation of its origin. We are not naively integrating it. Maybe lets write this in words/represent it in better ways to avoid confusion. I guess according to Ivanov $c_{2l+1}$ is associated with a log term but for our case due to the modified PN counting it is not.}
%For $c_{2\ell+1}=0$, there is no RG running of Love numbers. This happens for GR where $c_{2\ell+1}$ vanishes at all PN orders for spin-0 and spin-2 perturbations, and Love numbers do not get re-normalized by graviton corrections \cite{Ivanov:2022hlo}. However, for quadratic EFT, we have obtained non-vanishing $c_{2\ell+1}$ for both spin-0 and spin-2 perturbations, which shows that Love numbers, in this case, have an RG running. Also, note that the $r^{-(2\ell+1)}$-tail or $2\ell+1$-PN terms were obtained with a single insertion to the worldline (owing to the modified PN counting in higher curvature gravity) instead of $2\ell+1$ insertions as in GR \cite{Ivanov:2022hlo}.\\
\section{UV meets IR: Extracting the scalar Love numbers from EFT matching}
\label{sec6}
{\bf\textit{Main Idea:}} To extract the Love number, we adopt a two-pronged approach. First, we compute the one-point function of the perturbation field \(\delta\Phi\) within the EFT, which encapsulates the response of the system to an external source. Second, we directly solve the perturbation equation in the reconstructed background metric, which is the UV side in our context. By matching the results from these two complementary descriptions, the Love number is fixed unambiguously.
 %This is evident from the fact that UV metricWilsonian matching of thIn this c \textcolor{red}{AB: This is not the reason! ``This is because the UV metric in Quadratic Gravity has unknown UV charges, while the evaluation of Love numbers requires known metric coefficients". We should say about the UV-IR in the same coordinate matching to extract the Love number. This reconstructed metric is in the same coordinate as that of the source calculation. Kindly change it.}

\subsection{Spin-0 perturbation with EFT reconstructed metric}\label{sec8}
%\subsubsection{}
To extract Love numbers, it suffices to use the metric constructed from one worldline insertion only, i.e., the $1/m_p^2$ term only, i.e., Eq.(\ref {M1}), Eq.(\ref{M2}), and Eq.(\ref{M3}), since we are interested in the asymptotic limit (large $r$-limit). 

\subsubsection*{Scalar wave equation and radial ODE}

The Klein–Gordon equation $\Box\chi=0$ reads, upon separating
\[
\chi(t,r,\theta,\varphi) \;=\; e^{-i\omega t}\,Y_{\ell m}(\theta,\varphi)\,R(r).
\]
Therefore, in isotropic coordinates, the radial ordinary differential equation becomes
\[
R''(r)
\;+\;\Biggl[\frac{2}{r} \;+\;\frac{1}{2}\,\frac{d}{dr}\ln\bigl[f(r)g(r)\bigr]\Biggr]\,
R'(r)
\;-\;\frac{\ell(\ell+1)}{r^2}\,R(r)
\;=\;0.
\]
We focus on the static dipole mode, \(\omega=0\), \(\ell=1\), so the last term is \(-\,2/r^2\,R\). A general strategy to solve the above radial equation is outlined below.

\subsubsection*{``Yukawa-deformed'' Frobenius ansatz}

To capture the Yukawa tail, we use the ``Yukawa-deformed'' Frobenius-type ansatz
\begin{equation}
R_\ell(r)
=r^\ell\sum_{i=3}\Bigl(1+\frac{Md_i}{r^i\mu^28\pi m_p^2}\Big)\Big\{1-(1+\mu\,r)\,e^{-\mu r}\Big\}.\label{SourcePart}
\end{equation}
Since the $1/r^3$ term appears with just a single insertion in our case, there cannot be any term like $1/r$ or $1/r^2$ in our ansatz to begin with. %In the limit $\mu\to\infty$, $c_3=0$. Also, note that the growing mode is also Yukawa deformed, while at $r\to\infty$, the mode is as usual $R_{\ell}(r\to\infty)\sim r$ as required on physical grounds.
Furthermore, we have the freedom to add $r^{-(\ell+1)}$ to the Frobenius solution. In GR, the Frobenius solution is given by \cite{Ivanov:2022hlo},
\begin{equation}
R_\ell^{\textrm{GR}}=r^\ell\Bigg(1+c_2\Big(\frac{M}{2r}\Big)^2+c_4\Big(\frac{M}{2r}\Big)^4+...+c_{2n}\Big(\frac{M}{2r}\Big)^{2n}\Bigg)
\end{equation}
is the full solution since it is a pure polynomial and truncates at $n=\ell$ (a polynomial has infinite radius of convergence and becomes the full \emph{global} solution). This means that PN corrections to Love numbers and the Love numbers themselves vanish. In contrast, the Yukawa deformation spoils this algebraic relation and never truncates since the exponential introduces an essential singularity at $r\to\infty$. This means that the deformed Yukawa ansatz is not a full solution (it has a finite radius of convergence). Therefore, PN corrections to Love numbers and Love numbers need not vanish, and we add $r^{-(\ell+1)}$ at $r\to\infty$. Thus, the full solution becomes
\begin{equation}
    R^{\textrm{full}}_\ell(r)
=r^\ell\sum_{i=3}\Bigl(1+\frac{d_iR}{r^i}\Big)\Big\{1-(1+\mu\,r)\,e^{-\mu r}\big\}+k_\ell\frac{R}{r^{\ell+1}}\sum_{i=1}\Bigl(1+b_i\Big(\frac{r_s}{r}\Big)^i\Big)\Big\{1-(1+\mu\,r)\,e^{-\mu r}\big\} \label{full}
\end{equation}
where we have defined $\tfrac{M}{\mu^28\pi m_p^2}=R$ and $r_s=2GM$ is the Schwarzschild radius in $D=4$. We now define the two pieces of the differential operator
\begin{equation}
\mathcal L[R]
=\,\underbrace{R''+\tfrac{2}{r}R'-\tfrac{2}{r^2}R}_{L_0[R]}
\;+\;
\underbrace{\tfrac12\,( \ln(fg) )'\,R'}_{I[R]}
\;=\;0.
\end{equation}
We strip off the common factor \(e^{-\mu r}\) and expand at large \(r\) to identify the leading Yukawa tail \(e^{-\mu r}/r^2\), and we are left with the following two pieces.
\\\\
{\bf\textit{ Flat piece \(L_0[R]\)}:} A direct differentiation shows that,
\[
e^{+\mu r}\,L_0[R]
\;\supset\;
\,(d+k_\ell)M\;\frac{1}{r^2}
\;\;\Longrightarrow\;\;
L_0[R]\;\supset\;\,(d+k_\ell)M\;\frac{e^{-\mu r}}{r^2}.
\]\\
{\bf\textit{Log-derivative piece \(I[R]\)}:}
Similarly, one also finds that,
\[
(\ln fg)'
=\frac{d}{dr}\ln\bigl[f(r)g(r)\bigr]
\;=\;-X\Big(2\mu+\frac{2}{r}+\mu^2 r\Big)+\text{higher order terms},
\qquad
X=\frac{e^{-\mu r}}{r},
\]
and hence
\[
e^{+\mu r}\,I[R]
\;\supset\;
\,-\frac{(\tilde{c}_1+\tilde{c}_2)}{2}\;\frac{1}{r^2}
\;\;\Longrightarrow\;\;
I[R]\;\supset\;-\frac{(\tilde{c}_1+\tilde{c}_2)}{2}\;\frac{e^{-\mu r}}{r^2}.
\]
with no $d$-dependence at this order. Here, $\tilde{c}_1$ and $\tilde{c}_2$ are Yukawa deformed metric coefficients of $f(r)$ and $g(r)$ respectively. Readers are referred to the Table~(\ref{lll})  for these coefficients. \vspace{1mm}

\noindent
Collecting coefficients of $e^{-\mu r}/r^2$ tail, and equating to zero gives
\begin{equation}
    (d+k_\ell)M-\frac{(\tilde{c}_1+\tilde{c}_2)}{2}=0. \label{Love}
\end{equation}
The reason we set the coefficient of the $1/r^2$ term to zero is related to PN counting in our case. With one insertion, we get the term $1/r^3$, with two insertions $1/r^4$ and so on. Therefore, the $1/r^2$ term can be generated only with one insertion. More insertions will bring higher-order corrections. %Even with a $1/r^3$ term in the ansatz, one can check that higher-order terms, $1/r^4$ and $1/r^5$, are generated. 
But these terms can be combined with other $ d$'s in the ansatz, and then for the ODE to hold, we can demand that the coefficients of that order vanish. Since $1/r^2$ appears only with one insertion, it must vanish in order to satisfy the ODE (no higher insertions can produce that term, which we can combine with and demand to vanish identically).\vspace{1mm}

\begin{table}[htb!]
\centering
\begin{tabular}{lll}
\hline
$\tilde{c}_1$ & $\tilde{c}_2$ & Parameter condition \\
\hline
$2GM$      & $-\frac{48}{25}GM$        & $8\beta + 3\gamma = 0$ \\[6pt]
$2GM$     & $-2GM$        & $2\beta + \gamma = 0$ \\[6pt]
$2GM$      & $-2GM-\frac{GM(1-2k)}{5(3-8k)}$        & $\beta + k\gamma = 0\Big(\frac{3}{8}<k<\frac{1}{2}\Big)$ \\
\hline 
\end{tabular}
\caption{List of $\tilde{c}_1$ and $\tilde c_{2}$ following from (\ref{M1}), (\ref{M2}) and (\ref{M3}).} \label{lll}
\end{table} 

%To sum up, by construction, the ansatz (encoding PN corrections to the external field)
%\[
%R_\ell(r) \;=\; r^\ell \sum_{i=3}^{\infty} \biggl(1 + \frac{M\,d_i}{8\pi\,m_p^2\,\mu^2\,r^i}\biggr)
%\Bigl\{1 - (1 + \mu r)e^{-\mu r}\Bigr\}
%\]
%forbids any \(1/r\) or \(1/r^2\) term.  Hence:

%\begin{itemize}
%  \item Only the \(i=3\) insertion can source an \(\displaystyle \frac{e^{-\mu r}}{r^2}\) term in the ODE residual.  All other \(i\ge4\) generate tails of order \(\frac{e^{-\mu r}}{r^i}\) (or faster fall–off) under the action of the differential operator.
%  \item The condition that the coefficient of \(\displaystyle \frac{e^{-\mu r}}{r^2}\) vanishes yields a \emph{unique} matching equation fixing the UV charge \(\tilde{c}\) (see Appendix \ref{app_UV_fixing}).
%  \item Higher coefficients \(d_i\) for \(i\ge4\) only control subleading Yukawa tails,
%    \(\displaystyle \frac{e^{-\mu r}}{r^4}, \frac{e^{-\mu r}}{r^5}, \ldots,\)
%    which we consider only if we need higher–order accuracy.
%\end{itemize}
%Therefore, by starting the ansatz at \(1/r^3\) (which is fixed by PN counting), one ensures that once the \(\tfrac{e^{-\mu r}}{r^2}\) term is cancelled by fixing \(d\), \emph{no} higher–order piece can regenerate it.\vspace{1mm}

\subsubsection*{Calculating the scalar Love number $k_1$}

Let us now proceed with the calculation of the coefficient $d_i$ in the full Frobenius ansatz (\ref{full}). Since the part of the UV solution (as presented in (\ref{SourcePart})) giving the PN corrections to the source is unambiguously determined by EFT calculation, one can read off the coefficient $d_3$ by matching (\ref{SourcePart}) with the EFT calculation of one-point function (\ref{one_point_fn_spin_0}) as
\begin{align}
\begin{split}
    &d_3\;\equiv\;c_3 = \frac{1}{4}, \hspace{1.9cm}\text{if }8\beta+3\gamma=0,\\&
d_3\;\equiv\;c_3= 0, \hspace{2.0cm}\text{if }2\beta+\gamma=0,\\&
d_3\equiv c_3=1-2k,\hspace{1.3cm}\text{if }\beta+k\gamma=0.
\end{split}
\end{align}
Therefore, from (\ref{Love}), we find
\begin{align}
\begin{split}
    &k_1=-\frac{21}{100},\hspace{3.6cm}  \text{if }8\beta+3\gamma=0,\\&    
    k_1=0, \hspace{4.4cm} \text{if }2\beta+\gamma=0,\\&
    k_1=-\frac{(1-2k)(11-10k)}{10(1-k)}, \hspace{1.1cm} \text{if }\beta+k\gamma=0
    \end{split} \label{tilde_c_EFT}
\end{align}
where $\frac{3}{8}<k<\frac{1}{2}$. Therefore, the Love number $k_\ell$ does not vanish except for $2\beta+\gamma=0$. The upshot of the above calculations is that we obtain (in general) a \emph{scale-independent} non-zero scalar Love number $k_1$ which is given by (in terms of couplings as)
\begin{equation}
    k_1=\Bigg(-\frac{(1-2k)(11-10k)}{10(1-k)}\Bigg)M\sqrt{|\gamma|}.
\end{equation}
One immediately recovers the GR result that the scalar Love number vanishes, $k^{\rm GR}_1=0$ \cite{Ivanov:2022hlo,Ivanov:2022qqt} for $\gamma=0\,.$ \vspace{1mm} 

Before ending the section, we would like to make a comment. We have evaluated the Love number using our EFT reconstructed metric so that the matching remains consistent, as we use the same gauge (same coordinate systems) for the IR and UV computation of the one-point function. However, one can match gauge-invariant quantities in different gauges, which should be the case for the Love number and PN corrections to the source (i.e $c_{2l+1}$ coefficient). Keeping this in mind, in the Appendix \ref{app_UV_fixing}, we repeat the same exercise with an existing spherically symmetric black hole solution (in the Schwarzschild-like coordinate) \cite{Stelle:1977ry,Daas:2022iid} for quadratic gravity. 
%matches with that of our EFT reconstructed metric (\ref{4.36t}) after we fix the UV charges via Wilsonian matching as described in that section.

%

\section{Connection to Ladder Symmetry}\label{sec9}
In this section, we briefly discuss the connection to ladder symmetry in our context.
In the static, spherically–symmetric Schwarzschild background
\begin{equation}
\mathrm{d}s^2
= -\frac{\Delta}{r^2}\,\mathrm{d}t^2
  +\frac{r^2}{\Delta}\,\mathrm{d}r^2
  +r^2\mathrm{d}\Omega^2,
\qquad
\Delta = r(r - r_s)\,,
\end{equation}
a massless scalar perturbation $\phi_\ell(r)$ obeys
\begin{equation}\label{eq:scalar-eom}
\partial_r\!\bigl(\Delta\,\partial_r\phi_\ell\bigr)
- \ell(\ell+1)\,\phi_\ell = 0\,.
\end{equation}
One can define first–order ``ladder'' operators
\begin{equation}
D^+_\ell = -\Delta\,\partial_r + \tfrac{\ell+1}{2}(r_s - 2r)\,,
\qquad
D^-_\ell = \Delta\,\partial_r + \tfrac{\ell}{2}(r_s - 2r)\,,
\end{equation}
which satisfy the algebraic relations \cite{Hui:2021vcv}
\begin{equation}
\begin{aligned}
H_\ell 
&\equiv -\Delta\bigl(\partial_r(\Delta\,\partial_r)-\ell(\ell+1)\bigr)\,,\\
&= D^-_{\ell+1}D^+_\ell \;-\; \frac{(\ell+1)^2r_s^2}{4}\,
= D^+_{\ell-1}D^-_\ell \;-\; \frac{\ell^2r_s^2}{4}\,.
\end{aligned}
\end{equation}
These imply the “raising” and “lowering” relations
\begin{equation}
H_{\ell+1}\,D^+_\ell = D^+_\ell\,H_\ell\,,
\qquad
H_{\ell-1}\,D^-_\ell = D^-_\ell\,H_\ell\,,
\end{equation}
so that from any solution at multipole $\ell$ one can climb or descend to neighboring $\ell\pm1$ levels.  Moreover, one constructs the conserved “horizontal” charges
\begin{equation}
P_\ell = \Delta\,\partial_r\bigl(D^-_1 D^-_2\cdots D^-_\ell\,\phi_\ell\bigr)\,,
\qquad
\partial_r P_\ell = 0\,,
\end{equation}
which relate the behavior of $\phi_\ell$ at the horizon ($r\to r_s$) to that at infinity ($r\to\infty$) without explicit integration.  
In particular, the regular (growing‐branch) solution is annihilated by all $P_\ell$, enforcing its single–power law fall‐off and implying vanishing static response (Love number), while the decaying branch necessarily diverges at the horizon, consistent with expectation from the no‐hair theorem.  This ladder‐symmetry \cite{Hui:2021vcv,Hui:2022vbh,Rai:2024lho, Kehagias:2024rtz, Combaluzier-Szteinsznaider:2024sgb, Gounis:2024hcm, Charalambous:2024tdj} perspective thus provides a unified algebraic explanation of both the vanishing of Love numbers and the absence of linear hair in Schwarzschild-like metrics.\vspace{1mm}

Moreover, following \cite{Hui:2021vcv,Hui:2022vbh,Rai:2024lho}, it can be shown that starting from a general spherically symmetric static spacetime, the imposition of ladder symmetry enforces $g_{tt}\,g_{rr}=-1$. \textcolor{black}{In our case, the Yukawa-deformed part breaks this constraint; therefore, no such Ladder symmetry exists for our case (except for the special case when couplings satisfy $2\beta+\gamma=0$ and Ladder symmetry is restored, in which case the (scalar) Love numbers indeed vanish). This provides another way of looking at why Love numbers do not vanish (in general) in quadratic gravity. It is interesting that, even though our quadratic-gravity background does not admit the usual Ladder-operator structure ($f\cdot g\neq-1$), the Love number beta‐function remains zero, i.e, $\frac{d\lambda_\ell(\nu)}{d\nu}=0$}.\par
Presumably, in the worldline EFT, this follows from the fact that the extra massive spin-0 and spin-2 modes render all would-be IR divergences finite, so no $\ln r$ term can ever appear. The vanishing of the beta function in the absence of Ladder symmetry hints at a deeper UV mechanism—perhaps a hidden symmetry or duality in quadratic gravity—that enforces non-renormalization of the Love operators. It would be very interesting to uncover this structure in future work.

\section{Conclusion and Discussion}\label{sec10}
Motivated by the case study of calculating the running of Love numbers and computing the Love numbers explicitly by using the techniques of effective field theory, we further generalize it to showcase the effective field theory techniques in quadratic gravity to compute the Love number and its running. In the following, we list the main findings of our paper.

\begin{itemize}
\item \textcolor{black}{One of the principal focuses of this paper is devoted to reconstructing the metric using EFT techniques for higher curvature corrections. We demonstrate the obtained Yukawa-type metric for quadratic EFT, showcasing one-point functions of fields (scalar and tensor) in the static gauge, which couples to the worldline. Unlike GR, the Yukawa tails appear in 1PN because of the massive degrees of freedom ($m_1$ and $m_2$ respectively). Though the reconstructed metric is perturbative, obtaining the one-point functions required for its reconstruction is quite challenging because of the massive graviton modes in quadratic EFT. Though we manage to obtain closed-form analytic expressions for the integrals while demonstrating the reconstruction. These are a few new results to the best of our knowledge.} 
   \item Further, in this work, we have extended the classic calculation of tidal Love numbers for Schwarzschild black holes to a four-dimensional quadratic curvature effective field theory of gravity. In the UV side, we have derived an analytic Yukawa-deformed Frobenius solution for spin-0 perturbation.  Then by matching the UV-IR constructions, we unambiguously find the Love number for spin-0 perturbation, which turns out to be non-vanishing except for a certain choice of the higher curvature couplings.

\item Our results carry several important implications.  First, they demonstrate explicitly how higher-curvature corrections break the ladder‐symmetry protection $g_{tt}g_{rr}=-1$ that enforces zero Love numbers in GR.  The Yukawa tail introduced by the $R^2+C^2$ terms spoils the first‐order factorization of the static master equation, opening the door to genuine tidal response.  Second, the perturbative matching procedure we employed provides a clean worldline EFT interpretation of tidal charge computation, complete with non-running of the classical renormalization‐group of the Wilson coefficients $\lambda_\ell$. We anticipate that forthcoming gravitational-wave observations may have the sensitivity to probe or constrain this, opening a new window on the ultraviolet structure of gravity.   This framework can be readily generalized to higher multipoles, spin-induced effects, or theories with additional curvature invariants.

% Several caveats and avenues for future work deserve emphasis.  The EFT is known to propagate a massive spin-2 ghost; understanding whether a healthy UV completion (e.g.\ a stringy or nonlocal theory) qualitatively alters the tidal response is an open question.  Moreover, our analysis has been restricted to the static ($\omega=0$) limit and to the scalar sector.  Extending the matching to dynamical perturbations, vector modes, and higher multipoles will both enrich the phenomenology and test the robustness of the ladder‐symmetry argument’s breakdown.

\item Finally, we have also shown that quadratic‐curvature corrections generically endow black holes with a nonzero but \textit{scale-independent} tidal Love number. \textcolor{black}{Scale independence can be traced to the modified PN counting in quadratic EFT, where relevant tidal tails appear at tree-level and no logarithmic terms appear in the one-point function, rendering non-running of RG.} 
%Our combined analytic-numerical matching approach not only fixes the relevant Wilson coefficient but also illustrates how EFT techniques can be leveraged to extract observable signatures of beyond‐GR physics.  
\end{itemize}
\noindent
Several caveats and avenues for future work deserve emphasis. First of all it will be good to extend our computation beyond the linear order in higher curvature couplings. Second, the EFT is known to propagate a massive spin-2 ghost; understanding whether a healthy UV completion (e.g.\ a stringy or nonlocal theory) qualitatively alters the tidal response is an open question. \textcolor{black}{For example, our constraint on parameters $\beta,\gamma$ given by (\ref{parameter_constraints}) forbids the choice $\beta=1/3\mu^2$ and $\gamma=-1/\mu^2$
for a healthy theory with no tachyonic instability. Intriguingly, \cite{Antoniou:2024jku} includes this region of the $(\beta,\gamma)$ parameter while dealing with a classical black hole solution of quadratic gravity.  While their construction is formally valid at the level of the classical field equations, the underlying tachyonic instability we have identified calls into question the physical relevance of including this particular region.  In particular, one should examine whether higher‐order (e.g.\ cubic or quartic) curvature corrections or a nonperturbative completion can ameliorate the instability, and thus render the solution dynamically viable}. Moreover, our analysis has been restricted to the static ($\omega=0$) limit and to the scalar sector.  Extending the matching to dynamical perturbations, vector modes, and higher multipoles will both enrich the phenomenology and test the robustness of the ladder-symmetry argument’s breakdown.\vspace{1mm} 

Another direction is to examine the role of dissipation effects for black holes within theories beyond GR, which is non-vanishing for Kerr black holes in static external perturbation settings. Therefore, unlike in GR, modified gravity models, particularly those involving higher curvature corrections or additional degrees of freedom, may allow for non-trivial dissipative effects even under static external perturbations. It would be interesting to put forward a consistent matching procedure for dissipation numbers in such contexts, possibly identifying new signatures of non-GR theories that could be probed via GW observations. Another interesting direction is to examine the origin of the fine-tuning in the EFT description of PN dynamics beyond GR \cite{Charalambous:2025ekl}. In four dimensions, for GR, the cancellation of logarithmic corrections to static Love numbers is linked to a hidden Love symmetry enforcing polynomial structures in certain correlators. Whether this symmetry generalizes to non-GR theories, especially those with extra fields or higher-derivative terms, remains an open question. Such investigations, within the EFT framework, will also be interesting for higher dimensions or specific models like scalar-tensor or Gauss-Bonnet, which may uncover new patterns of fine-tuning, symmetry enhancement, or breakdowns. \vspace{1mm}

\textcolor{black}{Note that, in this work, we have restricted our UV computation to the spin-0 case only. One should try to generalize it for the spin-2 case, which involves computing gravitational perturbations using a modified version of the Teukolsky equation \cite{Li:2022pcy, 
Hussain:2022ins,Cano:2025zyk}. On the IR side, we have computed the PN corrections to the source both for the spin-0 and spin-2 electric-type perturbations. If one can perform a UV calculation for the gravitational case, then utilizing the UV and EFT matching, it will be possible to unambiguously obtain the Love number associated with gravitational perturbations. We leave this for future work.}

 On the phenomenological side, our nonvanishing Love number offers a concrete observational handle on quadratic curvature gravity.  Current and near-future gravitational-wave detectors probe the late-inspiral regime of compact binaries at radii $r\sim10\text{–}20\,M$, where the Yukawa correction is enhanced by $e^{-r/\ell_{\rm Yuk}}$.  Bounds on the phase shift induced by tidal interactions thus translate directly into constraints on the scales $m^{-1}_0\sim\sqrt{12\beta+4\gamma},\,\,m_2'^{-1}\sim\sqrt{-\gamma}$ and the magnitude of $\tilde c$.  A detailed Fisher analysis remains to be done, but our analytic expression (Eq. \ref{4.36t}) provides a ready starting point.
 \noindent
These studies can potentially clarify how non-GR theories encode the internal structure and response of compact objects or black holes and provide theoretically precise observables to distinguish them from GR. 

\section*{Acknowledgments}
The authors thank the speakers of ``Testing Aspects General Relativity-IV" for the illuminating discussion where this work is presented by one of the authors. S.G. (PMRF ID: 1702711) and S.P. (PMRF ID: 1703278) are supported by the Prime Minister’s Research Fellowship of the Government of India. The research of S.K. is funded by the National Post-Doctoral Fellowship (N-PDF: PDF/2023/000369)
from the ANRF (formerly SERB), Department of Science and Technology (DST), Government of
India. AB is supported by the Core Research Grant (CRG/2023/005112) by DST-ANRF of India Govt.  AB also acknowledges the associateship program of the Indian Academy of Science, Bengaluru.
%\newpage
\appendix
\section{Solving the Master Integrals for the Perturbative Reconstruction of the Metric}\label{app1}

The massless integral corresponding to the \(\nabla^2\Phi\,\partial_i\Phi\,\partial^i\Phi\) vertex (see Eq.~(\ref{fig2_integral})) can be recast as\footnote{Here, \(\int_{\vec{k}}=\int \frac{d^D k}{(2\pi)^D}\).}
\begin{equation}
    -\int_q \frac{e^{i q r}}{\boldsymbol{q}^2+m_1^2}\,\frac{1}{2}\int_{\vec{k}_1}\frac{k_1^2+(k_1+q)^2-q^2}{k_1^2\,(k_1+q)^2}\,.
\end{equation}
We begin by focusing on the \(k\)-integral. By applying an algebraic reduction procedure (for example, using \textsc{LiteRed} \cite{Lee:2012cn}, the integral simplifies to
\begin{equation}
    -q^2\,\int_{\vec{k}_1}\frac{1}{k_1^2\,(k_1+q)^2}\,.
\end{equation}

To evaluate this expression, we employ Feynman parametrization. Introducing the parameter \(x\in [0,1]\), the integral is rewritten as
\begin{equation}
    \mathcal{M}_{11} = \int_0^1 dx\, \int_{\vec{k}} \frac{d^D k}{(2\pi)^D}\,\frac{1}{\Bigl[k^2+2x\,k\cdot q+xq^2\Bigr]^2}\,.
\end{equation}
Defining the shifted momentum \(p=k+xq\), we obtain
\begin{equation}
    \mathcal{M}_{11} = \int_0^1 dx\, \int \frac{d^D p}{(2\pi)^D}\,\frac{1}{\Bigl[p^2+x(1-x)q^2\Bigr]^2}\,.
\end{equation}
The momentum integration over \(p\) is standard and yields
\begin{equation}
    \mathcal{M}_{11}(q) = \frac{1}{(4\pi)^{D/2}}\frac{\Gamma\Bigl(2-\frac{D}{2}\Bigr)}{\Gamma(2)}\frac{1}{\bigl(M^2\bigr)^{2-\frac{D}{2}}}\,,
\end{equation}
with \(M^2=x(1-x)q^2\). The remaining Feynman parameter integral
\begin{equation}
    I_x = \int_0^1 dx\,\bigl[x(1-x)\bigr]^{\frac{D}{2}-2}
\end{equation}
is recognized as the Beta function,
\begin{equation}
    I_x = B\Bigl(\frac{D}{2}-1,\frac{D}{2}-1\Bigr) = \frac{\Gamma\Bigl(\frac{D}{2}-1\Bigr)^2}{\Gamma(D-2)}\,.
\end{equation}
Substituting back, we arrive at
\begin{equation}
    \mathcal{M}_{11} = \frac{1}{(4\pi)^{D/2}}\frac{\Gamma\Bigl(2-\frac{D}{2}\Bigr)\Gamma\Bigl(\frac{D}{2}-1\Bigr)^2}{\Gamma(D-2)}\,(q^2)^{\frac{D}{2}-2}\,.
\end{equation}

Next, we consider the master integral \(\mathcal{D}_{11}\),
\begin{equation}
    \mathcal{D}_{11} = \int \frac{d^D k}{(2\pi)^D}\,\frac{1}{\Bigl[(k^2+m_1^2)((k+q)^2+m_1^2)\Bigr]}\,.
\end{equation}
Following the same Feynman parametrization strategy and shifting the integration variable via \(p=k+(1-x)q\), the integral takes the form
\begin{equation}
    \mathcal{D}_{11} = \int_0^1 dx\, \int \frac{d^D p}{(2\pi)^D}\,\frac{1}{\Bigl[p^2+\Delta\Bigr]^2}\,,
\end{equation}
with \(\Delta=m_1^2+x(1-x)q^2\). Performing the \(p\)-integration gives
\begin{equation}
    \mathcal{D}_{11}(q) = \frac{\Gamma\Bigl(2-\frac{D}{2}\Bigr)}{(4\pi)^{D/2}} \int_0^1 dx\,\Bigl[m_1^2+x(1-x)q^2\Bigr]^{\frac{D}{2}-2}\,.
\end{equation}
After some algebra, this \(x\)-integration may be expressed in terms of a hypergeometric function, leading to
\begin{equation}
    \mathcal{D}_{11} = \frac{\Gamma\Bigl(2-\frac{D}{2}\Bigr)}{(4\pi)^{D/2}}\Bigl(M^2\Bigr)^{\frac{D}{2}-2}\, {}_2F_1\!\Bigl(2-\frac{D}{2},\frac{1}{2};\frac{3}{2};\frac{q^2}{4m_1^2+q^2}\Bigr)\,,
\end{equation}
where \(M^2 = m_1^2+\frac{q^2}{4}\).

We now turn to the master integral \(\mathcal{D}_{01}\),
\begin{equation}
    \mathcal{D}_{01} = \int \frac{d^D k}{(2\pi)^D}\,\frac{1}{\bigl((k+q)^2+m_1^2\bigr)}\,.
\end{equation}
By setting \(k' = k+q\) and introducing a Schwinger parameterization,
\begin{equation}
    \frac{1}{k'^2+m_1^2} = \int_0^\infty dt\,e^{-t\,(k'^2+m_1^2)}\,,
\end{equation}
The integral is recast as
\begin{equation}
    \mathcal{D}_{01} = \int_0^\infty dt\,e^{-t\,m_1^2}\,\int d^D k'\,e^{-t\,k'^2}\,.
\end{equation}
Evaluating the Gaussian integral over \(k'\) yields
\begin{equation}
    \mathcal{D}_{01} = \pi^{D/2}\int_0^\infty dt\, t^{-D/2}\,e^{-t\,m_1^2}\,.
\end{equation}
Recognizing the \(t\)-integral as a Gamma function, we obtain
\begin{equation}
  \mathcal{D}_{01} = \frac{1}{(4\pi)^{D/2}}\Gamma\Bigl(1-\frac{D}{2}\Bigr)(m_1^2)^{\frac{D}{2}-1}\,.
\end{equation}

A particularly efficient strategy is to consider the master integral
\begin{equation}
  I = \int \frac{d^D k}{(2\pi)^D}\,\frac{1}{\Bigl(k^2+m_1^2\Bigr)\Bigl((k+q)^2+m_2^2\Bigr)}\,,
\end{equation}
where the parameters \(m_1\) and \(m_2\) can be chosen appropriately. Using Feynman parametrization, the integral is expressible as
\begin{equation}
    I = \frac{1}{(4\pi)^{D/2}}\Gamma\Bigl(2-\frac{D}{2}\Bigr)\int_0^1 dx\,\Bigl[x(1-x)q^2+(1-x)m_2^2+xm_1^2\Bigr]^{\frac{D}{2}-2}\,.
\end{equation}
For the case \(D=3\), the \(x\)-integral is evaluated as
\begin{equation}
    I_x = \frac{2}{q}\,\left[\arctan\!\Bigl(\frac{\boldsymbol{q}}{m_1-m_2}\Bigr)-\arctan\!\Bigl(\frac{2m_2q}{m_1^2-m_2^2+\boldsymbol{q}^2}\Bigr)\right]\,.
\end{equation}

Table~\ref{tab:cases} summarizes the various master integrals along with the corresponding parameter choices for \(m_1\) and \(m_2\).

\begin{table}[ht!]
    \centering
    \renewcommand{\arraystretch}{1.2}
    \begin{tabular}{|c|c|}
        \hline
        \textbf{Master Integral} & \textbf{Parameter Choices} \\
        \hline
        \(\mathcal{M}_{11}\) & \( m_1 = m_2 = 0 \) \\
        \hline
        \(\mathcal{D}_{11}\) & \( m_1 = m_2 \neq 0 \) \\
        \hline
        \(\mathcal{K}_{11}\) & \( m_1 = 0 \) or \( m_2 = 0 \) \\
        \hline
        \(\mathcal{J}_{11}\) &\( m_1 \neq m_2 \neq 0 \) \\
        \hline
    \end{tabular}
    \caption{Parameter choices for different classes of master integrals.}
    \label{tab:cases}
\end{table}
This gives
\begin{align}
\mathcal{M}_{01}&=0,\ \mathcal{M}_{11}=\frac{1}{8q},&\\
\mathcal{K}_{01}&=-\frac{m}{4\pi},\ \mathcal{K}_{11}=\frac{\arctan(q/m)}{4\pi q},&\\
\mathcal{D}_{01}&=-\frac{m}{4\pi},\ \mathcal{D}_{11}=\frac{1}{8q}-\frac{\arctan(2m/q)}{4\pi q},&\\
\mathcal{J}_{11}&=\frac{1}{8\pi}I_x.
\end{align}

\section{Useful Fourier Integrals}\label{app2}
In this appendix, we present detailed derivations of several three-dimensional Fourier integrals that appear in our analysis. In all cases, we denote $r=|\boldsymbol{x}|$.

\subsection{Fourier Inverse of $\boldsymbol{\frac{k^i}{k^4}}$}

We wish to evaluate
\begin{equation}\label{eq:fourier1}
I^i(\boldsymbol{x}) = \int \frac{d^3k}{(2\pi)^3}\, e^{i\boldsymbol{k}\cdot\boldsymbol{x}}\, \frac{k^i}{k^4}\,.
\end{equation}
A well-known property of Fourier transforms is that multiplication by $k^i$ in momentum space corresponds to differentiation in coordinate space:
\[
\int \frac{d^3k}{(2\pi)^3} \, e^{i\boldsymbol{k}\cdot\boldsymbol{x}}\, k^i\, F(k) = -i\partial_i \left( \int \frac{d^3k}{(2\pi)^3}\, e^{i\boldsymbol{k}\cdot\boldsymbol{x}}\, F(k) \right).
\]
It is also standard that
\begin{equation}
\int \frac{d^3k}{(2\pi)^3}\, \frac{e^{i\boldsymbol{k}\cdot\boldsymbol{x}}}{k^4} = -\frac{r}{8\pi}\,.
\end{equation}
Thus,
\begin{align}
    I^i(\boldsymbol{x}) = -i\,\partial_i\left(-\frac{r}{8\pi}\right)
= i\,\partial_i \left(\frac{r}{8\pi}\right)
= \frac{i}{8\pi}\,\partial_i r\,.
\end{align}
Since,
$\partial_i r = \frac{x^i}{r}$
we obtain
\begin{equation}\label{eq:result1}
\boxed{ I^i(\boldsymbol{x}) = \frac{i\,x^i}{8\pi\, r}\,. }
\end{equation}

\subsection{Fourier Inverse of $\boldsymbol{\frac{k^i}{k^2(k^2+m^2)}}$}

Next, we consider the integral
\begin{equation}\label{eq:fourier2}
J^i(\boldsymbol{x}) = \int \frac{d^3k}{(2\pi)^3}\, e^{i\boldsymbol{k}\cdot\boldsymbol{x}}\, \frac{k^i}{k^2(k^2+m^2)}\,.
\end{equation}
Define the scalar Fourier transform
\[
S(\boldsymbol{x}) = \int \frac{d^3k}{(2\pi)^3}\, \frac{e^{i\boldsymbol{k}\cdot\boldsymbol{x}}}{k^2(k^2+m^2)}\,.
\]
We first decompose the integrand using partial fractions:
\[
\frac{1}{k^2(k^2+m^2)} = \frac{1}{m^2} \left( \frac{1}{k^2} - \frac{1}{k^2+m^2} \right).
\]
With the standard results
\[
\int \frac{d^3k}{(2\pi)^3}\, \frac{e^{i\boldsymbol{k}\cdot\boldsymbol{x}}}{k^2} = \frac{1}{4\pi\,r}, \qquad
\int \frac{d^3k}{(2\pi)^3}\, \frac{e^{i\boldsymbol{k}\cdot\boldsymbol{x}}}{k^2+m^2} = \frac{e^{-mr}}{4\pi\,r}\,,
\]
we have
\[
S(\boldsymbol{x}) = \frac{1}{m^2}\left[\frac{1}{4\pi\,r} - \frac{e^{-mr}}{4\pi\,r}\right]
= \frac{1}{4\pi m^2}\,\frac{1-e^{-mr}}{r}\,.
\]
Since differentiation with respect to $x^i$ gives
\[
J^i(\boldsymbol{x}) = -i\,\partial_i S(\boldsymbol{x}),
\]
and noting that $S$ depends only on $r$, so that
\[
\partial_i S(\boldsymbol{x}) = S'(r)\,\frac{x^i}{r}\,,
\]
we compute
\[
S'(r) = \frac{d}{dr}\left[\frac{1-e^{-mr}}{4\pi m^2\,r}\right]
=\frac{1}{4\pi m^2}\left[\frac{m\,e^{-mr}\,r - (1-e^{-mr})}{r^2}\right]\,.
\]
Thus, we find
\[
J^i(\boldsymbol{x}) = -i\, \frac{x^i}{r} \,S'(r)
= -\frac{i\,x^i}{4\pi m^2\,r^3}\,\Bigl[m\,r\,e^{-mr} - (1-e^{-mr})\Bigr]\,.
\]
It is often preferable to rewrite the result as
\begin{equation}\label{eq:result2}
\boxed{ J^i(\boldsymbol{x}) = \frac{i\, x^i}{4\pi\, m^2\,r^3}\,\Bigl[ 1-e^{-mr}(1+m\,r)\Bigr]\,. }
\end{equation}

\subsection{Fourier Inverse of $\boldsymbol{\frac{k^i}{(k^2+m_1^2)(k^2+m_2^2)}}$}

We now evaluate
\begin{equation}\label{eq:fourier3}
K^i(\boldsymbol{x}) = \int \frac{d^3k}{(2\pi)^3}\, e^{i\boldsymbol{k}\cdot\boldsymbol{x}}\, \frac{k^i}{(k^2+m_1^2)(k^2+m_2^2)}\,,
\end{equation}
for $m_1\neq m_2$. Define the scalar transform
\[
T(\boldsymbol{x}) = \int \frac{d^3k}{(2\pi)^3}\, \frac{e^{i\boldsymbol{k}\cdot\boldsymbol{x}}}{(k^2+m_1^2)(k^2+m_2^2)}\,.
\]
Using partial fractions, we have:
\[
\frac{1}{(k^2+m_1^2)(k^2+m_2^2)}
=\frac{1}{m_2^2-m_1^2}\Bigl[\frac{1}{k^2+m_1^2}-\frac{1}{k^2+m_2^2}\Bigr]\,.
\]
Thus, using
\[
\int \frac{d^3k}{(2\pi)^3}\, \frac{e^{i\boldsymbol{k}\cdot\boldsymbol{x}}}{k^2+m^2} = \frac{e^{-mr}}{4\pi\,r}\,,
\]
we obtain
\[
T(\boldsymbol{x}) = \frac{1}{4\pi (m_2^2-m_1^2)}\,\frac{e^{-m_1r}-e^{-m_2r}}{r}\,.
\]
Then,
\[
K^i(\boldsymbol{x}) = -i\,\partial_i T(\boldsymbol{x})\,.
\]
Since $T(\boldsymbol{x})=T(r)$ we have
\[
\partial_i T(\boldsymbol{x}) = T'(r)\,\frac{x^i}{r}\,,
\]
with
\[
T'(r)=\frac{1}{4\pi(m_2^2-m_1^2)}\left[\frac{-m_1e^{-m_1r}+m_2e^{-m_2r}}{r}-\frac{e^{-m_1r}-e^{-m_2r}}{r^2}\right]\,.
\]
A little algebra shows that the result can be written in the compact form
\begin{equation}\label{eq:result3}
\boxed{ K^i(\boldsymbol{x}) = \frac{i\, x^i}{4\pi(m_2^2-m_1^2)\,r^3}\,\Bigl[ e^{-m_1r}(m_1r+1)- e^{-m_2r}(m_2r+1)\Bigr]\,. }
\end{equation}

\subsection{Degenerate Case: $\boldsymbol{m_1=m_2\equiv m}$}

In the limiting case where the masses coincide, we need to compute
\begin{equation}\label{eq:degenerate}
L^i(\boldsymbol{x}) = \int \frac{d^3k}{(2\pi)^3}\, e^{i\boldsymbol{k}\cdot\boldsymbol{x}}\, \frac{k^i}{(k^2+m^2)^2}\,.
\end{equation}
By taking the limit $m_2\to m_1=m$ in Eq.~\eqref{eq:result3} via L’Hôpital’s rule, one obtains
\[
L^i(\boldsymbol{x}) = \frac{i\, x^i}{8\pi}\,\frac{e^{-mr}}{r}\,.
\]
That is,
\begin{equation}\label{eq:result4}
\boxed{ L^i(\boldsymbol{x}) = \frac{i\, x^i}{8\pi}\,\frac{e^{-mr}}{r}\,. }
\end{equation}

\subsection{Fourier Inverse of $\displaystyle \frac{k^i}{k^2\,(k^2+m_1^2)\,(k^2+m_2^2)}$}
We consider the integral
\begin{equation}\label{eq:fourier4}
G^i(\boldsymbol{x}) = \int \frac{d^3k}{(2\pi)^3}\, e^{i\boldsymbol{k}\cdot\boldsymbol{x}}\, \frac{k^i}{k^2\,(k^2+m_1^2)\,(k^2+m_2^2)}\,.
\end{equation}
As in previous cases, define the scalar transform
\[
U(\boldsymbol{x}) = \int \frac{d^3k}{(2\pi)^3}\, \frac{e^{i\boldsymbol{k}\cdot\boldsymbol{x}}}{k^2\,(k^2+m_1^2)\,(k^2+m_2^2)}\,.
\]
Let us perform partial fraction decomposition:
\[
\frac{1}{k^2(k^2+m_1^2)(k^2+m_2^2)}
= \frac{A}{k^2} + \frac{B}{k^2+m_1^2} + \frac{C}{k^2+m_2^2},
\]
with
\[
A=\frac{1}{m_1^2m_2^2},
\quad
B=\frac{1}{m_1^2(m_1^2-m_2^2)},
\quad
C=\frac{1}{m_2^2(m_2^2-m_1^2)}.
\]
Thus
\[
G^i(\boldsymbol{x}) = A\!\int\frac{d^3k}{(2\pi)^3}\frac{k^i}{k^2}e^{i\boldsymbol k\cdot\boldsymbol x}
+ B\!\int\frac{d^3k}{(2\pi)^3}\frac{k^i}{k^2+m_1^2}e^{i\boldsymbol k\cdot\boldsymbol x}
+ C\!\int\frac{d^3k}{(2\pi)^3}\frac{k^i}{k^2+m_2^2}e^{i\boldsymbol k\cdot\boldsymbol x}.
\]
Using
\[
\int\frac{d^3k}{(2\pi)^3}\frac{k^i}{k^2}e^{i\boldsymbol k\cdot\boldsymbol x}
= i\,\frac{x^i}{4\pi r^3},
\qquad
\int\frac{d^3k}{(2\pi)^3}\frac{k^i}{k^2+\mu^2}e^{i\boldsymbol k\cdot\boldsymbol x}
= i\,\frac{x^i}{4\pi r^3}(1+\mu r)e^{-\mu r},
\]
one obtains
\[
G^i(\boldsymbol{x}) = \frac{i\,x^i}{4\pi r^3}
\Bigl[ A + B\,(1+m_1r)e^{-m_1r} + C\,(1+m_2r)e^{-m_2r} \Bigr].
\]
Substituting $A,B,C$ gives
\begin{equation}\label{eq:result5}
\boxed{
G^i(\boldsymbol{x}) = \frac{i\,x^i}{4\pi\,r^3}
\Bigl[\frac{1}{m_1^2m_2^2}
+ \frac{(1+m_1r)e^{-m_1r}}{m_1^2(m_1^2-m_2^2)}
+ \frac{(1+m_2r)e^{-m_2r}}{m_2^2(m_2^2-m_1^2)}\Bigr].
}
\end{equation}

\section*{Fourier Integrals with $\boldsymbol{k^ik^j}$ in the Numerator}\label{app3}

In this appendix, we derive several three‐dimensional Fourier integrals with $k^ik^j$, in the numerator  by relating them to the corresponding scalar transforms. The Fourier transform is defined as
\[
\tilde{f}(\boldsymbol{x}) = \int\frac{d^3k}{(2\pi)^3}\,e^{i\boldsymbol{k}\cdot \boldsymbol{x}}\, f(\boldsymbol{k})\,,
\]
and we denote \(r = |\boldsymbol{x}|\).

A useful identity for any spherically symmetric function \(F(r)\) is
\begin{equation}\label{eq:double_deriv}
\partial_i\partial_j F(r) = F''(r)\,\frac{x^i x^j}{r^2} + F'(r)\left(\frac{\delta^{ij}}{r} - \frac{x^i x^j}{r^3}\right)\,.
\end{equation}
Thus, if
\[
F(\boldsymbol{x}) = \int \frac{d^3k}{(2\pi)^3}\,e^{i\boldsymbol{k}\cdot \boldsymbol{x}}\,f(k^2)\,,
\]
then
\[
\int \frac{d^3k}{(2\pi)^3}\,e^{i\boldsymbol{k}\cdot\boldsymbol{x}}\, k^ik^j f(k^2) = - \partial_i\partial_j F(\boldsymbol{x})\,.
\]

\bigskip

\subsection{Fourier Inverse of $\displaystyle \frac{k^ik^j}{k^4}$}

We start with the scalar transform
\[
F_1(\boldsymbol{x}) = \int \frac{d^3k}{(2\pi)^3}\,\frac{e^{i\boldsymbol{k}\cdot\boldsymbol{x}}}{k^4}
= - \frac{r}{8\pi}\,.
\]
Then,
\[
I^{ij}(\boldsymbol{x}) = \int \frac{d^3k}{(2\pi)^3}\,e^{i\boldsymbol{k}\cdot\boldsymbol{x}}\frac{k^ik^j}{k^4} = -\partial_i\partial_j F_1(\boldsymbol{x})\,.
\]
Since 
\[
F_1(r) = -\frac{r}{8\pi}, \quad F'_1(r) = -\frac{1}{8\pi}, \quad F''_1(r)= 0\,,
\]
using \eqref{eq:double_deriv} we obtain
\[
I^{ij}(\boldsymbol{x}) = \frac{1}{8\pi}\left[\frac{\delta^{ij}}{r} - \frac{x^i x^j}{r^3}\right]\,.
\]
Thus,
\begin{equation}\label{eq:result_kikj_1}
\boxed{
\int \frac{d^3k}{(2\pi)^3}\,e^{i\boldsymbol{k}\cdot\boldsymbol{x}}\frac{k^ik^j}{k^4}
= \frac{1}{8\pi}\left[\frac{\delta^{ij}}{r} - \frac{x^i x^j}{r^3}\right]\,.
}
\end{equation}

\bigskip

\subsection{Fourier Inverse of $\displaystyle \frac{k^ik^j}{k^2(k^2+m^2)}$}

Define the scalar transform
\[
S(\boldsymbol{x}) = \int \frac{d^3k}{(2\pi)^3}\,\frac{e^{i\boldsymbol{k}\cdot\boldsymbol{x}}}{k^2(k^2+m^2)}
= \frac{1}{4\pi m^2}\,\frac{1-e^{-mr}}{r}\,.
\]
Then,
\[
J^{ij}(\boldsymbol{x}) = \int \frac{d^3k}{(2\pi)^3}\,e^{i\boldsymbol{k}\cdot\boldsymbol{x}}\,\frac{k^ik^j}{k^2(k^2+m^2)}
=-\partial_i\partial_j S(\boldsymbol{x})\,.
\]
Let
\[
S(r) = \frac{1-e^{-mr}}{4\pi m^2\,r}\,.
\]
Its derivatives are:
\[
S'(r)=\frac{1}{4\pi m^2}\,\frac{m\,r\,e^{-mr} - (1-e^{-mr})}{r^2}\,,\,
S''(r)=\frac{1}{4\pi m^2}\left[-\frac{m^2e^{-mr}}{r} - \frac{2m\,e^{-mr}}{r^2} + \frac{2(1-e^{-mr})}{r^3}\right]\,.
\]
Then by \eqref{eq:double_deriv}
\[
J^{ij}(\boldsymbol{x}) = -\Biggl\{ S''(r)\,\frac{x^ix^j}{r^2} + S'(r)\left(\frac{\delta^{ij}}{r} - \frac{x^ix^j}{r^3}\right) \Biggr\}\,.
\]
Thus, the result can be written in the derivative form as shown.

\bigskip

\subsection{Fourier Inverse of $\displaystyle \frac{k^ik^j}{(k^2+m_1^2)(k^2+m_2^2)}$}

For this case, define the scalar transform
\[
T(\boldsymbol{x}) = \int \frac{d^3k}{(2\pi)^3}\,\frac{e^{i\boldsymbol{k}\cdot\boldsymbol{x}}}{(k^2+m_1^2)(k^2+m_2^2)}
= \frac{1}{4\pi (m_2^2 - m_1^2)}\,\frac{e^{-m_1r} - e^{-m_2r}}{r}\,.
\]
Then,
\[
K^{ij}(\boldsymbol{x}) = \int \frac{d^3k}{(2\pi)^3}\,e^{i\boldsymbol{k}\cdot\boldsymbol{x}}\,
\frac{k^ik^j}{(k^2+m_1^2)(k^2+m_2^2)}
= -\partial_i\partial_j T(\boldsymbol{x})\,.
\]
\section{A general integral}
\label{app_gen_integral}
 We now discuss the general strategy for computing the $K$-family integrals. The $K$-family takes the following general form,
\begin{align}
    \begin{split}
        \boldsymbol{\zeta}_{m}(\gamma;r)\equiv \int_{\boldsymbol{q}}\,\frac{e^{i\boldsymbol{q}\cdot \boldsymbol{r}}}{|\boldsymbol{q}|^m(|\boldsymbol{q}|^2+\sigma_1^2)} \arctan\left(\frac{\gamma|\boldsymbol{q}|}{\sigma_2}\right)\,.
    \end{split}
\end{align}
Now taking derivative with respect to $\gamma$, we will get,
    \begin{align}
    \begin{split}
        \frac{d \,\boldsymbol{\zeta}_{m}(\gamma;r)}{d\gamma}&= \int_{\boldsymbol{q} }\,\frac{e^{i\boldsymbol{q}\cdot \boldsymbol{r}}}{|\boldsymbol{q}|^m(|\boldsymbol{q}|^2+\sigma_1^2)} \frac{|\boldsymbol{q}|\, \sigma _2}{\gamma ^2 |\boldsymbol{q}|^2+\sigma _2^2}\,,\\ &
        =\frac{4\pi \sigma_2}{r}\int_0^\infty dq \,\frac{\sin(q\,r)}{q^{m-2}(q^2+\sigma_1^2)(\gamma^2q^2+\sigma_2^2)}\,,\\ &
        =\frac{4\pi \sigma_2}{r(\sigma_2^2-\gamma^2\sigma_1^2)}\int_0^\infty \frac{dq\,\sin(q\,r)}{q^{m-2}}\left[\frac{1}{q^2+\sigma_1^2}-\frac{\gamma^2}{\gamma^2q^2+\sigma_2^2}\right]\,,\\ &
        =\frac{4\pi \sigma_2}{r(\sigma_2^2-\gamma^2\sigma_1^2)}\Big(S_{m}(\sigma_1,r)-\gamma^{m+1}S_{m}(\sigma_2,r/\gamma)\Big)\label{5.35j}
    \end{split}
\end{align}
where,
\begin{align}
    S_{m}(\sigma,r)=\int_{0}^{\infty}dz \frac{\sin(z\,r)}{z^{m-2}(z^2+\sigma^2)}\,.
\end{align}
For our case, we need the following $m$ values for which the integral should be evaluated: $m=-1,0,1,3$. Below we list down the corresponding results of the integrals.
\begin{align}
    \begin{split}
        & S_{-1}(\sigma,r)=-\frac{1}{2} \pi  \sigma ^2 e^{-r \sigma }\,,\\ &
        S_{0}(\sigma,r)=\sigma\,  \text{chi}(r \sigma ) \sinh (r \sigma )-\sigma  \cosh (r \sigma )\, \text{shi}(r \sigma )+\frac{1}{r}\,,\\ &
        S_{1}(\sigma,r)=\frac{1}{2} \pi  e^{-\sigma r},\\ &
        S_{2}(\sigma,r)=\frac{\cosh (r \sigma )\, \text{shi}(r \sigma )-\text{chi}(r \sigma )\, \sinh (r \sigma )}{\sigma }\,,\\ &
        S_{3}(\sigma,r)=\frac{\pi -\pi  e^{-\sigma r}}{2 \sigma^2}\,,
        \end{split}
\end{align}
where, \begin{align}
\begin{split}
    &\text{chi}(z):=\gamma_{E}+\log(z)+\int _0^z\frac{(\text{Cosh}(t)-1)}{t} dt\\&
    \text{shi}(z):=\int _0^z\frac{\text{Sinh}(t)}{t} dt.
    \end{split}
\end{align}
Now, integrating \eqref{5.35j} w.r.t $\gamma$ we have,
\begin{align}
    \begin{split}
        \boldsymbol{\zeta}_{m}(\gamma;r)=\frac{4\pi \sigma_2}{r}\int^\gamma d\gamma'\,\frac{1}{\sigma_2^2-\gamma'^2\sigma_1^2}\Big(S_{m}(\sigma_1,r)-\gamma'^{m+1}S_{m}(\sigma_2,r/\gamma')\Big)+C,
    \end{split}
\end{align}
where $C$ can be fixed by choosing an appropriate initial condition. We choose $\gamma=0$, for initial condition where we can do the integral $\boldsymbol{\zeta}_{m}(0;r)$. Let's show the exact results for some of the cases ($m=-1,1,3$),
\begin{align}
\begin{split}
\boldsymbol{\zeta}_{-1}(0;r) &= -\frac{2\pi \left(e^{r\sigma_1} \operatorname{Ei}(-r\sigma_1) - e^{-r\sigma_1} \operatorname{Ei}(r\sigma_1)\right)}{r\sigma_1}\,,\\
\\
\boldsymbol{\zeta}_1(0;r) &= \frac{2\pi \left(e^{r\sigma_1} \operatorname{Ei}(-r\sigma_1) - e^{-r\sigma_1} \operatorname{Ei}(r\sigma_1) - 2r\sigma_1 (\ln(r) + \gamma_{\text{E}} - 1)\right)}{r\sigma_1^3}\,, \\
\\
\boldsymbol{\zeta}_3(0,r) &= \frac{\pi}{9\sigma_1^5} \Biggl[ \frac{18 e^{-r\sigma_1} \left(\operatorname{Ei}(r\sigma_1) - e^{2r\sigma_1} \operatorname{Ei}(-r\sigma_1)\right)}{r} \\
&\qquad + \sigma_1^3 \left((6\gamma_{\text{E}} - 11)r^2 + 6r^2\ln(r)\right) + 36\sigma_1 (\ln(r) + \gamma_{\text{E}} - 1) \Biggr]\,.
\end{split}
\end{align}
Therefore the integral in \eqref{5.35j} becomes,
\begin{align}
\begin{split}
&\boldsymbol{\zeta}_{-1}(\gamma;r) = \frac{\pi}{4\sigma_1\sigma_2}\Biggl[\sigma_1^2 \ln\left(\frac{\sigma_2 - \gamma\sigma_1}{\gamma\sigma_1 + \sigma_2}\right)\\
&\qquad + \sigma_2^2 \left(e^{2 r \sigma_1} \operatorname{Ei}\left(-\frac{r(\gamma\sigma_1 + \sigma_2)}{\gamma}\right) - \operatorname{Ei}\left(r\left(\sigma_1 - \frac{\sigma_2}{\gamma}\right)\right)\right)\Biggr] + \boldsymbol{\zeta}_{-1}(0;r)\,, \\
\end{split}
\end{align}
\vspace{-1.0 cm}
\begin{align}
    \begin{split}
&\boldsymbol{\zeta}_1(\gamma;r) = -\frac{\pi}{4\sigma_1^3\sigma_2}\Biggl[2\sigma_1\sigma_2 \left(-\gamma e^{-\frac{r\sigma_2}{\gamma}} - r\sigma_2 \operatorname{Ei}\left(-\frac{r\sigma_2}{\gamma}\right)\right) \\
&\qquad + \sigma_2^2 e^{-r\sigma_1} \left(e^{2r\sigma_1} \operatorname{Ei}\left(-\frac{r(\gamma\sigma_1 + \sigma_2)}{\gamma}\right) - \operatorname{Ei}\left(r\left(\sigma_1 - \frac{\sigma_2}{\gamma}\right)\right)\right) \\
&\qquad + \sigma_1^2 e^{-r\sigma_1} \ln\left(\frac{\sigma_2 - \gamma\sigma_1}{\gamma\sigma_1 + \sigma_2}\right)\Biggr] + \boldsymbol{\zeta}_1(0;r)\,, \\
&\boldsymbol{\zeta}_{3}(\gamma;r) = \frac{\pi e^{-r(\frac{\sigma_2}{\gamma} + \sigma_1)}}{12 \sigma_1^5 \sigma_2^2}\Biggl[ \sigma_1^3 e^{r\sigma_1} \left(-2\gamma^3 + e^{\frac{r\sigma_2}{\gamma}}\left(2\gamma^3 - r^3\sigma_2^3 \operatorname{Ei}\left(-\frac{r\sigma_2}{\gamma}\right)\right) + \gamma r\sigma_2(\gamma - r\sigma_2)\right) \\
&\qquad - 6\sigma_1\sigma_2^2 e^{r\sigma_1} \left(\gamma + e^{\frac{r\sigma_2}{\gamma}}\left(r\sigma_2\operatorname{Ei}\left(-\frac{r\sigma_2}{\gamma}\right) - \gamma\right)\right) \\
&\qquad + 3\sigma_2^3 e^{\frac{r\sigma_2}{\gamma}} \left(e^{2r\sigma_1}\operatorname{Ei}\left(-\frac{r(\gamma\sigma_1 + \sigma_2)}{\gamma}\right) - \operatorname{Ei}\left(r\left(\sigma_1 - \frac{\sigma_2}{\gamma}\right)\right)\right) \\
&\qquad + e^{r(\frac{\sigma_2}{\gamma} + \sigma_1)} \ln\left(\frac{\sigma_2 - \gamma\sigma_1}{\gamma\sigma_1 + \sigma_2}\right) \\
&\qquad - 3\sigma_1^2\sigma_2 \left(e^{r\sigma_1} - 1\right) e^{\frac{r\sigma_2}{\gamma}} \ln\left(\frac{\sigma_2 - \gamma\sigma_1}{\gamma\sigma_1 + \sigma_2}\right) \Biggr] + \boldsymbol{\zeta}_3(0;r)\,.
\end{split}
\end{align}
In our case we need to evaluate the above integrals for $\gamma=1$, which can be obtained by taking $\gamma=1$ limit.
\section{List of integrals}\label{integrals}
\subsection*{${I}$-type integrals}
\begin{align}
\begin{split}
    %\tikzmarkin[disable rounded corners=true]{espqo}(0.7,-0.5)(-0.4,0.8)  
    &I_{1a}=-\frac{m_1^2}{64m_p^4}\int_q\frac{|q|\,\ e^{i\boldsymbol{q}\cdot \boldsymbol{r}}}{(\boldsymbol{q}^2+m_1^2)},\\&
    I_{1b}=-\frac{m_1^2}{64\pi m_p^4}\int_q\frac{e^{i\boldsymbol{q}\cdot \boldsymbol{r}}}{(\boldsymbol{q}^2+m_1^2)}\Big[2m_1+\Big(2m_1^2+\boldsymbol{q}^2\Big)\Big\{\frac{\arctan
    (q/m_1)}{q}\Big\}\Big],\\&
    I_{1c}=-\frac{m_1^2}{64\pi m_p^4}\int_q\frac{e^{i\boldsymbol{q}\cdot \boldsymbol{r}}}{(\boldsymbol{q}^2+m_1^2)}\Big[2m_1+\Big(2m_1^2+\boldsymbol{q}^2\Big)\Big\{\frac{\pi}{2q}-\frac{\arctan
    (2m_1/q)}{q}\Big\}\Big]\,,\\&
    I_2=-\frac{m_1^4}{64\pi m_p^4}\int_q\frac{e^{i\boldsymbol{q}\cdot \boldsymbol{r}}}{q^2(\boldsymbol{q}^2+m_1^2)}\Big[2m_1+\Big(2m_1^2+\boldsymbol{q}^2\Big)\Big\{\frac{\pi}{2q}-\frac{\arctan
    (2m_1/q)}{q}\Big\}\Big].%\tikzmarkend{espqo} 
    \label{4.15 m}\end{split}
\end{align}
After evaluating the integrals $I_{1a},I_{1b},I_{1c},I_2$, we get the following closed-form expressions
%\hfsetfillcolor{white!14}
%hfsetbordercolor{black}
\begin{align}
\begin{split}
    %\tikzmarkin[disable rounded corners=true]{az1}(14.9,0.4)(-0.4,0.8)
    \label{4.16y}  
      &I_{1a} =-\frac{m_1^2}{128\pi^2m_p^4}\Big[\frac{1}{r^2}-\frac{m_1\pi e^{-m_1r}}{2r}\Big]\,,\\&I_{1b}=-\frac{m_1^2}{64\pi m_p^4}\Bigg(\frac{\pi ^2 e^{-m_1 r} \left(m_1 r \left(-e^{2 m_1 r} \text{Ei}\left(-2 r m_1\right)+\log \left(2 m_1 r\right)+\gamma +2\right)+2\right)}{r^2}\Bigg)\,,\\&
    I_{1c}=\frac{d^2I_2}{dr^2},\\&
    I_2=\frac{-m_1}{64\pi m_p^4 r}\Bigg(\pi ^2 e^{-2 m_1 r} (e^{3 m_1 r} \text{Ei}\left(-3 r m_1\right)-2 e^{2 m_1 r} \left(2 m_1 r \text{Ei}\left(-2 r m_1\right)-1\right)\\&\hspace{4 cm}-e^{m_1 r} \left(\text{Ei}\left(-r m_1\right)+\log (3)\right)-2)\Bigg)\,.
      %\tikzmarkend{az1}
      \end{split}
\end{align}
%\newpage
\subsection*{${K}$-type integrals:}
\begin{align}
\begin{split}
    &K_{1a}=-\frac{m_2^2}{64m_p^4}\int_q\frac{|\boldsymbol{q}|\,\ e^{i\boldsymbol{q}\cdot \boldsymbol{r}}}{\boldsymbol q^2(\boldsymbol{q}^2+m_2^2)}\,,\\&
    K_{1b}=-\frac{m_2^2}{32\pi m_p^4}\int_q\frac{e^{i\boldsymbol{q}\cdot \boldsymbol{r}}}{\boldsymbol{q}^2(\boldsymbol{q}^2+m_2^2)}\Big[2m_1+\frac{\Big(2m_1^2+\boldsymbol{q}^2\Big)}{|\boldsymbol{q}|}{\arctan
    \left(\frac{|\boldsymbol{q}|}{m_1}\right)}\Big],\\&
    K_{1c}=-\frac{m_2^2}{32\pi m_p^4}\int_q\frac{e^{i\boldsymbol{q}\cdot \boldsymbol{r}}}{\boldsymbol{q}^2(\boldsymbol{q}^2+m_2^2)}\Big[2m_1+\frac{\Big(2m_1^2+\boldsymbol{q}^2\Big)}{|\boldsymbol{q}|}\arctan\left(\frac{|\boldsymbol{q}|}{2m_1}\right)\Big]\,,\\&
    K_{2a}=-\frac{m_2^2}{64m_p^4}\int_q\frac{|\boldsymbol{q}|\,\ e^{i\boldsymbol{q}\cdot \boldsymbol{r}}}{(\boldsymbol{q}^2+m_2^2)}\,,\\&
    K_{2b}=-\frac{m_2^2}{32\pi m_p^4}\int_q\frac{e^{i\boldsymbol{q}\cdot \boldsymbol{r}}}{(\boldsymbol{q}^2+m_2^2)}\Big[2m_1+\frac{\Big(2m_1^2+\boldsymbol{q}^2\Big)}{|\boldsymbol{q}|}{\arctan
    \left(\frac{|\boldsymbol{q}|}{m_1}\right)}\Big]\,,\\&
    K_{2c}=-\frac{m_2^2}{32\pi m_p^4}\int_q\frac{e^{i\boldsymbol{q}\cdot \boldsymbol{r}}}{(\boldsymbol{q}^2+m_2^2)}\Big[2m_1+\frac{\Big(2m_1^2+\boldsymbol{q}^2\Big)}{|\boldsymbol{q}|}{\arctan
    \left(\frac{|\boldsymbol{q}|}{2m_1}\right)}\Big]\,,\\&
    K_{3}=-\frac{m_2^4m_1^2}{16\pi m_p^4}\int_q\frac{e^{i\boldsymbol{q}\cdot \boldsymbol{r}}}{|\boldsymbol q|^3(\boldsymbol{q}^2+m_2^2)}\arctan\left(\frac{|\boldsymbol{q}|}{2m_1}\right)\,,\\& 
    K_{4a}=-\frac{m_2^2m_1^2}{64m_p^4}\int_q\frac{\,\ e^{i\boldsymbol{q}\cdot \boldsymbol{r}}}{|\boldsymbol{q}|(\boldsymbol{q}^2+m_2^2)}\,,\\&
    K_{4b}=-\frac{m_2^2m_1^2}{32\pi m_p^4}\int_q\frac{e^{i\boldsymbol{q}\cdot \boldsymbol{r}}}{q^2(\boldsymbol{q}^2+m_2^2)}\Big[2m_1+|\boldsymbol{q}|\arctan\left(\frac{|\boldsymbol{q}|}{m_1}\right)\Big]\,,\\&
    K_{4c}=-\frac{m_2^2m_1^2}{32\pi m_p^4}\int_q\frac{e^{i\boldsymbol{q}\cdot \boldsymbol{r}}}{q^2(\boldsymbol{q}^2+m_2^2)}\Big[2m_1+|\boldsymbol{q}|\arctan\left(\frac{|\boldsymbol{q}|}{2m_1}\right)\Big]\,,\\&
     K_{5}=-\frac{m_2^2m_1^2}{16\pi m_p^4}\int_q\frac{e^{i\boldsymbol{q}\cdot \boldsymbol{r}}}{|\boldsymbol{q}|(\boldsymbol{q}^2+m_2^2)}\Big[\arctan\left(\frac{|\boldsymbol{q}|}{m_1}\right)+\arctan\left(\frac{|\boldsymbol{q}|}{2m_1}\right)\Big]\,.
    \label{4.32n}\end{split}
\end{align}
\subsection*{$J$-type integrals}
\begin{align}
\begin{split}
    &J_{1a}=-\frac{m_1^2}{32m_p^4}\int_q\frac{|\boldsymbol{q}|\,\ e^{i\boldsymbol{q}\cdot \boldsymbol{r}}}{\boldsymbol{q}^2(\boldsymbol{q}^2+m_1^2)},\\&
    J_{1b}=-\frac{m_1^2}{32\pi m_p^4}\int_q\frac{e^{i\boldsymbol{q}\cdot \boldsymbol{r}}}{\boldsymbol{q}^2(\boldsymbol{q}^2+m_1^2)}\Bigg[m_1+m_2+\Bigg\{\Big(-m_1^2+m_2^2+\boldsymbol{q}^2\Big)\\&\frac{1}{q}\,\left[\arctan\!\Bigl(\frac{\boldsymbol{q}}{m_1}\Bigr)-\arctan\!\Bigl(\frac{\boldsymbol{q}}{m_2}\Bigr)\right]-\arctan\!\Bigl(\frac{2m_2|\boldsymbol{q}|}{-m_2^2+q^2}\Bigr)\Bigg\}\Bigg],\\&
    J_{1c}=-\frac{m_1^2}{32\pi m_p^4}\int_q\frac{e^{i\boldsymbol{q}\cdot \boldsymbol{r}}}{\boldsymbol{q}^2(\boldsymbol{q}^2+m_1^2)}\Bigg[m_1+m_2+\Bigg\{\Big(-m_1^2+m_2^2+\boldsymbol{q}^2\Big)\\&\frac{1}{q}\,\left[\arctan\!\Bigl(\frac{|\boldsymbol{q}|}{m_1-m_2}\Bigr)-\arctan\!\Bigl(\frac{2m_2|\boldsymbol{q}|}{m_1^2-m_2^2+\boldsymbol{q}^2}\Bigr)\right]\Bigg\}\Bigg].
   \end{split}
\end{align}
  \label{fig6} 
%\end{figure}
%The numerical procedure is as follows. We first integrate \(\displaystyle dr/d\rho=(r/\rho)\sqrt{g(r)}\) from \(\rho_{\min}\sim M\) down to \(\rho_{max}=500 M\), imposing \(r(\rho_{\max})=\rho_{\max}\) and then evaluate \(F(\rho)\), \(H(\rho)\) and the ansatz \(R(\rho)\) on a uniform \(\rho\) grid whose density (or number of sampling points) is taken to be approximately $N_\rho=6000$. We approximate derivatives by finite differencing and finally compute \(\mathcal{R}(\rho)\). %, and form \(\widetilde{\mathcal{R}}(\rho)\). 
%The best fit value of $c$ has been obtained by minimizing the $\rm L^2$ norm of the residual using the \textsc{scipy.optimize} module from Python library \textsc{SciPy}.

%The numerical profile (Fig.(\ref{fig6})) of \({\mathcal{R}}(\rho)\) shows it is effectively zero for \(\rho\gg M\), confirming that the ansatz successfully eliminates the \(e^{-\mu r}/r^2\) tail in isotropic coordinates (or solves the ODE at this order). The best fit value of $c$ is close to our calculated value of $-1/25=-0.04$ in Schwarzschild coordinates. This reiterates the fact that Love numbers as Wilson coefficients in EFT are gauge-invariant. Deviations of the residual near $\rho\sim M$ signal the breakdown of the asymptotic expansion.
\section{Scalar Love number for spherically symmetric solution of quadratic gravity}
\label{app_UV_fixing}
In this appendix, we consider a spherical symmetric black hole solution of the quadratic gravity found in \cite{Stelle:1977ry,Daas:2022iid} and repeat the computation performed in Section~(\ref{sec6}). We start with the following solution \cite{Stelle:1977ry,Daas:2022iid}, 
\[
\mathrm{d}s^2
\,=\,
-\,f(r)\,\mathrm{d}t^2
\;+\;
\frac{\mathrm{d}r^2}{g(r)}
\;+\;
r^2\,\mathrm{d}\Omega^2,
\]
with asymptotically flat metric functions \cite{Stelle:1977ry,Daas:2022iid}
\begin{align}
\begin{split}
f(r) &= 1 - \frac{2M}{r}
+ 2\,S_{2}\,\frac{e^{-m_{2}'\,r}}{r}
+ S_{0}\,\frac{e^{-m_{0}\,r}}{r},\\
g(r) &= 1 - \frac{2M}{r}
+ S_{2}\,\frac{(1 + m_{2}' r)\,e^{-m_{2}'\,r}}{r}
- S_{0}\,\frac{(1 + m_{0} r)\,e^{-m_{0}\,r}}{r},
\end{split}
\end{align}
where
\[
m_{2}'{}^{2} = \frac{1}{|\gamma|}, 
\qquad
m_{0}^{2} = \frac{1}{2\,(3\beta - |\gamma|)},
\]
and \(S_{2},S_{0}\) are free coefficients.  We again specialize to the following  three cases as discussed in the main text.
\subsection*{Choice I: \(\displaystyle \beta = \frac{3\,|\gamma|}{8}\)}
For \(\beta = 3|\gamma|/8\) one finds
\[
m_{2}' = \frac{1}{\sqrt{|\gamma|}},
\qquad
m_{0} = \frac{2}{\sqrt{|\gamma|}}
\;\;>\;\;
m_{2}'.
\]
Defining \(\mu \equiv m_{0}\) and absorbing \(2S_{2}\to G\,\tilde c_1\), the metric functions become
\begin{equation}
\begin{aligned}
f(r) &= 1 - \frac{2\,G\,M}{r}
+ G\,\tilde c_1\,\frac{e^{-\mu\,r}}{r},\\
g(r) &= 1 - \frac{2\,G\,M}{r}
+ G\,\tilde c_1\,\frac{(1 + \mu\,r)\,e^{-\mu\,r}}{r}.
\end{aligned}
\end{equation}
Since \(m_{0}>m_{2}'\), the spin-0 Yukawa term \(e^{-m_{0}r}/r\) decays faster than the spin-2 term \(e^{-m_{2}'r}/r\) at large \(r\) and may be dropped in the leading UV matching.

\subsection*{Choice II: \(\displaystyle \beta = \frac{|\gamma|}{2}\)}
For \(\beta = |\gamma|/2\) one obtains
\[
m_{2}' = \frac{1}{\sqrt{|\gamma|}},
\qquad
m_{0} = \frac{1}{\sqrt{|\gamma|}}
\,=\,m_{2}' \equiv \mu.
\]
In this case, both Yukawa exponents coincide, and the two corrections combine as
\begin{align}
f(r) &= 1 - \frac{2GM}{r}
+ (2\,S_{2}+S_{0})\,\frac{e^{-\mu\,r}}{r},\\
g(r) &= 1 - \frac{2GM}{r}
+ (S_{2}-S_{0})\,\frac{(1+\mu\,r)\,e^{-\mu\,r}}{r}.
\end{align}
Setting $2S_2+S_0=\tilde{c}_2$ and $S_2-S_0=\tilde{c}_3$, the metric takes the form
\begin{align}
f(r) &= 1 - \frac{2GM}{r}
+ G\tilde c_2\,\frac{e^{-\mu\,r}}{r},\\
g(r) &= 1 - \frac{2GM}{r}
+ G\tilde c_3\,\frac{(1 + \mu\,r)\,e^{-\mu\,r}}{r}.
\end{align}

\subsubsection*{Matching to EFT}

Collecting the coefficients of $e^{-\mu r}/r^2$ in $\mathcal L[R]$ using the Yukawa deformed Frobenius solution, where we define the three pieces of the differential operator $\mathcal L[R]$ (in Schwarzschild coordinates) as
\begin{equation}
\mathcal L[R]
=\,\underbrace{R''+\tfrac{2}{r}R'-\tfrac{2}{r^2}R}_{L_0[R]}
\;+\;
\underbrace{\tfrac12\,( \ln(fg) )'\,R'}_{I[R]}
\;\;
\underbrace{-\,\frac{2}{r^2}\Big(\frac{1}{g}-1\Big)\,R}_{II[R]}
\;=\;0.
\end{equation} 
gives
\begin{equation}
(d+k_\ell)M \;-\; \tilde{c}_1 \;+\; 2\tilde{c}_1
\;=\; 0
\quad\Longrightarrow\quad
(d+k_\ell)=-\frac{\tilde{c_1}}{M}.\label{c_UV}
\end{equation}
To proceed with the calculation of the UV charge, we recall that the part of the UV solution giving PN corrections to the source is unambiguously determined by the EFT calculation. Moreover, as determined in the previous section, the Love number $k_\ell=0$ only for the case $2\beta+\gamma=0$. From this, we can determine the UV charges. The result is summarized in the Table~(\ref{UV_charges_table}).
%The second relation gives the constraint on the parameter space of metric coefficient as
%\begin{equation}
 %   S_2=4S_0.
%\end{equation}
Note that even if the matching happens in two different coordinates (or gauges), since  EFT calculations are done in isotropic coordinates while the metric is in Schwarzschild coordinates, the result of the matching will remain unchanged. This follows from the fact that the Wilson coefficient in EFT (which gives Love numbers) is a manifestly gauge-invariant quantity. %Thus, the unique static dipole solution regular on the horizon and with the correct Yukawa‐tail is (with $8\beta+3\gamma=0$)
%\begin{equation}
%R_{\ell=1}(r)=r\Bigl(1-\frac{GM}{25\mu^2r^3}\Big)\Big\{1-(1+\mu\,r)\,e^{-\mu\,r}\Big\}+\mathcal{O}\Big(\frac{1}{r^4}\Big).
%\end{equation}

\begin{table}[htb!]
\centering
\begin{tabular}{lll}
\hline
Relation & $k_1$ & Parameter condition \\
\hline
$\tilde{c}_1 =-Mc_3 = -\dfrac{M}{25}$      & $\neq0$        & $8\beta + 3\gamma = 0$ \\[6pt]
$3\tilde{c}_3 - \tilde{c}_2 =0\implies S_2=4S_0$     & $0$        & $2\beta + \gamma = 0$ \\[6pt]
$\frac{3\tilde{c}_3 - \tilde{c}_2}{2M}\neq-Mc_3$      & $\neq0$        & $\beta + k\gamma = 0\Big(\frac{3}{8}<k<\frac{1}{2}\Big)$ \\
\hline
\end{tabular}
\caption{UV charges under different parameter constraints.}
\label{UV_charges_table}
\end{table}

\subsection{Numerical matching of the EFT and UV metric}
\begin{figure}[htbp]
  \centering
  \begin{subfigure}[b]{0.48\textwidth}
    \includegraphics[width=\textwidth]{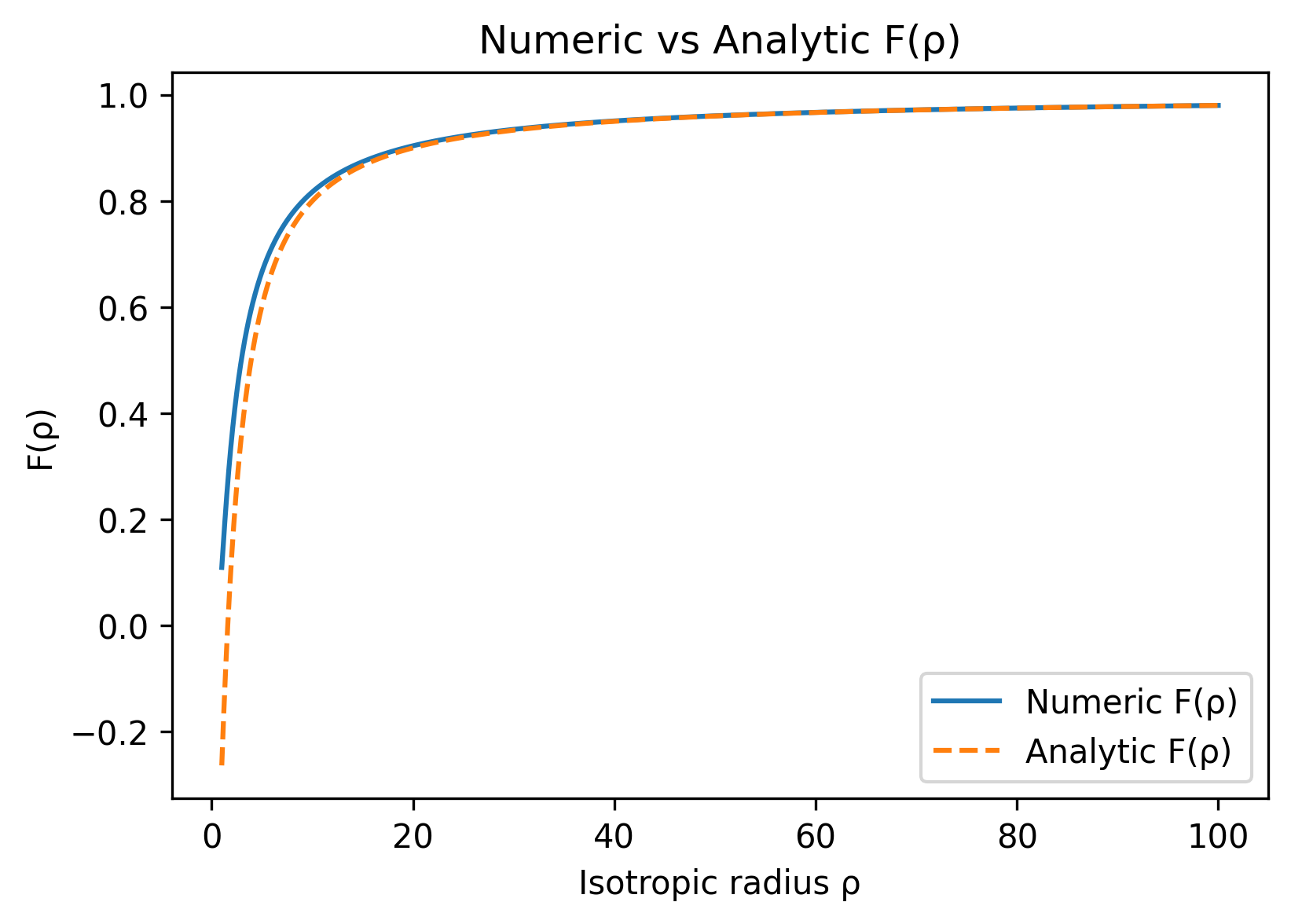}
    \caption{Numeric vs analytic $F(\rho)$}
    \label{fig:F_numeric_analytic}
  \end{subfigure}
  \hfill
  \begin{subfigure}[b]{0.48\textwidth}
    \includegraphics[width=\textwidth]{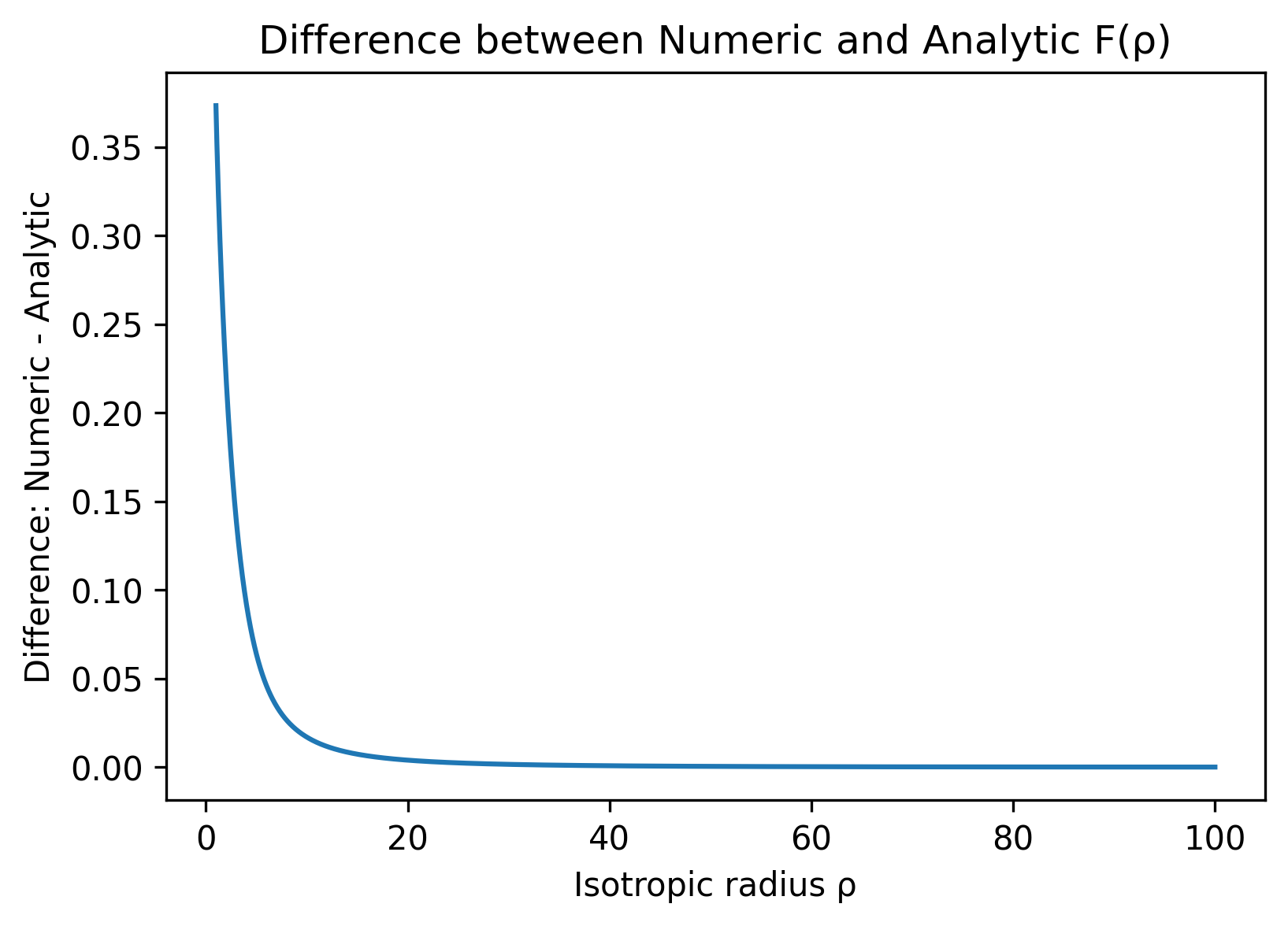}
    \caption{Difference: Numeric -- Analytic}
    \label{fig:F_difference}
  \end{subfigure}
  \caption{Comparison of the isotropic‐coordinate function $F(\rho)$ between UV metric with $\tilde{c}_1=-M/25$ and EFT reconstructed metric. (a) shows the numeric solution versus the analytic calculation (the EFT reconstructed metric), and (b) the residual. Both the analytic and numerical solutions overlap, and the residual between these two solutions is zero (both for large $\rho$), showing that they are in excellent agreement with each other in the asymptotic limit (which is the validity range for the EFT reconstructed metric).}
  \label{fig:F_comparison}
\end{figure}

\begin{figure}[htbp]
  \centering
  \begin{subfigure}[b]{0.48\textwidth}
    \includegraphics[width=\textwidth]{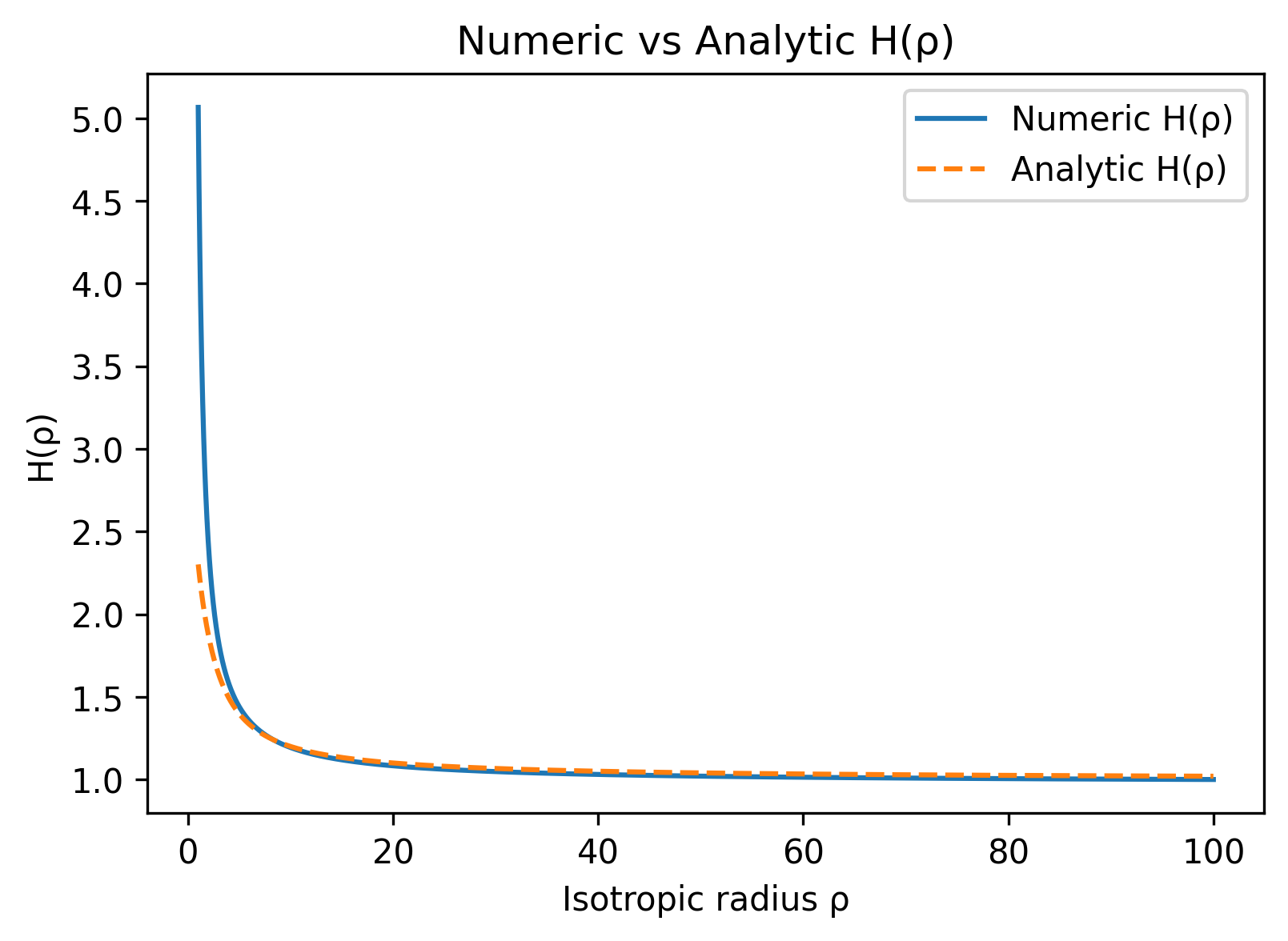}
    \caption{Numeric vs analytic $H(\rho)$}
    \label{fig:H_numeric_analytic}
  \end{subfigure}
  \hfill
  \begin{subfigure}[b]{0.48\textwidth}
    \includegraphics[width=\textwidth]{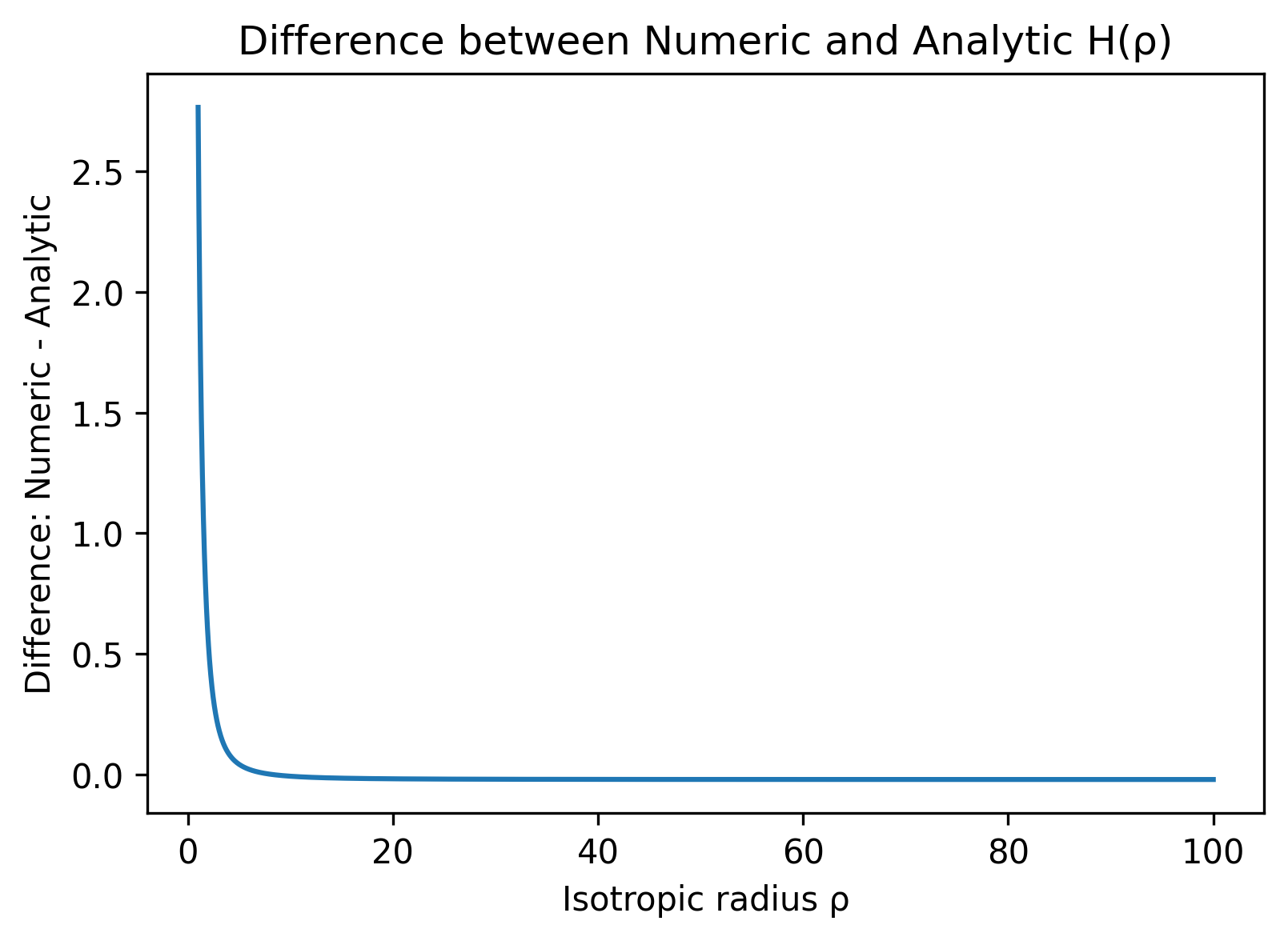}
    \caption{Difference: Numeric -- Analytic}
    \label{fig:H_difference}
  \end{subfigure}
  \caption{A similar plot for $H(\rho)$ showing excellent agreement between analytical and numerical solutions.}
  \label{fig:H_comparison}
\end{figure}

In this section, we match the EFT-reconstructed metric (called ``analytical'') (\ref{4.36t}) with the UV metric (presented in this appendix earlier) converted to isotropic coordinates with UV charge $\tilde{c}_1=-M/25$ obtained by Wilsonian matching. The purpose is to establish that the value obtained by Wilsonian matching is correct directly at the metric level, which should be the case. For this, we perform a numerical analysis. The results of numerical analysis are shown in Fig.~(\ref{fig:F_comparison}) and Fig.~(\ref{fig:H_comparison}). This unambiguously shows that our result is correct owing to the excellent matching of UV and EFT metric for large $\rho$ (here we have denoted the radial coordinate of the isotropic gauge by $\rho$).
\\
A few comments are in order:
\begin{itemize}
\item The background metric from UV theory (which is a fourth-order equation) has two free parameters $M$ and $\tilde{c}$ (in fact, three free parameters, two corresponding to the two massive modes), where $M$ is identified as the ADM mass, while $\tilde{c}$ is the Yukawa charge. In our EFT calculation, we only have one free parameter identified as $m$, identified as the ADM mass $M$. This is a direct consequence of the framework of EFT. In EFT of quadratic gravity, we are treating the higher curvature terms as a small perturbation (Wilson coefficient $\alpha \ll 1$) to Einstein's equation (a second-order ODE).    

The coefficients of these terms are parameters of the EFT itself and are suppressed by a high-energy scale. When we solve for a configuration, we start with the GR solution and then calculate the corrections to this solution, which are proportional to the EFT parameters $\beta, \gamma$. This determines how the ADM mass $M$ is modified by higher-derivative operators. Effectively, these higher-curvature corrections enter as the source on the right-hand side and not as a modification of the homogeneous part of the second-order operator. This means that under the perturbative expansion $g_{\mu\nu}=g_{\mu\nu}^{(0)}+\alpha\,h_{\mu\nu}+\mathcal{O}(\alpha^2)$, we have
    \begin{equation}
        \underbrace{\mathcal{D}^{(2)}}_\text{Lichnerowicz operator}[h_{\mu\nu}]=\underbrace{\mathcal{S}[g^{(0)}]}_\text{curvature-squared source}.
    \end{equation}
    In essence, the massive modes (corresponding to high energy) are suppressed by a cut-off scale $\Lambda$ which is order of their mass scale $m_0$. From the EFT perspective, they have been ``integrated out" while constructing the low-energy theory. Therefore, the equation of motion for light degrees of freedom (spin-2 massless graviton) remains effectively second-order but with modifications from the higher-derivative operators. However, one should not see this as EFT \emph{reducing} the integration constants, but rather as effectively not \emph{introducing} any new integration constant.

    \item The free parameter $\tilde{c}$ of the UV metric has been fixed by matching the Wilson coefficients (Love number). This can be directly fixed at the metric level by comparing the UV and EFT reconstructed metric (as discussed in this section). However, we emphasize that no matter which probe you choose (spin-0/1/2), the result after the matching will remain unchanged. This is because the Wilson coefficient is a gauge-invariant physical observable and is so $\tilde{c}$. Once we match the UV solution with EFT, the fixing is unique regardless of the probe and intermediate steps.
\end{itemize}
%\newpage
\bibliography{TLN}
\bibliographystyle{utphysmodb}
\end{document}